\documentclass[11pt,letterpaper]{article}
\usepackage[margin=1in]{geometry}
\usepackage[T1]{fontenc}
\usepackage{lmodern}
\usepackage{amsmath,amssymb,amsthm,mathtools}
\usepackage{microtype,booktabs,enumitem,longtable}
\usepackage{tikz}
\usetikzlibrary{arrows.meta,decorations.pathreplacing}
\definecolor{figureblue}{RGB}{35,76,108}
\definecolor{figureorange}{RGB}{156,78,20}
\tikzset{
  graph vertex/.style={circle,draw=black,fill=white,minimum size=6mm,
    inner sep=0pt,font=\normalsize},
  cut zero/.style={graph vertex},
  cut one/.style={graph vertex,fill=figureblue,text=white},
  graph edge/.style={draw=figureblue,line width=1pt},
  label change/.style={draw=figureorange,line width=1pt,dashed},
  figure text/.style={font=\normalsize,align=center}
}
\usepackage[colorlinks=true,linkcolor=blue,citecolor=blue,urlcolor=blue]{hyperref}
\usepackage[capitalize,noabbrev,nameinlink]{cleveref}
\hypersetup{pdftitle={Finding and Certifying Weighted Minimum Cuts with Cut Queries},
            pdfauthor={Troy Lee}}
\newtheorem{theorem}{Theorem}
\newtheorem{lemma}[theorem]{Lemma}
\newtheorem{proposition}[theorem]{Proposition}
\newtheorem{corollary}[theorem]{Corollary}
\theoremstyle{definition}
\newtheorem{definition}[theorem]{Definition}
\theoremstyle{remark}
\newtheorem{remark}[theorem]{Remark}
\theoremstyle{plain}
\newtheorem*{keystep}{Claim}
\AddToHook{env/lemma/begin}{\crefalias{theorem}{lemma}}
\AddToHook{env/proposition/begin}{\crefalias{theorem}{proposition}}
\AddToHook{env/corollary/begin}{\crefalias{theorem}{corollary}}
\AddToHook{env/definition/begin}{\crefalias{theorem}{definition}}
\AddToHook{env/remark/begin}{\crefalias{theorem}{remark}}
\newcommand{\R}{\mathbb R}
\newcommand{\cert}{\operatorname{cert}}
\newcommand{\iso}{\operatorname{iso}}
\newcommand{\vol}{\operatorname{vol}}
\newcommand{\ind}{\mathbf 1}
\newcommand{\cT}{\mathcal T}
\newcommand{\cA}{\mathcal A}
\newcommand{\cut}{c_G}
\newcommand{\OPT}{\operatorname{OPT}}
\newcommand{\sub}{\operatorname{sub}}
\newcommand{\out}{\mathrm{out}}
\newcommand{\inn}{\mathrm{in}}
\newcommand{\paths}{\mathcal P}
\newcommand{\Jprot}{J_{\mathrm{prot}}}
\newcommand{\Jaux}{J_{\mathrm{aux}}}
\newcommand{\lean}[1]{\nolinkurl{#1}}
\title{Finding and Certifying Weighted Minimum Cuts\\with Cut Queries}
\author{Troy Lee \thanks{Centre for Quantum Software and Information, University of Technology Sydney. \url{troyjlee@gmail.com}}}
\date{}
\begin{document}
\pagenumbering{roman}
\maketitle

\begin{abstract}
\normalsize
We study the deterministic and nondeterministic cut query complexity
of minimum cut in undirected graphs with arbitrary nonnegative edge weights.  For the nondeterministic
case, we show that every $n$-vertex graph has an $O(n\log^2(2n))$-query
certificate of its exact global minimum-cut value, such that any other nonnegatively weighted
graph on the same vertex set that agrees on these queries has the same minimum cut value.
A key ingredient is a way to certify a maximum $s$--$t$ flow value with only $O(n)$ cut queries.

For the deterministic case we give an algorithm that finds an exact global
minimum cut using $O(n^{3/2}\log^3(2n)(\log\log(2n))^2)$ cut queries.  Previously no $O(n^{2-\varepsilon})$
deterministic bound was known for arbitrary nonnegatively weighted graphs.
The algorithm follows Karger's two-respecting tree approach.  The key new ingredient
is a deterministic construction of a weak skeleton of the input graph $G$: an explicit weighted
graph $H$ in which every minimum cut of $G$ is a $5/4$-approximate minimum cut.
\end{abstract}

\thispagestyle{empty}
\clearpage
\pagenumbering{arabic}

\section{Introduction}\label{sec:introduction}
In this paper we study the problem of finding the global minimum-cut
value of an undirected graph with arbitrary nonnegative real edge
weights using cut queries.  A cut query to a graph $G=(V,w)$ specifies
some subset $S\subseteq V$ and returns the total weight
$c_G(S)=w(S,V\setminus S)$ of edges with exactly one endpoint in $S$.
Write $n=|V|\geq2$ and
$\lambda(G)=\min_{\varnothing\ne S\subsetneq V}c_G(S)$.
We give new upper bounds for computing $\lambda(G)$ in both the deterministic
and nondeterministic cut query models.  As usual in the cut query setting, we count only oracle calls; exact
arithmetic and offline computation are unrestricted.

Minimum cuts have repeatedly shown themselves to have a great deal
of structure that allows for efficient descriptions.  One result of
this form concerns the \emph{cut dimension}.  For a graph $G$ with
$m$ positive-weight edges, represent each minimum cut by the
characteristic vector of its edges in $\{0,1\}^m$.  The cut dimension
of $G$ is the dimension of the span of these vectors over all minimum
cuts.  Graur, Pollner, Ramaswamy, and Weinberg~\cite{GPRW20} introduced
the cut dimension as a lower-bound technique for the deterministic
cut-query complexity of global minimum cut.  They proved an
$n+1$ upper bound for the analogous dimension of minimum $s$--$t$
cuts, even in directed weighted graphs, but left open the question of how large
the cut dimension of global minimum cuts could be.  Lee, Li, Santha, and
Zhang~\cite{LLSZ21} then showed that the cut dimension of global
minimum cuts is at most $2n-3$, and that this bound is tight.
Faced with this obstruction to proving larger lower bounds, they
introduced the $\ell_1$-approximate cut dimension, a generalization
with the potential to give stronger lower bounds.  They showed an infinite family
of $n$-vertex graphs, with $n=3k+1$, whose $\ell_1$-approximate cut dimension is $2n-2$
and left showing larger lower bounds as an open question.

Later, Auza and Lee~\cite{AL21} showed that, up to an additive term
of at most one, the $\ell_1$-approximate cut dimension equals the
\emph{linear-query} certificate complexity of minimum cut.  A linear
query is more general than a cut query and may ask for any linear combination of the edge weights.
Here we study certificates restricted to cut queries, resulting in a potentially
even larger quantity that is still a lower bound on deterministic cut query
complexity.  A \emph{cut-query certificate} for a graph $G$ is a set of queries $\mathcal Q$ such
that every nonnegatively weighted graph $G'$ on the same vertex set
that agrees with $G$ on all queries in $\mathcal Q$ has the same
minimum-cut value.  The certificate queries may be chosen with full
knowledge of $G$.  Certifying that the minimum-cut value is at most
$\lambda(G)$ is easy: just query a shore realizing it.  The hard
part is certifying that no shore can achieve a smaller value.

We show that the cut-query certificate complexity of minimum cut is
$O(n\log^2(2n))$.   Consequently, neither cut-query certificate complexity nor 
$\ell_1$-approximate cut dimension can yield a larger deterministic query lower
bound for minimum cut.  This is yet another way in which minimum cuts are simple: 
a nearly-linear number of cut queries suffices to fix the minimum cut value in a graph.
The key ingredient in our proof is a way to certify the existence of an $s$--$t$ flow of a
given feasible value using just $O(n)$ cut queries.

With another lower bound route blocked off, we turn to deterministic algorithms
for finding a global minimum cut.  In the randomized case, this question has
essentially been solved.  We know randomized
$\widetilde O(n)$ cut query algorithms for both simple unweighted
graphs~\cite{RSW18} and weighted graphs~\cite{MN20}; for simple graphs the
bound was improved to $O(n)$ in~\cite{AEG+22}.

In the deterministic case, a simple graph can be reconstructed with
$O(n^2/\log n)$ cut queries by simulating the additive queries of
Grebinski and Kucherov~\cite{GK00}, and this was the best bound known
until Anand, Saranurak, and Wang~\cite{ASW} gave an
$\widetilde O(n^{5/3})$ algorithm.  Kenneth-Mordoch~\cite{KM26}
recently improved this to $O(n^{3/2}\log^2(2n))$ cut queries.

For arbitrary nonnegative real weights, however, the best previously
known upper bound was the trivial one from reconstructing the graph,
which uses $\Theta(n^2)$ cut queries, far from the lower bound of
$2n-2$.
We give a deterministic algorithm using
$O(n^{3/2}\log^3(2n)(\log\log(2n))^2)$ cut queries for the weighted case,
within a $\log(2n)(\log\log(2n))^2$ factor of the current best unweighted bound.
With the same number of queries the algorithm can return every
minimum cut of $G$, not just one.  

The recent deterministic algorithms for simple graphs~\cite{ASW,KM26}
use maximum-flow and minimum $s$--$t$-cut algorithms.  Our algorithm
instead follows the two-respecting-cut approach of
Karger~\cite{Karger00}.  Karger's nearly-linear time randomized algorithm finds
$O(\log n)$ spanning trees of the input graph such that, with high
probability, some minimum cut contains at most two edges of one of
these trees; such a cut \emph{two-respects} the tree.  This reduces
the problem to finding a minimum cut among those that two-respect a
given tree.  Gawrychowski, Mozes, and Weimann~\cite{GMW}, building
on Mukhopadhyay and Nanongkai~\cite{MN20}, gave a simple deterministic
algorithm for this fixed-tree problem.  While they did not explicitly state
the algorithm in the cut query model, it is easily adapted to this model: we show in
\cref{app:2respecting} how to implement their reduction with
$O(n\log n)$ deterministic cut queries, even when the supplied tree
has edges absent from the input graph.  Thus the remaining problem
is to find a small family of trees such that a minimum cut
two-respects at least one of them.

Karger obtains the trees by sparsifying the graph with random
sampling and using a spanning-tree packing~\cite{Karger00}.
We do not know how to construct such a sparsifier with a comparable
deterministic cut-query bound.  Instead, we construct a weaker
object, a \emph{weak skeleton}.  After handling zero minimum cuts
separately, we use $O(n^{3/2}\log^3(2n)(\log\log(2n))^2)$ cut queries to construct
an explicit nonnegative weighted graph $H$ on $V(G)$ such that
$\lambda(H)\geq\lambda(G)$ and every minimum cut shore $S^*$ of $G$
satisfies $c_H(S^*)\leq\tfrac54\lambda(G)$.  In particular, every minimum cut shore
$S^*$ of $G$ is a $5/4$-approximate minimum cut of $H$.
We then construct a family of $O(\log n)$
spanning trees of $H$ such that every $5/4$-approximate minimum
cut of $H$ crosses at most two edges of some tree in the family
(\cref{lem:tree-cover}).  This step uses only the
explicit graph $H$ and requires no further queries to $G$.
Searching these trees with cut queries to $G$ finds an exact minimum
cut in another $O(n\log^2 n)$ queries.  

\Cref{sec:certificate-overview} gives an overview of the certificate proof.
\Cref{sec:skeleton-overview} gives a roadmap of the deterministic
algorithm, and \cref{sec:conversion-overview} and \cref{sec:discovery-overview}
explain how we construct a weak skeleton via a tree cut sparsifier.
Complete proofs
are given in \cref{sec:preliminaries,sec:certificates,sec:skeletons,sec:hierarchy,sec:refinement,sec:discovery,app:2respecting}.

\paragraph{Lean Formalization}
As part of this work we have developed a library for cut-query
algorithms in Lean~4~\cite{Lee26}, available at
\url{https://github.com/troyjlee/cut-query-complexity}.  The library
formalizes the main certificate and deterministic cut-query theorems of
this paper together with their central ingredients, as well as many
previous cut-query results on global minimum cut, $s$--$t$ minimum cut
and all-pairs minimum cut.  We hope that this library will make it
easier for others working on cut-query algorithms to formalize their
results in Lean.  \Cref{app:lean} records the correspondence between
the named results of this paper and the Lean statements, including
differences in formulation and the few claims that are not formalized.

\section{Certifying a lower bound with few cut values}
\label{sec:certificate-overview}

Define $\cert(G)$ to be the minimum size of a family $\mathcal Q$
of cuts such that
\begin{equation}\label{eq:certificate-definition}
 c_{G'}(Q)=c_G(Q)\quad(Q\in\mathcal Q),\qquad w_{G'}\geq0
 \quad\Longrightarrow\quad \lambda(G')=\lambda(G).
\end{equation}
The competing graph $G'$ has the same vertex set and may put positive
weight on any pair, including a nonedge of $G$.
Equivalently, $\cert(G)$ is the nondeterministic cut-query complexity
of determining $\lambda(G)$ on input $G$: the query shores may be
provided by an all-knowing prover with knowledge of $G$, but their 
answers must determine the minimum-cut value uniquely.
In particular $\cert(G)$ is a lower bound on the number of cut
queries that any deterministic algorithm computing $\lambda$ on
nonnegatively weighted graphs makes on input $G$: the shores it asks
on $G$ are answered identically by every consistent $G'$, so its
output, which is $\lambda(G')$, equals $\lambda(G)$.

\begin{theorem}[Near-linear certificates]\label{thm:certificates}
Every nonnegatively weighted undirected graph $G$ on $n\geq2$ vertices satisfies
\[
 \cert(G)=O(n\log^2(2n)).
\]
The certificate can be chosen to contain a shore $S^*$ that remains
a minimum cut in every consistent nonnegatively weighted graph on
$V(G)$.  If $\lambda(G)=0$, the single query $S^*$ suffices.
\end{theorem}

Fix $\lambda=\lambda(G)$ and query an optimum shore for the
upper bound.  If $\lambda=0$, this already certifies the optimum
by nonnegativity.  Otherwise, the difficulty is to force every
nontrivial cut to have value at least $\lambda$.

For $s$--$t$ cuts there is a clear strategy to lower bound their value: a feasible $s$--$t$ flow of value $c$ certifies that every
$s$--$t$ cut has value at least $c$.  A first thought is to exhibit such a flow edge-by-edge, but in some weighted graphs 
every maximum $s$--$t$ flow uses $\Omega(n^2)$ edges, making this strategy too expensive.

Our key ingredient is a way to certify the existence of a feasible
$s$--$t$ flow of value $c$ using $O(n)$ cut queries.  We describe
the flow implicitly through an optimal dual solution to an auxiliary
minimum-cost flow problem.  The dual solution is represented by
one potential per vertex, and a carefully chosen tree $T$ encodes
these potentials.  Complementary slackness then forces every
non-tree edge to carry its full capacity in a direction determined
by $T$.  We learn the total incoming and outgoing non-tree flow at
each vertex using a constant number of cut queries.  These totals
determine the remaining divergence that the tree must supply so
that the full flow has divergence $c$ at $s$, $-c$ at $t$, and zero
elsewhere.  We explicitly learn the tree-edge capacities with
$O(n)$ cut queries and check that the uniquely determined tree
flow respects them.  We describe this certificate in the next
subsection.

We certify that every nontrivial cut has value at least $\lambda$
by combining the flow certificate with the isolating-cut framework of Li and
Panigrahi~\cite{LP25}.  The obstacle in adapting their terminal
reduction is that it relies on expansion guarantees, which we do not
know how to certify cheaply with cut queries.  We instead use a terminal-halving
argument whose key guarantee is witnessed by a feasible flow, so each
halving step can be certified with $O(n)$ cut queries.
Specifically, given a terminal set $U$, we retain at most half
its terminals in $R\subseteq U$ and certify a feasible flow sending $\lambda/2$
units from each terminal in $U\setminus R$ to $R$, thereby
certifying a lower bound of $\lambda$ for every shore avoiding
$R$ that contains at least two terminals of $U$.
Our construction is described in \cref{sec:li-panigrahi}.

\subsection{A tree describes a certifiable flow}

A \emph{signed flow} on $V$ is a function $f:V\times V\to\R$
satisfying $f_{uv}=-f_{vu}$ for all $u,v\in V$, where $f_{uv}>0$
means flow from $u$ to $v$.  The flow is \emph{feasible} in $G$ if
$|f_{uv}|\leq w_{uv}$ for all $u,v\in V$.  Its \emph{divergence}
at a vertex $v$ is the net outflow
\[
 b_v=\sum_{u\in V}f_{vu},
\]
the sum over $u$ of the flow from $v$ to $u$, so that a source has
positive and a sink negative divergence.  Every feasible flow satisfies
$c_G(S)\geq|\sum_{v\in S}b_v|$ for every cut.

The key novelty in our global mincut certificate is the following lemma.
\begin{lemma}[Flow certificates]
\label{lem:flow}
If a divergence vector $b\in\R^V$ admits a feasible flow in $G$,
then some labelled spanning tree $T$ of the complete graph on $V$
determines at most $4n-1$ cut queries certifying that $b$ admits a
feasible flow in every consistent graph.  The queried shores depend
only on $T$ and the fixed vertex labels.  In particular, an $s$--$t$
minimum-cut shore and value have a certificate of at most $4n$ queries.
For fixed disjoint nonempty terminal sets $A,B\subseteq V$, their
minimum separating-cut value has an $O(n)$ cut-query certificate.
\end{lemma}

To prove \cref{lem:flow}, label the vertices $0,\ldots,n-1$.
For each pair $\{u,v\}$ with $u<v$, assign the auxiliary cost
$a_{uv}=2^{un+v}$, and put $a_{vu}=-a_{uv}$.
The sum of the costs around any simple cycle is nonzero for every
choice of signs: the largest power of two exceeds the sum of all
smaller powers.  We then choose a feasible flow $f$ with divergence
$b$ minimizing $\sum_{u<v}a_{uv}f_{uv}$.

By complementary slackness for minimum-cost flows (see
Bertsekas~\cite[Sections~1.2.3 and~2.3]{Bertsekas91}),
there are vertex potentials $p_v$ such that, for the
\emph{reduced costs} $\rho_p(u,v)=a_{uv}-p_u+p_v$,
a negative reduced cost forces $f_{uv}=w_{uv}$ and a positive
reduced cost forces $f_{uv}=-w_{uv}$.
Thus every unsaturated edge has zero reduced cost.
The pairs with zero reduced cost form a forest: around any
cycle of such pairs, the identities $a_{uv}=p_u-p_v$ would
telescope to a zero signed cost sum, contrary to our choice
of costs.

The vertex potentials have one additive degree of freedom on each
component of the zero-reduced-cost forest: a common shift preserves
all potential differences, and hence all reduced costs, within that
component.  We use this freedom to extend the forest to a spanning
tree while preserving complementary slackness.
Choose a component and a pair crossing it with minimum absolute
reduced cost, and add the same constant to every vertex potential
in the component until that pair has reduced cost zero.  By this
choice of crossing pair, no reduced cost reverses sign, so $f$ still
satisfies the same optimality conditions.
Each step merges components; repeating produces a spanning tree $T$
of pairs with zero reduced cost.
Root $T$ at any vertex $r$ and normalize $p_r=0$.
The equations $p_u-p_v=a_{uv}$ along tree edges then determine
all vertex potentials, and hence every non-tree reduced cost.
Thus $T$ encodes the direction in
which each non-tree pair is saturated.  This applies even to pairs
absent from $G$, and $T$ may itself contain zero-capacity edges.
The full construction and proof appear in \cref{sec:flow-proof}.

It remains to check that this prescribed non-tree flow can be
completed on $T$ to have divergence $b$.  First query all singleton
cuts and the two-vertex cuts for tree edges.  By the well-known fact
that for two disjoint sets $A,B$
\begin{equation}\label{eq:overview-rectangle}
 2w_G(A,B)=c_G(A)+c_G(B)-c_G(A\cup B),
\end{equation}
these $n+(n-1)$ queries determine the capacity of every tree edge.
For each vertex $v$, let $O_v$ be its outgoing non-tree neighbors
as prescribed by $T$, and query $O_v$ and $O_v\cup\{v\}$.
The same identity gives the total outgoing non-tree flow at $v$.
Its total incoming non-tree flow is the singleton cut value minus
the incident tree capacities and the outgoing non-tree flow.
Thus at most $n+(n-1)+2n=4n-1$ queries determine all tree capacities
and the non-tree divergence $b^{\mathrm{nt}}$.

\begin{figure}[!htbp]
\centering
\begin{tikzpicture}[>=Latex]
  \begin{scope}
    \node[figure text] at (2.3,2.15)
      {(a) Tree and prescribed directions};
    \foreach \v/\x/\y in {0/0/1.05,1/1.35/0,2/0/-1.05,
                           3/4.6/1.05,4/3.25/0,5/4.6/-1.05}
      \node[graph vertex] (a\v) at (\x,\y) {$\v$};
    \foreach \u/\v in {0/1,1/2,1/4,3/4,4/5}
      \draw[graph edge] (a\u) -- (a\v);
    \draw[black!65,line width=0.6pt,dashed,->] (a0) to[bend right=35] (a2);
    \draw[black!65,line width=0.6pt,dashed,->] (a3) to[bend left=35] (a5);
    \draw[black!65,line width=0.6pt,dashed,->] (a3) to[bend right=18] (a0);
    \draw[black!65,line width=0.6pt,dashed,->] (a2) to[bend right=18] (a5);
    \node[figure text] at (2.3,-2.05)
      {$n+(n-1)+2n=4n-1$ queries};
  \end{scope}
  \begin{scope}[xshift=8.1cm]
    \node[figure text] at (2.3,2.15)
      {(b) Recover one tree-edge flow};
    \filldraw[fill=figureblue!6,draw=figureblue!45,rounded corners=5pt]
      (-0.48,-1.48) rectangle (1.95,1.48);
    \node[figure text] at (1.35,1.12) {$A$};
    \foreach \v/\x/\y in {0/0/1.05,1/1.35/0,2/0/-1.05,
                           3/4.6/1.05,4/3.25/0,5/4.6/-1.05}
      \node[graph vertex] (b\v) at (\x,\y) {$\v$};
    \foreach \u/\v in {0/1,1/2,3/4,4/5}
      \draw[graph edge] (b\u) -- (b\v);
    \draw[label change,->] (b1) -- node[above] {$e$} (b4);
    \node[figure text] at (2.3,-2.05)
      {$f^T_{14}=\displaystyle\sum_{i\in A}q_i,\qquad q=b-b^{\mathrm{nt}}$};
  \end{scope}
\end{tikzpicture}
\caption{A flow certificate described by a tree. Solid edges form $T$;
dashed arrows in (a) show selected saturated non-tree pairs.
Deleting $e=\{1,4\}$ leaves the component $A=\{0,1,2\}$.
Its residual divergence determines the signed tree flow from $1$ to $4$;
$b^{\mathrm{nt}}$ denotes the non-tree divergence.}
\label{fig:flow-tree}
\end{figure}
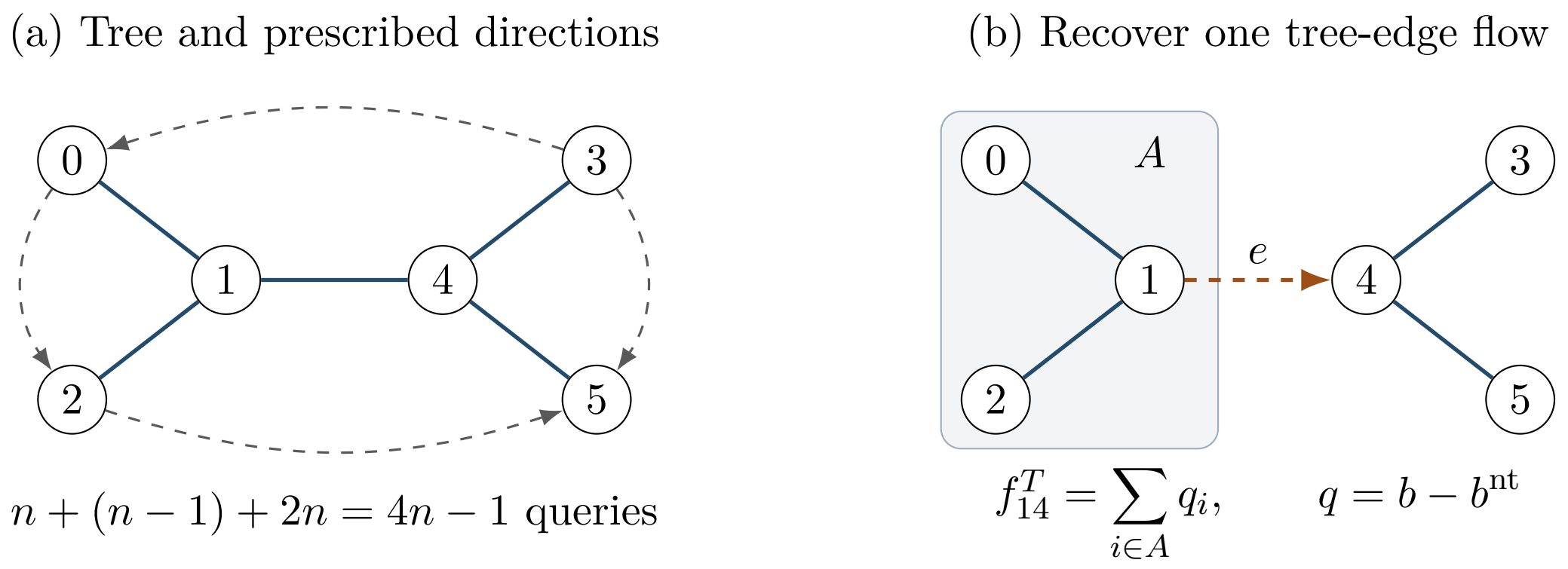

The residual divergence $q=b-b^{\mathrm{nt}}$ sums to zero and
determines a unique flow on $T$.  If deleting a tree edge $uv$ leaves
the component $A$ containing $u$, its flow from $u$ to $v$ must be
$\sum_{x\in A}q_x$.  Checking the absolute values of these $n-1$
flows against their measured tree capacities completes verification.
Figure~\ref{fig:flow-tree} illustrates this recovery from the tree.

In any graph $G'$ consistent with these answers, the tree capacities
and the measured non-tree divergence are unchanged.  Saturate its
non-tree pairs in the directions fixed by $T$, using their capacities
in $G'$, and add the same tree flow.  The resulting flow is feasible
and has divergence $b$.  This also handles new positive edges,
since $T$ already prescribes a direction for every non-tree pair.

For the $s$--$t$ conclusion of \cref{lem:flow}, certify a
maximum flow and query a matching minimum-cut shore.
This gives both bounds on the optimum using at most $4n$
queries.  If the optimum is zero, the shore query alone
suffices by nonnegativity.

\subsection{Isolating cuts and terminal halving}
\label{sec:li-panigrahi}

A cut \emph{separates} a terminal set $U$ if each shore
contains at least one vertex of $U$.  Start with $U=V$, so
every nontrivial cut separates $U$.
We maintain that any nontrivial cut of value below $\lambda$
in a consistent graph must separate $U$.
Each round has two steps.  First, we certify that every cut
isolating a single terminal of $U$ has value at least $\lambda$.
Any remaining cut below $\lambda$ therefore has at least two
terminals on each shore.  Second, we certify a feasible flow
that forces every cut of value below $\lambda$ with at least two
terminals of $U$ on each shore to separate a smaller terminal set
$R\subseteq U$, with $|R|\leq |U|/2$.
We then replace $U$ by $R$ and repeat.
Eventually $U$ is a singleton, which no cut can separate,
establishing the desired lower bound.

\paragraph{Certifying isolating cuts.}
For a terminal set $U$ and $v\in U$, an isolating cut separates
$v$ from $U\setminus\{v\}$.  We certify all these optimum values
using $O(n\log|U|)$ queries (\cref{lem:isolating}).
We combine the isolating-cut framework of Li and Panigrahi~\cite{LP25}
with the flow certificates above.
Give the terminals distinct binary labels.  For each bit, query a
minimum cut separating its zero and one classes and certify its
optimality by a flow.  Intersect the
chosen shores containing a terminal $v$ to obtain a cell $P_v$.
The cells are disjoint, and an uncrossing argument places an
inclusion-minimal optimum isolating shore inside $P_v$.  Because
the bit cuts are certified, this containment holds in every
consistent graph.

We can now certify the local flow problem in every cell.  Contract
the outside of $P_v$ to a single sink $t_v$.  A cut query to the
contracted graph is one cut query to $G$: a shore $X\subseteq P_v$
has value $c_G(X)$, and a shore containing $t_v$ has value
$c_G(P_v\setminus(X\setminus\{t_v\}))$.  A consistent graph contracts
to a consistent local graph.  The total size of these local networks
is $O(n)$ because the cells are disjoint, so their flow certificates
and optimum-shore queries cost $O(n)$ altogether; the
$\lceil\log_2|U|\rceil$ bit cuts account for the logarithm.
This is where the isolating-cut framework makes many lower-bound
witnesses affordable.

\paragraph{Halving the terminal set.}
For $|U|\geq2$, there is a nonempty
$R\subset U$, $|R|\leq|U|/2$, such that every discarded terminal
simultaneously sends $\lambda/2$ units to $R$, with zero divergence
outside $U$ (\cref{lem:halving}).  Certifying this divergence forces
\begin{equation}\label{eq:overview-halving}
 c_{G'}(S)\geq\frac{\lambda}{2}|S\cap U|
 \qquad\text{whenever }S\cap R=\varnothing.
\end{equation}

To obtain the flow, we split off the nonterminals.  For multigraphs,
the splitting-off theorems of Lov\'asz and Mader~\cite{Lov76,Mad78}
allow this: a nonterminal can be removed, replacing pairs of its
edges $su,sv$ by edges $uv$, without lowering the edge connectivity
between any two remaining vertices.  We need a fractional form for
arbitrary real capacities, which we prove directly
(\cref{lem:splitting}).  The resulting graph $J$
is supported on $U$, is dominated by $G$ on every cut, and retains
value at least $\lambda$ on every $U$-separating cut.  Choose
$X\subseteq U$ maximizing $c_J(X)$ and let $R$ be the smaller of $X$
and $U\setminus X$.  Every terminal outside $R$ has at least half its
incident weight going to $R$, since otherwise moving it across
would improve the maximum cut.  Its weighted degree is at least $\lambda$.
It follows that $c_G(S)\geq c_J(S)\geq
(\lambda/2)|S\cap U|$ for every shore avoiding $R$.
Max-flow/min-cut turns these inequalities into the simultaneous
flow in $G$.  Only its divergence is certified; the split-off graph
and maximum cut are chosen with full input knowledge.

Equation~\eqref{eq:overview-halving} shows that a cut below
$\lambda$ with at least two terminals of $U$ on each shore
must separate $R$: a shore avoiding $R$ would have value at
least $\lambda$.

Each of the at most $\lceil\log_2n\rceil$ rounds costs
$O(n\log|U|)$ queries for isolating cuts and $O(n)$ for the
halving flow, giving $O(n\log^2(2n))$ in total.
If $\lambda=0$, the queried optimum shore alone suffices by
nonnegativity.

\section{The deterministic algorithm: a roadmap}
\label{sec:skeleton-overview}

The cut certificate for $G$ of \cref{sec:certificate-overview} already yields
a \emph{skeleton}, a sparse graph $H$ with $O(n \log^2 n)$ edges and the
same vertex set as $G$
such that $\lambda(H)=\lambda(G)$ and a shore $S^*$ that realizes a minimum cut of $G$ also realizes a minimum cut of $H$
(\cref{prop:bridge}).  The certificate, however, is chosen with full knowledge
of $G$.  For our deterministic algorithm we settle for a weaker
object.

\begin{definition}[Weak skeleton]\label{def:weak-skeleton}
Let $G$ be a graph with $\lambda=\lambda(G)>0$.  An explicit
nonnegative weighted graph $H$ on $V(G)$ is a \emph{weak skeleton}
of $G$ if $\lambda(H)\geq\lambda$ and every minimum-cut shore $S^*$
of $G$ satisfies $c_H(S^*)\leq\tfrac54\lambda$.  In particular every
minimum cut of $G$ is a $5/4$-approximate minimum cut of $H$.
\end{definition}

Once $H$ is explicit, we follow Karger's nearly linear-time
algorithm~\cite{Karger00}: a tree-packing argument deterministically
produces $O(\log n)$ spanning trees of $H$ such that every
$5/4$-approximate minimum cut of $H$, and in particular $S^*$,
two-respects one of them (\cref{lem:tree-cover}).   We then adapt the algorithm of
Gawrychowski, Mozes, and Weimann~\cite{GMW} to the cut-query model to find a minimum
two-respecting cut of $G$ for each tree with $O(n\log n)$ cut queries
(\cref{thm:2respecting,app:2respecting}).  These two steps require
only modest adaptations of prior work.  The main novelty is
deterministically constructing the weak skeleton, which we focus on in this overview.

\paragraph{The pipeline.}
After testing connectivity of the positive-weight support of $G$ with
$O(n\log n)$ queries, and returning a zero cut if it is disconnected,
we may assume $\lambda(G)>0$.  The weak skeleton is then constructed
in two stages, with a parameter $\alpha$ to be chosen.
\begin{enumerate}[label=(\roman*),leftmargin=*]
\item \emph{Learn} (\cref{sec:discovery-overview}).  Using
 $O((n^2/\alpha)\log^3(2n)(\log\log(2n))^2)$ cut queries, contract vertex pairs that
 no minimum cut of $G$ separates, and find an explicit tree with
 $O(n)$ vertices that approximates every cut of the contracted graph
 within a factor $\alpha$.  Such a tree is an $\alpha$-tree cut
 sparsifier; the precise definitions follow in
 \cref{sec:roadmap-definitions}.  This stage works for every $\alpha$
 between $C\log^2(2n)(\log\log(2n))^2$ and $\sqrt n$, for an absolute
 constant $C$.
\item \emph{Convert} (\cref{sec:conversion-overview}).  Using
 $O(n\alpha\log^3(2n)(\log\log(2n))^2)$ further cut queries, turn that tree into a
 weak skeleton of the contracted graph, and lift it back to $G$.
\end{enumerate}
Balancing the two stages at $\alpha=\Theta(\sqrt n)$ gives the
following theorem.  For the finitely many $n$ with $\sqrt n$ below the
threshold $C\log^2(2n)(\log\log(2n))^2$, the graph is simply
reconstructed with $O(n^2)=O(1)$ queries.

\begin{theorem}[Deterministic weighted minimum cut]\label{thm:all-mincuts}
There is a deterministic algorithm that, given cut-query access to an
$n$-vertex undirected graph $G$, $n\geq2$, with arbitrary nonnegative
real edge weights, returns $\lambda(G)$ and every minimum-cut shore of
$G$ using
\[
 O\!\left(n^{3/2}\log^3(2n)(\log\log(2n))^2\right)
\]
cut queries.  If $\lambda(G)=0$, it returns the partition of $V$ into
the components of the positive support, whose proper unions are exactly
the minimum-cut shores.
\end{theorem}

Two features of the construction let the algorithm return every
minimum cut rather than a single one: a weak skeleton approximates
every minimum cut of $G$ at once, and the two-respecting search can
report every cut at a known level rather than a single lightest one
(\cref{thm:all-2respecting}).  The proof is given in
\cref{sec:complete-algorithm}.

The rest of this section fixes the definitions and notation needed for the weak skeleton construction.

\subsection{Quotients, hierarchies, and tree cut sparsifiers}\label{sec:roadmap-definitions}

\paragraph{Quotients and safe partitions.}
For a partition $\mathcal P$ of $V$, the \emph{quotient graph}
$G/\mathcal P$ has one vertex for each part of $\mathcal P$.  The weight between
distinct parts $A,B$ is the total weight $w_G(A,B)$ of edges joining
them in $G$; edges within a part are discarded.  Every shore
$\mathcal S\subseteq\mathcal P$ therefore satisfies
\[
 c_{G/\mathcal P}(\mathcal S)
 =c_G\!\left(\bigcup_{A\in\mathcal S}A\right),
\]
so one cut query to the quotient is one cut query to $G$.
A partition $\mathcal P$ of $V$ is \emph{safe} if no minimum cut
of $G$ separates two vertices in the same part.  Equivalently,
every minimum-cut shore of $G$ is a union of parts of $\mathcal P$,
and $\lambda(G/\mathcal P)=\lambda(G)$.  The learning stage contracts
only vertex pairs that no minimum cut separates, so its contracted
classes form a safe partition.

\paragraph{Tree cut sparsifiers.}
Let $T$ be a nonnegatively weighted tree with $V(G)\subseteq V(T)$.
The vertices in $V(G)$ are the \emph{terminals}; the other vertices
of $T$ are \emph{Steiner vertices}.  A shore $S\subseteq V(G)$ is
represented in $T$ by any vertex set $X$ with $X\cap V(G)=S$, and its
\emph{representation cost} is the cheapest such representation,
\[
 R_T(S)=\min_{\substack{X\subseteq V(T)\\X\cap V(G)=S}}c_T(X).
\]
We drop the subscript when the tree is clear from context.
A tree whose representation costs approximate every cut of $G$ is a
tree cut sparsifier~\cite[Definition~2.4]{ADK26}.

\begin{definition}[Tree cut sparsifier]\label{def:tree-cut-sparsifier}
Let $G=(V,w)$ and $\alpha\geq1$.  A nonnegatively weighted tree
$T$ with $V\subseteq V(T)$ is an \emph{$\alpha$-tree cut sparsifier}
of $G$ if
\begin{equation}\label{eq:overview-all-cut-hierarchy}
 c_G(S)\leq R_T(S)\leq\alpha c_G(S)
 \qquad\text{for every }S\subseteq V.
\end{equation}
\end{definition}

\paragraph{Hierarchies and canonical capacities.}
The tree cut sparsifiers we work with have a special form
introduced by R\"acke~\cite{Racke02}.  A \emph{hierarchy} of $G$ is a rooted tree
$T$ with leaf set $V(G)$ and no unary internal nodes, so that it has
fewer than $2n$ nodes.  Each nonroot node represents its descendant
leaf cluster $A$.  The hierarchy is \emph{canonical} if the parent
edge of each such node carries the capacity $d_A=c_G(A)$.  Any tree
cut sparsifier of $G$ can be made canonical with $O(n)$ queries
without weakening either bound in \eqref{eq:overview-all-cut-hierarchy} as described in
\cref{sec:conversion-overview}.

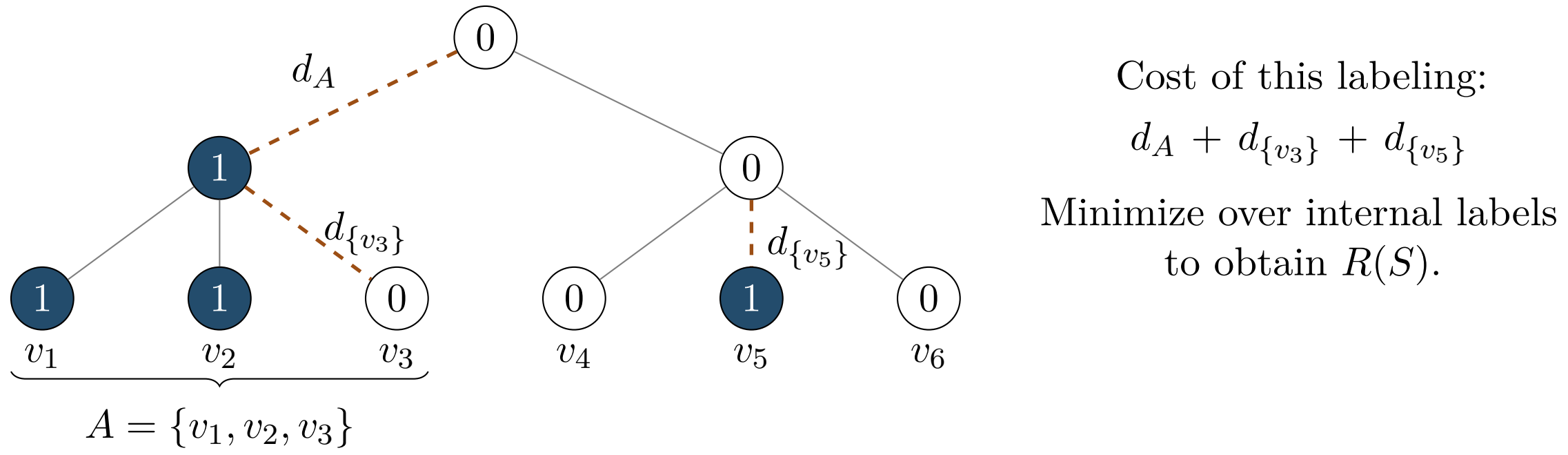
\begin{figure}[!htbp]
\centering
\begin{tikzpicture}
  \node[cut zero] (root) at (4.25,2.5) {$0$};
  \node[cut one] (left) at (1.7,1.25) {$1$};
  \node[cut zero] (right) at (6.8,1.25) {$0$};
  \foreach \i/\x/\s/\b in {1/0/one/1,2/1.7/one/1,3/3.4/zero/0,
                            4/5.1/zero/0,5/6.8/one/1,6/8.5/zero/0}
    \node[cut \s,label=below:{$v_{\i}$}] (leaf\i) at (\x,0)
      {$\b$};
  \draw[label change] (root) -- node[above left] {$d_A$} (left);
  \draw[gray] (root) -- (right);
  \foreach \i in {1,2} \draw[gray] (left) -- (leaf\i);
  \draw[label change] (left) -- node[right] {$d_{\{v_3\}}$} (leaf3);
  \foreach \i in {4,6} \draw[gray] (right) -- (leaf\i);
  \draw[label change] (right) -- node[right,pos=0.72] {$d_{\{v_5\}}$} (leaf5);
  \draw[decorate,decoration={brace,mirror,amplitude=4pt}]
    (-0.3,-0.7) -- (3.7,-0.7)
    node[midway,below=6pt] {$A=\{v_1,v_2,v_3\}$};
  \node[figure text,text width=5.3cm] at (12.05,1.2)
    {Cost of this labeling:\\[3pt]
     $d_A+d_{\{v_3\}}+d_{\{v_5\}}$\\[7pt]
     Minimize over internal labels\\
     to obtain $R(S)$.};
\end{tikzpicture}
\caption{Representing $S=\{v_1,v_2,v_5\}$ in a hierarchy.
Leaf labels are fixed by $S$; internal labels are chosen freely.
Each edge leading to a cluster $B$ has capacity $d_B=c_G(B)$.
The dashed edges join unequal labels and contribute to the cost.
The displayed labeling is one candidate in the minimization defining $R(S)$.}
\label{fig:hierarchy-cut}
\end{figure}

In a hierarchy, representing $S$ means labeling the leaves by
$\ind_S$ and extending the binary labels to internal nodes; $R_T(S)$
is the minimum total capacity of edges whose endpoint labels
disagree.  \cref{fig:hierarchy-cut} shows the cost of one
labeling.  Every original edge crossing $S$ crosses at least one of
the cluster boundaries in any labeling, so canonical capacities give
\begin{equation}\label{eq:overview-canonical-lower}
 c_G(S)\leq R_T(S)\qquad\text{for every }S\subseteq V(G)
\end{equation}
with no further assumption on the tree.  The algorithm will
repeatedly build explicit graphs $J$ from learned data, for instance
a learned subgraph of $G$, and need a canonical hierarchy of $J$ whose
representation costs approximate the cuts of $J$.  Since $J$ is
explicit, this is an offline computation: it costs no queries, and we
place no limit on its running time.  In this setting an algorithm of
R\"acke and Shah~\cite{RS14} gives, for every nonnegative weighted
graph $J$ on $r$ vertices, a canonical hierarchy with
$R(S)\leq a(r)\,c_J(S)$ for all $S$, where
$a(r)=O(\log(2r)\log\log(2r))$ (\cref{lem:offline-hierarchy}); since
offline computation is unrestricted, their construction may use exact
rather than approximate sparsest cuts, which is where these
logarithms come from.

\section{From a tree cut sparsifier to a weak skeleton}
\label{sec:conversion-overview}

The conversion stage proves the following theorem.

\begin{theorem}[From a tree cut sparsifier to a weak skeleton]\label{thm:conversion-overview}
Given cut-query access to a graph $G$ on $n$ vertices with
$\lambda(G)>0$ and an explicit $B$-tree cut sparsifier of $G$ with
$O(n)$ vertices, there is a deterministic algorithm that constructs a
weak skeleton of $G$ using
\[
 O\!\left(nB\log^3(2n)(\log\log(2n))^2\right)
\]
cut queries.
\end{theorem}

The full proof is in \cref{sec:hierarchy,sec:refinement}; this section
explains the ideas.  Throughout, $G$ denotes the current quotient, with
$\lambda=\lambda(G)>0$, and we are given a $B$-tree cut sparsifier of
$G$ with $O(n)$ vertices.  We first bring it into normal form: make
every terminal a leaf, contract non-terminal leaves into their
neighbours, root the tree at an internal node, and suppress unary
nodes, so that it is a hierarchy; then query the cut value $c_G(A)$
of each of its $O(n)$ clusters and make it the canonical capacity
$d_A$.  None of these steps increases a representation cost: a
labeling of the old tree restricts or extends to a labeling of the
new tree of no greater cost, and the edge above a cluster $A$
represents $A$ by itself, so its old capacity is at least
$R(A)\geq c_G(A)$.  Let $T$
be the resulting canonical hierarchy and write $R(S)=R_T(S)$.

Of the sparsifier guarantee \eqref{eq:overview-all-cut-hierarchy},
we keep the lower bound $c_G(S)\leq R(S)$ for every shore, which
\eqref{eq:overview-canonical-lower} gives for free, and from the
upper bound we retain only what it says about minimum cuts:
\begin{equation}\label{eq:overview-promise}
 R(S^*)\leq B\lambda
 \qquad\text{for every minimum-cut shore }S^*\text{ of }G.
\end{equation}
We call \eqref{eq:overview-promise} the \emph{promise} with bound
$B$.  Dropping the upper bound for all other cuts is deliberate: the
refinement in \cref{sec:conversion-refine} produces intermediate
trees that satisfy a promise with a smaller $B$ but need not have
all-cut distortion bounded by that smaller $B$.
Neither the minimum cuts nor $\lambda$ are known; all parameters
below depend only on the public bound $B$.  The analysis fixes one
minimum-cut shore $S^*$; since the construction never reads $S^*$,
every conclusion drawn for it holds for each minimum cut
simultaneously.

We first give a direct construction costing $O(nB^2\log n)$ queries.
We then show that a coarser version of the same construction halves
the promise bound with $O(nB\log^3(2n)(\log\log(2n))^2)$ queries,
linear in the current bound $B$.  Repeating it while the bound is
at least a polylogarithmic threshold, and finishing with the direct
construction below that threshold, costs a geometric series dominated
by its first term, which gives \cref{thm:conversion-overview}.

\subsection{A direct construction}

We want an explicit graph $H$ with $\lambda(H)\geq\lambda$ and
$c_H(S^*)\leq\tfrac54\lambda$.  We take $H$ to be a weighted sum of
two graphs, $F$ and $\Jprot$, described next.  The role of $F$ is to
approximate $G$: it is learned with cut queries, and its error on a cut
$S$ grows with the representation cost $R(S)$, so it is accurate at
the minimum cuts, which satisfy the promise, but may undervalue cuts
of large representation cost.
The role of $\Jprot$ is to compensate: it is built from the tree
without queries, and its cut values track $R(S)$, so it restores every
cut that $F$ may undervalue while adding little at $S^*$.

\paragraph{Ingredient 1: a completion accurate on cheap cuts.}
Fix an accuracy parameter $0<\eta<1$.  We learn a nonnegative graph
$F$ that agrees with $G$ on every cluster cut and nearly agrees on
every symmetric difference of two clusters:
\begin{equation}\label{eq:overview-interactions}
 c_F(A)=c_G(A),\qquad
 |c_F(A\mathbin\triangle B)-c_G(A\mathbin\triangle B)|\leq2\eta\min(d_A,d_B)
 \quad\text{for all clusters }A,B.
\end{equation}
Clusters are laminar, so $A\mathbin\triangle B$ is the union of two
disjoint clusters or the difference of two nested ones.  Writing $L_G$
for the Laplacian, so that $c_G(S)=\ind_S^{\mathsf T}L_G\ind_S$, and
expanding $c_G(A\cup B)$ or $c_G(B\setminus A)$ as a quadratic form
gives
\[
 \ind_A^{\mathsf T}L_G\ind_B
  =\begin{cases}
   \tfrac12\bigl(c_G(A\cup B)-c_G(A)-c_G(B)\bigr) & A\cap B=\varnothing,\\[2pt]
   \tfrac12\bigl(c_G(A)+c_G(B)-c_G(B\setminus A)\bigr) & A\subseteq B.
  \end{cases}
\]
Since $F$ matches every cluster cut, \eqref{eq:overview-interactions}
says that these \emph{cluster interactions} satisfy
$|\ind_A^{\mathsf T}(L_F-L_G)\ind_B|\leq\eta\min(d_A,d_B)$.
For each cluster $A$, the cut
value $d_A=c_G(A)$ is the total weight of the edges leaving $A$.
Charging each such edge to its endpoint inside $A$, or instead to its
endpoint outside $A$, gives two additive measures on vertices of total
mass $d_A$,
\[
 \mu_A^{\rm in}(X)=w_G(A\cap X,V\setminus A),
 \qquad
 \mu_A^{\rm out}(X)=w_G(A,X\setminus A),
\]
both queryable using the rectangle identity
\eqref{eq:overview-rectangle}.  For fixed
$A$ the interaction with $B$ is one of $-\mu_A^{\rm out}(B)$,
$\mu_A^{\rm in}(B)$, or $d_A-\mu_A^{\rm out}(B)$, according as $B$ is
disjoint from, inside, or contains $A$.  It therefore suffices to
learn each boundary measure of $A$ to within $\eta d_A$ on every
cluster $B$.

In depth-first leaf order every cluster is an interval of leaves, so
each such mass is a difference of two prefix masses, and it suffices
to control the prefix function of each boundary measure.  Fix a
cluster $A$ and one of its two boundary measures, and put
$q=\lceil2/\eta\rceil$.  Binary searches locate the $q$ prefixes at
which the measure crosses each multiple of $d_A/q$; recording the
prefixes on both sides of each crossing handles a single heavy edge.
Any nonnegative measure agreeing with these checkpoint values has
every prefix mass within $d_A/q$, hence every interval mass within
$2d_A/q\leq\eta d_A$.  Each prefix mass costs two cut queries, so the
two measures of a cluster cost $O(\eta^{-1}\log n)$ queries and all
$O(n)$ clusters cost $O(n\eta^{-1}\log n)$, regardless of the
hierarchy's depth.  Finally, $F$ is any nonnegative weighted graph whose cluster
cuts equal $d_A$ and whose boundary measures take the recorded
checkpoint values.  These are linear equalities in the weights of
$F$, and $G$ itself satisfies them, so the system is feasible.

The error on a cut is controlled by its representation cost.  Fix the
minimum labeling of $S$ and let $A_1,\ldots,A_m$ be the clusters whose
parent edge it cuts, so that $R(S)=\sum_i d_{A_i}$ and, since each
$d_{A_i}\geq\lambda$, $m\leq R(S)/\lambda$.  Walking from the root to
a leaf $v$, the label changes exactly at the cut edges above the
clusters containing $v$, so up to a constant vector
$\ind_S=\sum_i\sigma_i\ind_{A_i}$ with $\sigma_i\in\{-1,1\}$.
Laplacians annihilate constant vectors, so expanding
$c(S)=\ind_S^{\mathsf T}L\ind_S$ for $F$ and $G$ and applying
\eqref{eq:overview-interactions} to each of the $m^2$ terms gives
\begin{equation}\label{eq:overview-error}
 |c_F(S)-c_G(S)|
 =\Bigl|\sum_{i,j}\sigma_i\sigma_j\,
   \ind_{A_i}^{\mathsf T}(L_F-L_G)\ind_{A_j}\Bigr|
 \leq\sum_{i,j}\eta\,d_{A_i}
 =\eta\,m\,R(S)\leq\eta R(S)^2/\lambda .
\end{equation}
At a minimum cut the promise makes the error at most $\eta B^2\lambda$.  The
same bound identifies the difficulty: a cut with a costly
representation might become artificially cheap in $F$.  The next
ingredient takes care of this.

\paragraph{Ingredient 2: a protection graph to lower bound costly cuts.}
The second ingredient makes sure that no cut of $H$ dips below
$\lambda$, the minimum cut value of $G$.  By
\eqref{eq:overview-error}, the cuts at risk are those with large
representation cost $R(S)$, so we want to add a graph whose cut values
track $R(S)$.  Now $R(S)$ is itself a cut function, the minimum cut of
the tree $T$ separating the leaves in $S$ from the rest, but $T$ cannot
be added to $F$ directly because its internal nodes are not vertices of
$G$.  We therefore construct, without queries, an explicit graph
$\Jprot$ on the leaves of $T$ that mirrors the tree's cuts:
\begin{equation}\label{eq:overview-protection}
 R(S)\leq c_{\Jprot}(S)\leq\rho R(S)\qquad\text{for every }S,
\end{equation}
for an absolute constant $\rho$.  By Theorem~3.7 of Englert et
al.~\cite{EGKRTT}, the capacitated tree $T$ admits a constant-quality
vertex cut sparsifier on its leaves.  Since $T$ is explicit, we can
choose such a sparsifier deterministically by finite linear
feasibility, without further queries to $G$.  \Cref{lem:guard} gives a
self-contained proof with $\rho=42$.

\paragraph{Ingredient 3: the combination.}
Set $H=F+\sqrt\eta\,\Jprot$.  For every nontrivial cut $S$, write
$x=\sqrt\eta\,R(S)/\lambda$.  Nonnegativity of $F$ together with
\eqref{eq:overview-error} gives $c_F(S)\geq\max\{0,1-x^2\}\lambda$,
and \eqref{eq:overview-protection} gives
$\sqrt\eta\,c_{\Jprot}(S)\geq x\lambda$.  Hence
\begin{equation}\label{eq:overview-combination}
 \frac{c_H(S)}{\lambda}\geq\max\{0,1-x^2\}+x\geq1 .
\end{equation}
If $x\geq1$ the protection term alone pays; if $x<1$ then
$1-x^2+x\geq1$.  So $\lambda(H)\geq\lambda$.  For the promised
optimum, \eqref{eq:overview-error} and
\eqref{eq:overview-protection} give
\[
 \frac{c_H(S^*)}{\lambda}\leq1+\eta B^2+\rho\sqrt\eta\,B .
\]
Choosing $\eta=1/(100\rho^2B^2)$ makes the right side at most
$1+1/(100\rho^2)+1/10\leq5/4$.  As $S^*$ was an arbitrary minimum
cut, together with $\lambda(H)\geq\lambda$ this makes $H$ a weak
skeleton of $G$ (\cref{def:weak-skeleton}), using
$O(n\eta^{-1}\log n)=O(nB^2\log n)$ queries.

\subsection{Refining the promise}\label{sec:conversion-refine}

With $B=\sqrt n$ the direct construction would cost $\Omega(n^2\log n)$
queries.  It spends $\Theta(B^2)$ checkpoints per cluster because it must
approximate the optimum to additive $O(\lambda)$: the completion
error at $S^*$ is $\eta B^2\lambda$, which forces
$\eta\approx1/B^2$.

A \emph{refinement} round takes a canonical hierarchy of $G$ that
satisfies the promise with bound $B$ and outputs a new canonical
hierarchy of $G$ that satisfies the promise with bound $B/2$: every
minimum cut, still unidentified, has representation cost at most
$B\lambda/2$ in the new tree.  Since the output is a hierarchy and
not a skeleton, a round asks less of the completion.  The completion
need only track $G$ at $S^*$ to within a small fraction of $B\lambda$
rather than within $\lambda$, so we allow a larger error,
$\eta\approx1/B$, compute an offline hierarchy of the resulting
explicit graph, and query the original graph on the new cluster
boundaries.  The transfer inequality below shows that this halves the
promise bound at a cost linear in $B$.  Throughout, $a=a(n)=O(\log(2n)\log\log(2n))$ is the
approximation factor of the offline canonical hierarchy of R\"acke and
Shah: every explicit nonnegative weighted graph on $n$ vertices has a
canonical hierarchy with $R(S)\leq a\,c(S)$ for every shore $S$
(\cref{sec:roadmap-definitions}).

\begin{keystep}[Halving step; informal version of \cref{lem:refinement-round}]
Given a canonical hierarchy of $G$ with $O(n)$ clusters satisfying the
promise \eqref{eq:overview-promise} with bound $B\geq1/\gamma$, one can
compute a canonical hierarchy of $G$ with $O(n)$ clusters satisfying
the promise with bound $B/2$, using $O(na^2B\log n)$ cut queries,
where $\gamma=1/(100a^2\rho^2)$.
\end{keystep}

To see this, build $F$ and $\Jprot$ as above with $\eta=\gamma/B$, but
form $H=F+\gamma\Jprot$, with a heavier contribution from $\Jprot$ than in the
final skeleton.  Compute an offline hierarchy of this explicit $H$
with factor $a$; with the $H$-capacities $c_H(A)$ its representation
cost $R_H$ satisfies $R_H(S^*)\leq a\,c_H(S^*)$.  Then query the
original capacities $c_G(A)$ of its clusters to make it canonical for
$G$.  Since $c_H(S)\geq\gamma R(S)$ for every $S$, the completion
error \eqref{eq:overview-error} can be written in terms of $H$ alone:
\[
 c_G(S)\leq c_H(S)+\frac{\eta}{\gamma^2\lambda}\,c_H(S)^2 .
\]
Fix a labeling of $S^*$ that is optimal for the $H$-capacities, let
$A_1,\ldots,A_m$ be the clusters it cuts, and put $h_i=c_H(A_i)$, so
that $\sum_i h_i=R_H(S^*)$.  Reusing the same labeling with the
$G$-capacities, applying the displayed inequality to each $A_i$, and
using $\sum_i h_i^2\leq(\sum_i h_i)^2$ transfers it to the new
representation cost:
\begin{equation}\label{eq:overview-transfer}
 \frac{R_{\rm new}(S^*)}{\lambda}
 \leq\frac1\lambda\sum_i c_G(A_i)
 \leq x+\frac{\eta}{\gamma^2}x^2,
 \qquad x=\frac{R_H(S^*)}{\lambda}\leq a\bigl(1+\eta B^2+\gamma\rho B\bigr).
\end{equation}
There is again no depth factor.  With $\gamma=1/(100a^2\rho^2)$,
$\eta=\gamma/B$, and $B\geq1/\gamma$, the right side of
\eqref{eq:overview-transfer} is at most $B/2$.  The cost is that of
the completion, $O(n\eta^{-1}\log n)=O(na^2\rho^2B\log n)$.
Intermediate graphs $H$ serve only to produce better hierarchies;
they need not be weak skeletons themselves.

\paragraph{Adding up the queries.}
Repeat the halving step while the promise bound is at least
$1/\gamma=O(\log^2(2n)(\log\log(2n))^2)$.  The successive costs are geometric in
$B$ and sum to $O(na^2B\log n)$.  Once the remaining bound is
$b<1/\gamma$, the direct construction costs
$O(nb^2\log n)\leq O(nB\log n/\gamma)=O(na^2B\log n)$ as well,
since $b\leq B$.  Altogether, converting a supplied hierarchy with
bound $B$ costs $O(nB\log^3(2n)(\log\log(2n))^2)$ queries.
Every parameter above is computed from the public bound $B$ and
known quantities; the unknown $S^*$ and $\lambda$ enter only the
analysis, so the final $H$ is a weak skeleton of $G$.  This
completes the overview of \cref{thm:conversion-overview}.

\section{Learning a tree cut sparsifier}
\label{sec:discovery-overview}

The learning stage proves the following theorem.

\begin{theorem}[Learning a tree cut sparsifier]\label{thm:tree-sparsifier-discovery}
There is an absolute constant $C$ such that, for every nonnegatively
weighted graph $G$ on $n\geq2$ vertices with $\lambda(G)>0$ and every
$C\log^2(2n)(\log\log(2n))^2\leq\alpha\leq\sqrt n$, a deterministic algorithm uses
\[
 O\!\left(\frac{n^2}{\alpha}\log^3(2n)(\log\log(2n))^2\right)
\]
cut queries to find a safe partition $\mathcal P$ of $V(G)$ and an
explicit $\alpha$-tree cut sparsifier of the quotient $G/\mathcal P$
with at most $8n$ vertices.
\end{theorem}

The full proof is in \cref{sec:tree-sparsifier-proof}; this section
explains the ideas.  The algorithm proceeds in
phases, each of which either learns a tree cut sparsifier of the
current quotient or safely contracts it to at most half as many
vertices.  We call this the \emph{sparsify-or-shrink} principle.
After describing the weighted degree normalization applied in every
phase, we state the dichotomy as a lemma conditional on an auxiliary graph
(\cref{sec:discovery-dichotomy}), then show how to learn that
auxiliary graph with cut queries (\cref{sec:discovery-auxiliary}),
and finally add up the queries.

Throughout, $G$ denotes the current quotient, $N$ its number of
vertices, and $\lambda=\lambda(G)>0$.  We maintain a known
nontrivial cut of value $\Lambda\geq\lambda$.

\subsection{Weighted degree normalization}

Let $d_v$ be the \emph{weighted degree} of vertex $v$, i.e., the sum of the weights of the edges incident to
$v$.  All weighted degrees are learned once, with $n$ cut queries at
the start of the algorithm, and are then maintained without further
queries, since every contraction we perform is of an edge whose
weight we know.  We also maintain $\Lambda\leq\min_v d_v$: initially
$\Lambda$ is the minimum weighted degree, witnessed by a singleton
cut, and $\Lambda$ only ever decreases, to the weighted degree of a
newly contracted vertex or to the value of a newly found nontrivial
cut.  At the start of each phase, and after every change to
$\Lambda$, we normalize the current quotient so that its weighted
degrees satisfy
\begin{equation}\label{eq:overview-normalized}
 \lambda\leq \Lambda\leq d_v\leq(N-1)\Lambda\quad\text{for every } v.
\end{equation}
Whenever a vertex has $d_v>(N-1)\Lambda$, some edge incident to $v$
has weight greater than $\Lambda\geq\lambda$.  No minimum cut can
cross such an edge, so contracting it is safe.  We find such an edge
with $O(\log N)$ queries by binary search, repeatedly retaining the
half of the candidate neighbors with larger average incident edge
weight.  If the heavy edge is $\{u,v\}$, contracting $u,v$ gives new
weighted degree $d_u+d_v-2w_G(u,v)$; all other weighted degrees stay
unchanged.  Decrease $N$ by one, lower $\Lambda$ to the new vertex's
weighted degree if that is smaller, and repeat until
\eqref{eq:overview-normalized} holds.

The invariant \eqref{eq:overview-normalized} is used in three ways.  It makes the degree spread
polynomial in $N$, so that later scan counts and bucket counts are
$O(\log N)$ regardless of the smallest positive edge weight.  It
forces the vertex count to halve whenever $\Lambda$ drops by a
factor $4N$ (\cref{sec:discovery-dichotomy}).  And it is cheap:
weighted degrees are queried only once, every subsequent search
costs $O(\log n)$ queries and contracts a pair, and there are at most
$n-2$ contractions, so all normalization calls together cost
$O(n\log n)$ queries.  Every original minimum cut survives.

\subsection{Sparsify or shrink}\label{sec:discovery-dichotomy}

A phase with parameter $2\leq k<N$ seeks a tree cut sparsifier of
quality
\[
 B=O(k\log^2(2N)(\log\log(2N))^2).
\]
Suppose we have an explicit auxiliary graph $\Jaux$ on at most $2N$ vertices whose terminals are the vertices
of $G$ and which may have additional vertices.  For a shore
$S\subseteq V(G)$, define its minimum extension cost in $\Jaux$ by
\[
 \Psi(S)=\min_{\substack{X\subseteq V(\Jaux)\\X\cap V(G)=S}}c_{\Jaux}(X).
\]
The approximation guarantee we require has both a multiplicative and
an additive error:
\begin{equation}\label{eq:overview-aux-bounds}
 c_G(S)\leq\Psi(S)\leq A c_G(S)+\varepsilon,
 \qquad A=O(k\log(2N)\log\log(2N)),\qquad \varepsilon=\frac{\Lambda}{8N}.
\end{equation}

\begin{keystep}[Dichotomy; informal version of \cref{lem:general-certificate}]
Given $\Jaux$ satisfying \eqref{eq:overview-aux-bounds}, one phase
either outputs a $B$-tree cut sparsifier of $G$ with
$B=O(k\log^2(2N)(\log\log(2N))^2)$ using $O(N)$ further queries, or
safely contracts $G$ to at most $\lfloor N/2\rfloor$ vertices.
\end{keystep}

Compute offline the minimum cut of $\Jaux$ that separates two vertices of
$G$:
\[
 \mu=\min_{\varnothing\ne S\subsetneq V(G)}\Psi(S),
 \qquad \Psi(S_{\mathrm{aux}})=\mu.
\]
Requiring $S$ to be nontrivial excludes cuts separating only added
vertices.  Since $\Jaux$ is explicit, computing $\mu$ and $S_{\mathrm{aux}}$ requires
no queries to $G$.

\emph{Case $\mu\geq2\varepsilon$: sparsify.}
Every nontrivial shore $S$ satisfies
$2\varepsilon\leq\Psi(S)\leq A c_G(S)+\varepsilon$.
Thus $\varepsilon\leq A c_G(S)$, and the additive error is absorbed:
$\Psi(S)\leq2A c_G(S)$.  Apply the offline construction of R\"acke
and Shah to the explicit graph $\Jaux$, obtaining a hierarchy of all
its vertices with distortion $a=O(\log(2N)\log\log(2N))$.  Leave the labels of the added
vertices free and compress the tree to the terminal set $V(G)$ by
contracting non-terminal leaves into their neighbours and, on each
terminal-free path of degree-two vertices, all but a
minimum-capacity edge; this preserves every minimum extension of the
terminal labels.  Minimizing the hierarchy inequalities over the free
labels then gives a tree $T'$ with $O(N)$ clusters and
$\Psi(S)\leq R_{T'}(S)\leq a\Psi(S)$ for every $S\subseteq V(G)$.
Finally use $O(N)$ cut queries to replace each cluster capacity by its
$G$-cut value.  This lowers capacities while keeping every $G$-cut
dominated, so the resulting tree, with representation cost $R_G$
under these canonical $G$-capacities, satisfies
\begin{equation}\label{eq:overview-initial-bound}
 c_G(S)\leq R_G(S)\leq Bc_G(S)\quad(S\subseteq V(G)),
 \qquad B=2aA=O(k\log^2(2N)(\log\log(2N))^2).
\end{equation}
This is the desired tree cut sparsifier.

\emph{Case $\mu<2\varepsilon$: shrink.}
Query $S_{\mathrm{aux}}$ to obtain a smaller cut value
$\Lambda'=c_G(S_{\mathrm{aux}})\leq\mu<\Lambda/(4N)$, and normalize the vertex
degrees again to satisfy \eqref{eq:overview-normalized} with this
bound.  Suppose an old vertex remained a singleton in the resulting
quotient on $m\leq N$ vertices.  Its weighted degree would be
unchanged and at least the old $\Lambda$, whereas normalization gives
$d_v\leq(m-1)\Lambda'<\Lambda$, even if the bound decreased further.
Every final supervertex therefore contains at least two old vertices,
so $m\leq\lfloor N/2\rfloor$.  Each failed phase halves the vertex
count, independently of the weight spread.

Thus within $O(\log n)$ phases we either obtain the desired tree cut
sparsifier or reach $N\leq k$, when we reconstruct the quotient with
$O(N^2)$ queries and compute its tree cut sparsifier offline.

\subsection{Learning the auxiliary graph}\label{sec:discovery-auxiliary}

It remains to construct $\Jaux$ satisfying \eqref{eq:overview-aux-bounds}
with few queries.

\begin{keystep}[Auxiliary graph; informal version of \cref{lem:general-auxiliary}]
An explicit graph $\Jaux$ satisfying \eqref{eq:overview-aux-bounds}
can be learned with $O(Nk\log^2(2N)+(N^2/k)\log(2N))$ cut queries.
\end{keystep}

\paragraph{The weighted rounding obstacle.}
Learning $G$ exactly
costs $\Theta(N^2)$ queries, but the quotient of $G$ on a partition
of $V(G)$ into $p=O((N/k)\log N)$ parts costs only $O(p^2)$, and it
gives the exact value of every cut whose shore is a union of parts.
To estimate $c_G(S)$ for an arbitrary shore $S$ from the quotient we
\emph{round} $S$: replace it by a union $\widehat S$ of parts.  The
two shores disagree on the vertices of $S\mathbin\triangle\widehat S$:
those of $S\setminus\widehat S$ leave the shore and those of
$\widehat S\setminus S$ join it.  An edge changes between cut and
uncut only if exactly one of its endpoints is in
$S\mathbin\triangle\widehat S$, so the error of the estimate
$c_G(\widehat S)$ is at most the total weighted degree of these
vertices, their \emph{volume}
$\vol(S\mathbin\triangle\widehat S)=\sum_{v\in S\triangle\widehat S}d_v$.
We therefore want a partition, fixed before any cut is seen, such that every $S$ can be rounded with
$\vol(S\mathbin\triangle\widehat S)$ at most a bounded multiple of
$c_G(S)$; this multiple becomes the
multiplicative error $A$.  The rounding is then built into the
explicit graph: $\Jaux$ has one added vertex per part, labeling the
added vertices chooses $\widehat S$, and minimizing over their labels
rounds $S$ automatically (Step~4 below).  The question is what a label
change should cost.  Charging it to the full weighted degree of a
vertex of $S\mathbin\triangle\widehat S$ is prohibitively expensive:
if both endpoints of a learned edge of weight $M$ lie in
$S\mathbin\triangle\widehat S$, they contribute $2M$ to
$\vol(S\mathbin\triangle\widehat S)$ although the edge is cut by
$\widehat S$ exactly when it is cut by $S$, and $M$ may be far larger
than $c_G(S)$.  We therefore learn a subgraph $K$ of heavy edges exactly,
keep it on the original vertices, and charge label changes only to
degrees in the residual graph $R=G-K$, writing $r_v$ for the weighted
degree of $v$ in $R$ and $\vol_R(X)=\sum_{v\in X}r_v$.  The target is
that every shore $S$ admits a union $\widehat S$ of parts with
\begin{equation}\label{eq:overview-residual-rounding}
 \vol_R(S\mathbin\triangle\widehat S)\leq k(1+a)c_G(S)+Nt,
\end{equation}
where $a=O(\log(2N)\log\log(2N))$ is the offline hierarchy factor and $t$ is a
\emph{residual degree floor}: vertices of residual degree at most $t$
are simply charged their degree, which contributes at most $Nt$ in
total.  Choosing $t=\Lambda/(16N^2)$ makes $2Nt$ equal to the
additive slack $\varepsilon=\Lambda/(8N)$ allowed in
\eqref{eq:overview-aux-bounds}.  The four steps below first shape the
residual so that such a partition exists, then build it, then
assemble $\Jaux$.

\paragraph{Step 1: learn heavy edges so the residual is spread.}
We learn an exact subgraph $K$ so that, in the residual graph $R=G-K$,
every vertex either has residual weighted degree $r_v\leq t$ or has
every incident residual edge of weight at most $r_v/k$.  Thus a vertex
with substantial residual degree has its remaining weight spread
across at least $k$ neighbors.  Whenever $r_v>t$, find the edges
heavier than a $1/(2k)$ fraction of its current residual degree and
move them into $K$.  Rescan only after that degree halves.
Normalization bounds the ratio between the initial degree and $t$ by
$O(N^3)$, so there are $O(\log N)$ scans per vertex, regardless of the
smallest positive edge weight.  The cost is $O(Nk\log^2(2N))$, and
afterwards the \emph{row} of every vertex $v$, the list of residual
weights $w_R(v,u)$ over all $u$, satisfies
\[
 r_v\leq t\quad\text{or}\quad
 w_R(v,u)\leq r_v/k\quad\text{for every }u.
\]
Queries to $R$ are simulated by subtracting the known contribution of
$K$ from queries to $G$.

\paragraph{Step 2: learn the rows of a small center set.}
Next build a center set $C$ of size $O((N/k)\log N)$ with the
following property: every noncenter with residual degree above $t$
sends at least $r_v/k$ of its residual weight to $C$.  The row of
each center is learned as it is chosen, at $O(N)$ queries per row.
This is possible
because of Step 1: no single neighbor of such a vertex carries more
than a $1/k$ fraction of its residual weight, so a center set covering
a $1/k$ fraction of every such row exists, and a greedy argument finds
one of the stated size.  It can be found deterministically with
$O((N^2/k)\log(2N))$ queries.  Each vertex $v$ not yet covered carries
a weight $p_v=\prod_{c\in C}(1-kw_R(v,c)/(2r_v))$, which stays above
$1/2$ while $v$ is uncovered, and a candidate center $u$ scores
\[
 \sigma(u)=\frac k2\sum_{v\text{ uncovered}}\frac{p_v}{r_v}\,w_R(v,u).
\]
The scores sum to at least $(k/4)\sum_vp_v$, so some center decreases
the potential $\sum_vp_v$ by an $\Omega(k/N)$ fraction.  To find one
with cut queries, round the coefficients $p_v/r_v$ down to powers of
two.  The degree floor bounds their ratio by a polynomial in $N$, so
there are $O(\log N)$ buckets $A_j$ with common coefficient
$\theta_j$, and the total score of a candidate group $W$ is, up to a
factor $2$, $\sum_j\theta_j\sum_{v\in A_j}\sum_{u\in W}w_R(v,u)$.
Each inner double sum costs $O(1)$ cut queries, even when $A_j$ and
$W$ overlap (\cref{lem:bucketed-centers}), and binary search over
candidate groups finds a good center.  Thus center selection uses only
cut queries, without knowing the remaining individual edge weights.
This step is a deterministic weighted analogue of the choice of
centers in star contraction~\cite{AEG+22}.

\paragraph{Step 3: a rounding partition that moves little residual volume.}
Let $H_C^R$ contain the residual edges incident to $C$, all now
known.  Using the offline construction of R\"acke and Shah, build an
$a$-tree cut sparsifier of the explicit graph $K+H_C^R$, where
$a=O(\log(2N)\log\log(2N))$.  A tree can be compressed around a terminal set
while preserving every minimum extension of its labels: prune
branches without terminals, and on each terminal-free degree-two
path retain only a minimum-capacity edge.  Apply this with the
centers as terminals, adding one arbitrary vertex so that the terminal
set is nonempty.  The compression sends every vertex of $G$ to a node
of the compressed tree, and the vertices sent to the same node form a
part.  This is the rounding partition: it is fixed before any cut is
seen, and it has one part per node of the compressed tree, hence
$O(1+(N/k)\log N)$ parts.

Every cut $S$ can be rounded to a union $\widehat S$ of these parts
without changing its center labels, with
$c_{K+H_C^R}(\widehat S)\leq a c_{K+H_C^R}(S)$.  For the changed set
$X=S\mathbin\triangle\widehat S$, center coverage gives the target
bound \eqref{eq:overview-residual-rounding}.  Indeed, a changed vertex with residual degree above $t$ charges its
residual degree to its edges to centers, which carry at least a $1/k$
fraction of it.  Since no center changes labels, these edges cross
$S$ or $\widehat S$.  The remaining changed vertices contribute at
most $Nt$.

\begin{figure}[!htbp]
\centering
\begin{tikzpicture}
  \node[figure text] at (4.25,3.25)
    {Learned graph $K$: original vertices};
  \foreach \i/\x/\s/\b in {1/0/one/1,2/1.7/one/1,3/3.4/zero/0,
                            4/5.1/one/1,5/6.8/one/1,6/8.5/zero/0}
    \node[cut \s,label=above:{$v_{\i}$}] (original\i) at (\x,2.2) {$\b$};
  \foreach \i/\j in {1/2,2/3,3/4,4/5,5/6}
    \draw[graph edge] (original\i) -- (original\j);
  \node[cut one,label=left:{$p_1$}] (partA) at (0.85,0) {$1$};
  \node[cut one,label=below:{$p_2$}] (partB) at (4.25,0) {$1$};
  \node[cut zero,label=right:{$p_3$}] (partC) at (7.65,0) {$0$};
  \foreach \i/\p in {1/A,2/A,4/B,6/C}
    \draw[gray] (original\i) -- (part\p);
  \draw[label change] (original3) -- node[left] {$r_{v_3}$} (partB);
  \draw[label change] (original5) -- node[left] {$r_{v_5}$} (partC);
  \draw[graph edge] (partA) -- node[above] {$w_G(P_1,P_2)$} (partB);
  \draw[graph edge] (partB) -- node[above] {$w_G(P_2,P_3)$} (partC);
  \draw[graph edge] (partA) .. controls (2.55,-1.35) and (5.95,-1.35) ..
    node[below=2pt] {$w_G(P_1,P_3)$} (partC);
  \node[figure text] at (4.25,-1.85)
    {Quotient of $G$: one vertex per part};
  \node[figure text,text width=5.2cm] at (12.05,2.2)
    {Fixed original labels $S$\\[3pt]$c_K(S)$};
  \node[figure text,text width=5.2cm] at (12.05,0.95)
    {Changed labels: $v_3,v_5$\\[3pt]$r_{v_3}+r_{v_5}$};
  \node[figure text,text width=5.2cm] at (12.05,-0.35)
    {Part labels: $\widehat S=P_1\cup P_2$\\[3pt]$c_G(\widehat S)$};
\end{tikzpicture}
\caption{The auxiliary graph for parts $P_1=\{v_1,v_2\}$,
$P_2=\{v_3,v_4\}$, and $P_3=\{v_5,v_6\}$, with quotient vertices
$p_1,p_2,p_3$.
Each connector has capacity $r_v$; dashed connectors join unequal
labels. The displayed cut has total cost
$c_K(S)+c_G(\widehat S)+r_{v_3}+r_{v_5}$.
Minimizing over part labels gives $\Psi(S)$.
Thus only residual weighted degrees pay for label changes, while $K$ stays
on the original vertices.}
\label{fig:residual-auxiliary}
\end{figure}
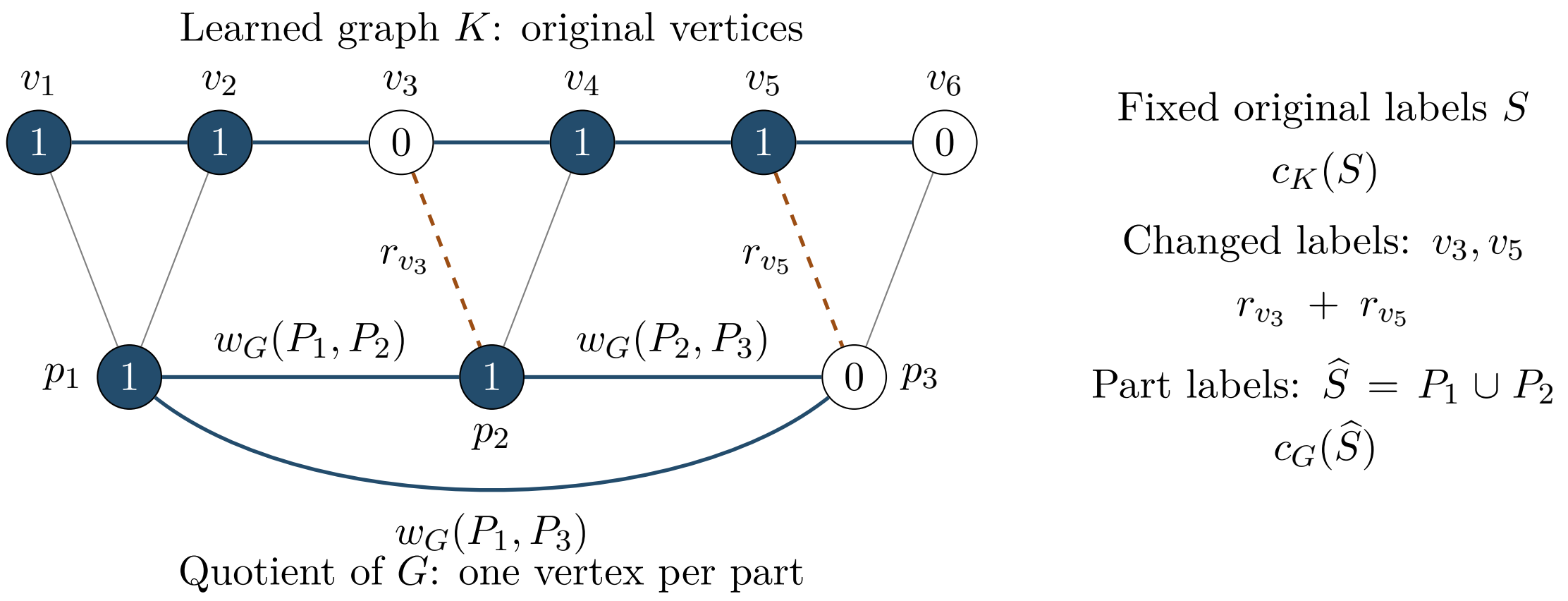

\paragraph{Step 4: assemble the auxiliary graph.}
As announced, $K$ stays on the original vertices and only the
residual is rounded.  Learn the
quotient of the whole graph $G$ on the partition of Step 3 with
$O(p^2)$ queries, $p$ the number of parts, and place it on one new
vertex per part.  Join each original vertex to its part vertex with
capacity $r_v$.  Keep $K$ on the original vertices.  For fixed
original labels $S$, the minimum extension cost in this explicit
auxiliary graph $\Jaux$ is
\begin{equation}\label{eq:overview-auxiliary}
 \Psi(S)=c_K(S)+
 \min_{\widehat S\text{ a union of parts}}
 \bigl(c_G(\widehat S)+\vol_R(S\mathbin\triangle\widehat S)\bigr).
\end{equation}
Figure~\ref{fig:residual-auxiliary} separates the three
contributions.

We verify \eqref{eq:overview-aux-bounds}.  For the lower bound, fix
any union of parts $\widehat S$.  The explicit $c_K(S)$ term accounts
for all learned edges, and changing the labels of
$S\mathbin\triangle\widehat S$ changes an $R$-cut by at most the
residual volume moved, so
$c_R(S)\leq c_R(\widehat S)+\vol_R(S\mathbin\triangle\widehat S)
\leq c_G(\widehat S)+\vol_R(S\mathbin\triangle\widehat S)$.
Adding $c_K(S)$ gives $c_G(S)\leq\Psi(S)$.  For the upper bound, take
$\widehat S$ to be the rounded cut of Step 3 and write
$X=S\mathbin\triangle\widehat S$, so that
$\Psi(S)\leq c_K(S)+c_G(\widehat S)+\vol_R(X)$, and bound the middle
term by the rounding guarantee of Step 3 and the same volume argument:
\[
 c_G(\widehat S)=c_K(\widehat S)+c_R(\widehat S)
 \leq a\,c_G(S)+c_R(S)+\vol_R(X).
\]
Hence $\Psi(S)\leq(1+a)c_G(S)+2\vol_R(X)$, and
\eqref{eq:overview-residual-rounding} gives
$A=(1+a)(2k+1)=O(k\log(2N)\log\log(2N))$ and $\varepsilon=2Nt$.  Unlike the
contraction classes of normalization, which replace the input by a
smaller quotient and must preserve every minimum cut, the rounding
partition need not be safe: it is used only to construct $\Jaux$, and
$\Psi$ optimizes over its labels separately for each cut.

\subsection{The query bound}

Write $L=\lceil\log_2(2N)\rceil$ and $K=\lceil\log_2(2L)\rceil$,
and choose an absolute constant $D$ such that a successful phase
has quality at most $DkL^2K^2$.
Set $k=\lfloor\alpha/(DL^2K^2)\rfloor$; the lower bound on $\alpha$
ensures $k\geq2$.  If $k\geq N$, reconstruct the current quotient
and compute its offline hierarchy, whose quality is at most
$\alpha$.  Otherwise the phase of \cref{sec:discovery-dichotomy}
either succeeds with quality at most $\alpha$ or halves $N$.

The rounding partition has $p\leq N$ parts and
$p=O((N/k)L)$, so quotient reconstruction costs
$O(p^2)=O(N^2L/k)$ queries.  Including the residual preparation,
center rows, and the test, a phase uses
\[
 O\!\left(NkL^2+\frac{N^2L}{k}\right)
 =O\!\left(N\alpha+\frac{N^2L^3K^2}{\alpha}\right)
\]
queries.  A final direct reconstruction costs at most $O(N\alpha)$.
The phase sizes decrease geometrically, and all normalization
calls together use $O(n\log(2n))$ queries.  Summing gives
$O(n\alpha+(n^2/\alpha)\log^3(2n)(\log\log(2n))^2)$, which is
$O((n^2/\alpha)\log^3(2n)(\log\log(2n))^2)$ because $\alpha^2\leq n$.
The contracted classes form a safe partition, proving the
claimed guarantees of \cref{thm:tree-sparsifier-discovery}.
Combined with \cref{thm:conversion-overview}, this completes the
overview of \cref{thm:all-mincuts}; the lift of the quotient skeleton
back to $V(G)$ and the final two-respecting search are in
\cref{sec:complete-algorithm}.

\label{page:overview-end}

\section{Preliminaries and query primitives}\label{sec:preliminaries}

An algorithm receives the vertex set and the exact cut oracle, and
only oracle calls are charged.  Computation on known quantities is
free, including exhaustive search and linear programming on explicit
graphs, and arithmetic is exact.  Weights are arbitrary nonnegative
reals, with no assumption on their magnitudes or ratios; parallel
edges are aggregated and loops ignored, and in the unweighted case
the input is a simple graph.  A certificate may be chosen with full
knowledge of $G$ but must be sound for every nonnegative graph
agreeing with the queried values, and its size is its number of
queries.  All algorithms are deterministic, with fixed tie-breaking,
and always correct.

We write $d_v=c_G(\{v\})$ for the weighted degree of $v$, the sum
of its incident edge weights, and $\vol_G(X)=\sum_{v\in X}d_v$.  For disjoint sets $A,B$,
\begin{equation}\label{eq:rectangle}
 2w_G(A,B)=c_G(A)+c_G(B)-c_G(A\cup B).
\end{equation}
Thus an individual edge weight or a partial-star total costs at most
three cut queries.  Once the weighted degrees are known, a row mass costs at
most two further queries:
\begin{equation}\label{eq:row-mass}
 2w_G(v,X)=d_v+c_G(X)-c_G(X\cup\{v\})
 \qquad(v\notin X).
\end{equation}
For general $X$, remove $v$ before applying this formula.
The values of empty and full cuts are known to be zero.

Two other elementary inequalities will be used repeatedly:
\begin{equation}\label{eq:cut-triangle}
 c_G(S\mathbin\triangle T)\leq c_G(S)+c_G(T),\qquad
 |c_G(S)-c_G(T)|\leq\vol_G(S\mathbin\triangle T).
\end{equation}
Both follow by checking each nonnegative edge contribution.
If an explicit subgraph $K$ with original edge weights has been learned,
cut queries to $G-K$ are simulated by subtracting $c_K(S)$ from
$c_G(S)$.  A query to a contracted graph is one original query on
a union of its classes.

\begin{lemma}[Relative-heavy row search]\label{lem:row-search}
Suppose a residual row has known total mass $M>0$ and queryable
nonnegative group masses.  For $b\geq1$, all its entries exceeding
$M/b$ can be found, with exact values, using $O(b\log(2n))$ cut queries.
\end{lemma}
\begin{proof}
Arrange the possible neighbors in a balanced binary tree.  Explore the
children of a group only when its mass exceeds $M/b$.  At each depth
fewer than $b$ disjoint groups pass this test.  Their children cost
$O(b)$ row-mass queries, each implemented by \eqref{eq:row-mass}.
Every entry above the threshold is reached at a leaf, and there are
$O(\log(2n))$ depths.  A row of mass zero needs no search.
\end{proof}

\begin{lemma}[Positive-support forest]\label{lem:forest}
A maximal spanning forest of the positive support can be learned
deterministically in $O(n\log(2n))$ cut queries.
\end{lemma}
\begin{proof}
Maintain the components of a partial forest.  Query the cut of a
component $A$.  If it is positive, find a positive edge from $A$ to
its complement by two balanced searches using \eqref{eq:rectangle}:
first locate its endpoint in $A$, then its endpoint outside $A$.
Learn the weight and merge the two components.  A component with
zero cut is finished and remains finished.  Initially there are $n$
components, each merge creates one new component, and each component
is tested at most once before being merged or finished.  There are
$O(n)$ tests and at most $n-1$ edge searches.
\end{proof}

In particular, disconnected positive support yields a zero optimum
within this query bound.  All skeleton constructions below use
$\lambda=\lambda(G)>0$.  The graph Laplacian is denoted by $L_G$, so that
$c_G(S)=\ind_S^{\mathsf T}L_G\ind_S$.

\section{Near-linear certificates}\label{sec:certificates}

A queried optimum shore supplies the upper bound.  We build the
matching lower bound from feasible divergences: first for one
terminal problem, then for all isolating cuts of a terminal set,
and finally for every nontrivial cut by halving that set.  Every
query in this section may depend on the entire input graph.

\subsection{Certifying flows and isolating cuts}\label{sec:flow-proof}

Label the vertices $0,\ldots,n-1$, orient each pair from its smaller
to its larger endpoint, and let $D$ be the incidence matrix, with the
convention that $Df$ is net outflow.  We also write a signed flow
antisymmetrically as $f_{uv}=-f_{vu}$, with $f_{uv}>0$ meaning flow
from $u$ to $v$.  Feasibility means $|f_{uv}|\leq w_{uv}$.

\begin{proof}[Proof of \cref{lem:flow}]
\emph{Fixed costs and a minimum-cost flow.}
For $u<v$ give the pair $uv$ the auxiliary cost
$a_{uv}=2^{un+v}$, and put $a_{vu}=-a_{uv}$.  These costs are fixed
before $G$ and $b$.  Every nonempty signed sum of distinct pair costs is
nonzero: its largest power of two exceeds the sum of all smaller
powers.  In particular, the signed cost around any simple undirected
cycle is nonzero.

Choose a flow $f$ minimizing
\[
 \sum_{u<v}a_{uv}f_{uv}
 \qquad\text{subject to}\qquad Df=b,\quad -w\leq f\leq w.
\]
A minimizer exists because the feasible polytope is nonempty and
compact.  A residual arc $u\to v$ is available whenever
$f_{uv}<w_{uv}$ and has cost $a_{uv}$.  There is no negative-cost
residual cycle, since sending a sufficiently small circulation around
one would decrease the objective while preserving feasibility and
divergence.

Attach a new source with zero-cost arcs to all vertices and let
$\ell_v$ be the shortest-path distance to $v$ in this residual graph.
These distances exist: removing nonnegative-cost closed subwalks
leaves a simple path, and there are only finitely many simple paths.
Set $p_v=-\ell_v$ and define the reduced costs
\[
 \rho_p(u,v)=a_{uv}-p_u+p_v.
\]
Every residual arc has nonnegative reduced cost.  Applying this to
both orientations gives the standard complementary-slackness
conditions (see Bertsekas~\cite[Section~1.2]{Bertsekas91}):
\begin{equation}\label{eq:flow-complementarity}
 \rho_p(u,v)<0\ \Longrightarrow\ f_{uv}=w_{uv},
 \qquad
 \rho_p(u,v)>0\ \Longrightarrow\ f_{uv}=-w_{uv}.
\end{equation}

\emph{A tree determines the saturation directions.}
Consider the graph of all pairs with zero reduced cost, including
pairs of capacity zero.  It is a forest: a cycle of zero reduced-cost
pairs would have zero signed cost, since the potentials telescope.
If this forest is disconnected, let $A$ be one of its components.
Among the ordered pairs $u\in A$, $v\notin A$, choose one minimizing
$|\rho_p(u,v)|$, and put $\tau=\rho_p(u,v)\ne0$.  Add $\tau$ to
every potential in $A$.  Reduced costs within $A$ or its complement
are unchanged; a reduced cost directed out of $A$ decreases by
$\tau$.  By the choice of $|\tau|$, no nonzero reduced cost changes
to the opposite sign, and the chosen pair becomes zero.  All old
zero pairs remain zero, so the zero forest strictly grows.
Equation~\eqref{eq:flow-complementarity} still holds for the same flow.
Repeating produces a spanning tree $T$ of zero reduced-cost pairs.

Normalize the final potential to be zero at vertex $0$.  It is now
the unique potential $p^T$ satisfying
\[
 p^T_0=0,\qquad p^T_u-p^T_v=a_{uv}\quad(uv\in T),
\]
obtained by summing the fixed costs along tree paths.  Thus $p^T$
depends only on $T$.  Write $\rho_T=\rho_{p^T}$.  Its zero pairs are
exactly the tree edges, and \eqref{eq:flow-complementarity} specifies
the flow on every non-tree pair:
\begin{equation}\label{eq:tree-prescribed-flow}
 f_{uv}=-\operatorname{sgn}(\rho_T(u,v))\,w_{uv}
 \qquad(uv\notin T).
\end{equation}
The tree need not lie in the positive support of $G$, and some of its
edges may themselves be saturated.

\emph{The queries and their verification.}
For each vertex $v$, the tree determines the outgoing non-tree set
\[
 O_T(v)=\{u\ne v:uv\notin T,\ \rho_T(v,u)<0\}.
\]
Query all singleton shores, all two-vertex shores $\{u,v\}$ with
$uv\in T$, and the two shores $O_T(v)$ and $O_T(v)\cup\{v\}$ for
each $v$.  This is at most $n+(n-1)+2n=4n-1$ cut queries.  By
\eqref{eq:rectangle}, the singleton and two-vertex answers determine
every tree capacity.  The divergence $b^{\mathrm{nt}}_v$ of the
prescribed non-tree flow is twice its outgoing capacity minus its
total incident capacity.  Substituting \eqref{eq:row-mass} gives
\begin{equation}\label{eq:tree-nontree-divergence}
 b^{\mathrm{nt}}_v
 =c_G(O_T(v))-c_G(O_T(v)\cup\{v\})
   +\sum_{u:uv\in T}w_{uv},
\end{equation}
which is determined by the answers.

In a consistent graph $G'$, retain $f$ on the tree and use
\eqref{eq:tree-prescribed-flow} with the new capacities on every
non-tree pair.  Tree capacities are unchanged, so this flow is
feasible.  Equation~\eqref{eq:tree-nontree-divergence} preserves its
non-tree divergence, and hence its full divergence is still $b$.
This also handles new positive edges: their prescribed directions
were already determined by $T$.

The verifier does not need the original tree-edge flow values.
Given $b$ and the answers, its tree flow is the unique one with
divergence $b-b^{\mathrm{nt}}$.  Explicitly, if deleting $uv$ from
$T$ leaves the component $A$ containing $u$, then its value from
$u$ to $v$ is
\[
 \sum_{x\in A}(b_x-b^{\mathrm{nt}}_x).
\]
Check that the residual divergences sum to zero and that each of
these $n-1$ values respects its measured tree capacity.  These checks
are sufficient for feasibility, as follows by successively removing
leaves of the tree.  Thus $T$ fixes all queried shores and all
non-tree directions; $b$ and the answers determine the tree flow.

For distinct $s,t$, take the divergence of a maximum $s$--$t$ flow
and add a query for a matching minimum shore $S^*$.  Its answer
determines $b_s=c_G(S^*)$, $b_t=-c_G(S^*)$, and all other entries
zero.  The feasible divergence supplies the lower bound in every
consistent graph, and the queried shore supplies the equal upper
bound.  The certificate is described by $(T,S^*)$ and uses at most
$4n$ queries.

For the terminal-set assertion, add a source joined to every
vertex of $A$ and a sink joined to every vertex of $B$, each added
edge with a fixed capacity greater than $\sum_e w_e$.  A maximum
flow in this augmented network has value $\nu$ equal to the minimum
$A$--$B$ separating-cut value in $G$.  Its restriction to the
original edges is a feasible flow in $G$ whose divergence $b$ is
nonnegative on $A$, nonpositive on $B$, zero elsewhere, and sums to
$\nu$ over $A$.  Certify $b$ by the construction above, with at most
$4n-1$ queries to $G$.  In every consistent graph $b$ admits a
feasible flow, so every shore $S$ with $A\subseteq S$ and
$S\cap B=\emptyset$ satisfies $c(S)\geq\sum_{v\in S}b_v=\nu$.  Query
one original minimum $A$--$B$ cut for the matching upper bound.
Alternatively, the construction can be run on the augmented network
itself, whose added capacities are constants; each of its cut
queries is then a bounded number of cut queries to $G$
(\cref{rem:local-flow}).
\end{proof}

The certified object is the divergence $b$, not the chosen flow $f$.
The proof constructs a possibly different capacity-respecting flow
for each consistent graph.  This is exactly the lower-bound witness
needed for terminal cuts, even when the competitor adds new edges.

\begin{remark}
\label{rem:local-flow}
The same argument applies to a network on a subset $W\subseteq V$
whose unknown edges are the original edges within $W$, together
with a constant number of auxiliary vertices and fixed-capacity
edges.  It uses $O(|W|+1)$ cut queries on $G$: the unknown part of
each local cut is $w_G(A,W\setminus A)$ for some $A\subseteq W$,
which takes at most three original cut queries by
\eqref{eq:rectangle}; the fixed contributions are added exactly.
The fixed capacities need not dominate the capacities of a competitor.
\end{remark}

For a terminal set $R\subseteq V$, $|R|\geq2$, and $v\in R$, put
\[
 \iso_G(v;R)=\min\{c_G(S):S\cap R=\{v\}\}.
\]

\begin{lemma}[Isolating-cut certificates]
\label{lem:isolating}
For every $R\subseteq V$ with $|R|\geq2$, all the values
$\iso_G(v;R)$, $v\in R$, can be certified simultaneously
with $O(n\log|R|)$ cut queries.
\end{lemma}

\begin{proof}
Use the isolating-cut construction of Li and
Panigrahi~\cite{LP25}.  Give the terminals distinct binary
labels of length $k=\lceil\log_2|R|\rceil$.  For each bit,
choose a minimum cut separating the terminals with bit zero
from the terminals with bit one.  Certify that cut's optimality
by \cref{lem:flow}, using $O(nk)$ queries in total.  Each chosen
cut is therefore still optimal for its terminal partition in
every consistent graph.

For $v\in R$, let $P_v$ be the intersection of the chosen
bit-cut shores containing $v$.  The cells $P_v$ are disjoint
and satisfy $P_v\cap R=\{v\}$.  An inclusion-minimal minimum
isolating shore for $v$ lies in $P_v$.  Indeed, if such a shore
$S$ is not contained in a bit-cut shore $T$ containing $v$,
submodularity gives
\[
 c_G(S\cap T)+c_G(S\cup T)\leq c_G(S)+c_G(T).
\]
The intersection still isolates $v$, and the union still
separates the two terminal classes for this bit.  Their
optimality inequalities force equality, making $S\cap T$ a
strictly smaller minimum isolating shore, a contradiction.
The same argument applies in every consistent graph because
the bit cuts remain optimal there.

For each $v\in R$, let $G_v$ be the graph obtained from $G$ by
contracting $V\setminus P_v$ to a single vertex $t_v$, and certify a
minimum $v$--$t_v$ cut of $G_v$ by \cref{lem:flow}, using at most
$4(|P_v|+1)$ queries.  Each cut query to $G_v$ is one cut query to
$G$: a shore $X\subseteq P_v$ has value $c_{G_v}(X)=c_G(X)$, and a
shore containing $t_v$ has value $c_G(P_v\setminus(X\setminus\{t_v\}))$.
A graph consistent with these queries on $G$ contracts to a graph
consistent with them on $G_v$, so the certificate remains valid
there: in every consistent graph, every shore $S\subseteq P_v$ with
$v\in S$ has value at least the certified minimum.  Since the cells
are disjoint, the certificates of all the contracted graphs together
cost
\[
 O\Bigl(\sum_{v\in R}(|P_v|+1)\Bigr)=O(n).
\]
The containment property gives the required lower bounds on
the global isolating-cut values.  For the upper bounds, query
one original minimum isolating shore for each terminal.
Together with the bit-cut certificates, this uses
$O(n\log|R|)$ queries.

Alternatively, the local network can be formed explicitly on
$P_v\cup\{t_v\}$: query the boundary capacity
$d_v^{\mathrm{out}}(u)=w_G(\{u\},V\setminus P_v)$ of each $u\in P_v$,
at most $n$ partial-star queries in all, and join $u$ to $t_v$ by an
edge of this fixed capacity.  The capacity of every shore
$S\subseteq P_v$ in this network equals $c_G(S)$, also in every
consistent graph, since the boundary totals were fixed, and the flow
certificate of \cref{rem:local-flow} again costs $O(|P_v|+1)$ queries.
\end{proof}

\subsection{Halving the terminal set}

Throughout this subsection put $\lambda=\lambda(G)>0$.
All flows use the original vertex set, even when the terminal
set becomes smaller.

Call $S\subseteq V$ \emph{$U$-separating} if $S\cap U\neq\emptyset$
and $U\not\subseteq S$.  For a symmetric weighting $h$ write $c_h$ for
its cut function.

\begin{lemma}[Fractional splitting off]\label{lem:splitting}
Let $U\subseteq V$ and suppose $c_G(S)\geq\lambda$ for every
$U$-separating $S$.  There is a nonnegative weighted graph $J$ on $V$,
with weight zero on every pair having an endpoint outside $U$, such
that
\[
 c_J(S)\leq c_G(S)\quad\text{for every }S,\qquad
 c_J(S)\geq\lambda\quad\text{for every $U$-separating }S.
\]
\end{lemma}
\begin{proof}
Let $\mathcal H$ be the set of symmetric nonnegative weightings $h$ on
$V$ with zero diagonal satisfying both displayed families of
inequalities.  It contains $w_G$ and is closed.  It is bounded, since
$h(x,y)\leq c_h(\{x\})\leq c_G(\{x\})$.  Choose $h\in\mathcal H$
minimizing the weight $\Phi(h)=\sum_{z\notin U}\sum_yh(z,y)$ at
nonterminals, and let $H$ be the corresponding graph.  Call a pair
with positive $H$-weight an edge, and call a $U$-separating set
\emph{tight} if its $H$-cut equals $\lambda$.  Complements of tight
sets are tight.

Suppose, for a contradiction, that some nonterminal $x$ has an edge.
Let $D$ be a symmetric weighting with zero diagonal that is negative
only on edges, with $c_D(S)\leq0$ for every $S$ and $\Phi(D)<0$.  If no
tight set $S$ had $c_D(S)<0$, then $h+\varepsilon D\in\mathcal H$ for
all small $\varepsilon>0$, because there are finitely many sets;
this would contradict minimality.  We use two such directions.
\begin{enumerate}[label=(\roman*)]
 \item Deleting an edge $xu$: $c_D(S)$ is $-1$ if $S$ separates $x$
 from $u$ and $0$ otherwise, and $\Phi(D)\leq-1$.  Hence, after taking
 a complement, some tight set contains $u$ but not $x$.
 \item Splitting distinct edges $xu,xv$ into $uv$: then
 $c_D(S)\in\{0,-2\}$, with $c_D(S)=-2$ exactly when $S$ separates $x$
 from both $u$ and $v$, and $\Phi(D)=-2$ (the weight at $x$ drops by
 $2$, while at $u$ and at $v$ the lost edge to $x$ is offset by the new
 edge $uv$, whether or not they are terminals).  Hence some tight set
 contains $u$ and $v$ but not $x$.
\end{enumerate}

Among the tight sets that avoid $x$ and contain a neighbor of $x$,
choose $X$ of maximum size, containing the neighbor $u$.  The set
$X\cup\{x\}$ is $U$-separating because $x\notin U$, so
\[
 \lambda\leq c_H(X\cup\{x\})=\lambda+d_H(x)-2w_H(x,X).
\]
Thus $w_H(x,X)\leq d_H(x)/2$.  Since $w_H(x,X)\geq H(x,u)>0$, some
neighbor $v$ of $x$ lies outside $X$.  By (ii) there is a tight set
$Y$ avoiding $x$ and containing $u$ and $v$.

If $X\cap Y$ and $X\cup Y$ are both $U$-separating, submodularity
gives $c_H(X\cup Y)\leq2\lambda-c_H(X\cap Y)\leq\lambda$.  Then
$X\cup Y$ is a tight set avoiding $x$, containing $u$, and larger than
$X$ because it contains $v$, a contradiction.  Otherwise either
$X\cap Y$ contains no terminal or $X\cup Y$ contains all terminals.
In both cases $X\setminus Y$ contains a terminal outside $Y$ and misses
the terminals of $Y$, so it is $U$-separating, and symmetrically so is
$Y\setminus X$.  The identity
\[
 c_H(X)+c_H(Y)=c_H(X\setminus Y)+c_H(Y\setminus X)
 +2w_H\bigl(X\cap Y,V\setminus(X\cup Y)\bigr)
\]
then gives $w_H(X\cap Y,V\setminus(X\cup Y))\leq0$, contradicting
$u\in X\cap Y$, $x\notin X\cup Y$, and $H(u,x)>0$.  Hence no nonterminal
has an edge, and $J=H$ satisfies the lemma.
\end{proof}

\begin{lemma}[A flow to half the terminals]
\label{lem:halving}
For every $U\subseteq V$ with $|U|\geq2$, there is a nonempty
$R\subset U$ with $|R|\leq |U|/2$ and a feasible flow whose
divergence satisfies
\[
 b_u=\lambda/2\quad(u\in U\setminus R),\qquad
 b_v=0\quad(v\notin U),\qquad
 b_v\leq0\quad(v\in R).
\]
Consequently, $O(n)$ cut queries certify the inequalities
\begin{equation}
 \label{eq:halving}
 c_{G'}(S)\geq\frac{\lambda}{2}|S\cap U|
 \qquad\text{for every }S\text{ with }S\cap R=\emptyset.
\end{equation}
\end{lemma}

\begin{proof}
Every $U$-separating set is a nontrivial cut, so \cref{lem:splitting}
applies and gives $J$.  For $u\in U$ the singleton $\{u\}$ is
$U$-separating because $|U|\geq2$, so $d_J(u)\geq\lambda$.  Since $J$
has no weight outside $U$, $c_J(S)=c_J(S\cap U)$ for every $S$; in
particular $c_J(U\setminus X)=c_J(X)$ for $X\subseteq U$.

Choose $X\subseteq U$ maximizing $c_J(X)$, and let $R$ be the smaller
of $X$ and $U\setminus X$.  Then $|R|\leq|U|/2$, and $R$ also
maximizes $c_J$.  For $u\in U\setminus R$, moving $u$ into $R$ cannot
increase the cut:
\[
 c_J(R\cup\{u\})=c_J(R)+d_J(u)-2w_J(u,R)\leq c_J(R).
\]
Hence $w_J(u,R)\geq d_J(u)/2\geq\lambda/2$.  Also
$c_J(R)\geq c_J(\{u\})\geq\lambda>0$ for any $u\in U$, so $R$ is
neither empty nor all of $U$.  For every $S$ with $S\cap R=\emptyset$,
\begin{equation}\label{eq:halving-cut}
 c_G(S)\geq c_J(S)\geq w_J(S\cap U,R)
 \geq\frac\lambda2|S\cap U|.
\end{equation}

Add a source $s$ joined to every $u\in U\setminus R$ with capacity
$\lambda/2$, and a sink $t$ joined to every vertex of $R$ with a
fixed capacity $M>(\lambda/2)|U\setminus R|$.  Any cut containing a
vertex of $R$ on its source side already has capacity at least $M$.
Every remaining source--sink cut has source side $\{s\}\cup S$ with
$S\cap R=\emptyset$, and capacity
\[
 c_G(S)+\frac\lambda2\bigl|(U\setminus R)\setminus S\bigr|
 \geq\frac\lambda2\bigl|(U\setminus R)\cap S\bigr|
 +\frac\lambda2\bigl|(U\setminus R)\setminus S\bigr|
 =\frac\lambda2|U\setminus R|
\]
by \eqref{eq:halving-cut}.  By max-flow/min-cut, a maximum flow
saturates every source edge.  Its restriction to $G$ is a feasible
flow with the required divergences.

Finally certify this fixed divergence vector by
\cref{lem:flow}.  A shore disjoint from $R$ has total
divergence $(\lambda/2)|S\cap U|$, giving
\eqref{eq:halving} in every consistent graph.
\end{proof}

\begin{proof}[Proof of \cref{thm:certificates}]
If $\lambda(G)=0$, one zero-weight nontrivial cut suffices.
Otherwise start with $U_0=V$.  At every stage with
$|U_i|\geq2$, certify all isolating-cut values for $U_i$
by \cref{lem:isolating}, and use \cref{lem:halving} to
choose and certify $U_{i+1}\subset U_i$ with
$1\leq|U_{i+1}|\leq|U_i|/2$.  Stop when $U_L$ is a
singleton.  Also query one minimum cut of $G$.
There are at most $\lceil\log_2 n\rceil$ rounds, and the
total number of queries is
\[
 O\left(n\sum_{i=0}^{L-1}\log(2|U_i|)\right)
 =O(n\log^2(2n)).
\]

Suppose that a consistent graph has a nontrivial cut $S$
of capacity less than $\lambda$.  Both shores meet $U_0$.
If both meet $U_i$, neither contains exactly one terminal
$v\in U_i$, since it would then have capacity at least
\[
 \iso_{G'}(v;U_i)=\iso_G(v;U_i)\geq\lambda.
\]
Thus both shores contain at least two terminals of $U_i$.
By \eqref{eq:halving}, neither can be disjoint from
$U_{i+1}$.  Induction implies that both meet the singleton
$U_L$, a contradiction.  Hence every nontrivial cut of
$G'$ has capacity at least $\lambda$.  The queried minimum
cut gives the reverse inequality.
\end{proof}

The split-off graphs and the maximum cuts used to choose
retained terminals are selected with knowledge of $G$.  Only
the resulting divergence vectors are certified; the graphs
$J$ need not remain valid for a consistent graph.
\Cref{lem:splitting} is a fractional form of the splitting-off
theorems of Lov\'asz~\cite{Lov76} and Mader~\cite{Mad78}: it
preserves only connectivity across the terminal set, which is all
the maximum-cut argument uses, and it needs neither integrality nor
the preservation of pairwise connectivities.  The proof above works
directly with arbitrary real capacities and
requires no rational approximation or integrality hypothesis.

\section{Weak skeletons and tree covers}\label{sec:skeletons}

We now pass from certifying an optimum to an explicit graph in which
it can be found.  The certificate construction gives an exact
completion after its measurements are supplied.  For an algorithm
choosing its own measurements, the weaker condition below suffices:
a short tree cover of the weak skeleton reduces the final search to
$O(n\log^2 n)$ further queries to the original graph.

Recall from \cref{def:weak-skeleton} that, for a graph $G$ with
$\lambda=\lambda(G)>0$, a weak skeleton is an explicit nonnegative
weighted graph $H$ on $V(G)$ with
\begin{equation}\label{eq:target}
 \lambda(H)\geq\lambda,\qquad c_H(S^*)\leq\tfrac54\lambda
 \quad\text{for every minimum-cut shore $S^*$ of $G$.}
\end{equation}
Thus the positive support of $H$ is connected.  Edges absent from $G$
may be used in $H$.  The tree cover below uses only that every
minimum cut of $G$ is a $5/4$-approximate minimum cut of $H$.

\begin{remark}[Sparse output]\label{rem:sparse}
The number of edges of $H$ never enters a query bound.  If an
explicit graph $H_0$ satisfies $\lambda(H_0)\geq\lambda$ and
$c_{H_0}(S^*)\leq\tfrac{121}{100}\lambda$ for every minimum cut $S^*$,
deterministic spectral sparsification~\cite{BSS}, scaled so that
$L_{H_0}\preceq L_H\preceq\tfrac{101}{100}L_{H_0}$, gives an $O(n)$-edge
weak skeleton $H$ without further queries, since
$c_H(S^*)\leq(12221/10000)\lambda<(5/4)\lambda$.
\end{remark}

\subsection{Certificates give sparse consistent skeletons}

\begin{proposition}[A certificate-to-skeleton connection]\label{prop:bridge}
Let $\mathcal Q$ be a $q$-query minimum-cut certificate for $G$ that
includes an optimum shore $S^*$.  There is a nonnegative graph $F$
with at most $q$ edges, consistent with every answer, such that
\[
 \lambda(F)=c_F(S^*)=\lambda(G).
\]
Given the certificate shores and answers, such an $F$ can be computed
without further queries.  If $\lambda(G)>0$, then $F$ satisfies
\eqref{eq:target} for the certified shore $S^*$, which is all the
tree cover of \cref{lem:tree-cover} needs to find $S^*$; $F$ need not
approximate the other minimum cuts of $G$.
\end{proposition}
\begin{proof}
Let $A$ have the queried cut-incidence vectors as rows, and let
$a=Aw_G$.  The system $Ax=a$, $x\geq0$, is feasible.  Choose a
feasible solution with minimal support.  If its supported columns
were linearly dependent, moving along a nonzero dependence in one
of its two directions until a positive coordinate became zero would
preserve feasibility and reduce support.  Thus its support has size
at most $\operatorname{rank}(A)\leq q$.  This solution can be found
by linear programming, or by enumerating supports and solving their
linear feasibility problems, using only $A,a$.

Every feasible solution has minimum-cut value $\lambda(G)$ by the
certificate property.  Since $S^*$ was queried, its value is also
fixed to $\lambda(G)$.  This proves the last claim.  The support of the computed completion need not be known
to be contained in the support of $G$.
\end{proof}

The certificate in \cref{thm:certificates} already includes $S^*$.
It therefore gives a graph-dependent collection of $O(n\log^2 n)$
measurements sufficient for constructing an explicit graph in which
the certified minimum cut can be found.  The
algorithmic question is how to select measurements from oracle
answers alone.  The next sections answer this by constructing a
weak skeleton directly.  They do not invert the certificate proof
or recover its terminal sets and flow witnesses.

\subsection{Covering near-minimum cuts by a few trees}

We use the minimum-spanning-tree inequality behind fractional tree
packing, in the following elementary form.  For a set $E$ of pairs
and a cut $S$, write $\delta_E(S)$ for the pairs of $E$ with one
endpoint in $S$.  Call $E$ \emph{spanning} if $\delta_E(S)\neq\emptyset$
for every nontrivial cut $S$.

\begin{lemma}[Light spanning edge sets]\label{lem:light}
Let $H$ be a nonnegative weighted graph on $V$ with
$\lambda(H)\geq\lambda>0$, and let $\ell\geq0$ be lengths on the pairs
of $V$.  There is a spanning set $E$ of positive-weight pairs of $H$
with
\begin{equation}\label{eq:light}
 \sum_{e\in E}\ell_e\leq\frac2\lambda\sum_{\{a,b\}}w_H(a,b)\,\ell_{ab}.
\end{equation}
\end{lemma}
\begin{proof}
We prove a statement for partitions.  Let $\mathcal P$ be a partition
of $V$ into $k\geq1$ blocks such that every nonempty proper union of
blocks has $H$-cut at least $\lambda$.  Say that $E$ \emph{connects}
$\mathcal P$ if every such union is crossed by a pair of $E$.  We show
that, if $\ell$ is nonnegative on the positive-weight pairs crossing
$\mathcal P$, there is a set $E$ of at most $k-1$ positive-weight
pairs connecting $\mathcal P$ with
\[
 \sum_{e\in E}\ell_e\leq\frac2\lambda\,
 \sum_{\{a,b\}\text{ crossing }\mathcal P}w_H(a,b)\,\ell_{ab}.
\]
The lemma is the case of singleton blocks.

For $k=1$ take $E=\emptyset$.  Let $k\geq2$.  A single block is a
proper union of blocks, so its cut is at least $\lambda>0$ and some
positive-weight pair crosses $\mathcal P$.  Among these, choose $uv$
with the least length $\ell_0=\ell_{uv}\geq0$.  Merge the blocks of
$u$ and $v$ to obtain $\mathcal P'$ with $k-1$ blocks.  Unions of
blocks of $\mathcal P'$ are unions of blocks of $\mathcal P$, and the
lengths $\ell'=\ell-\ell_0$ are nonnegative on the positive-weight
pairs crossing $\mathcal P'$, since those pairs also cross
$\mathcal P$.  By induction there is a set $E'$ of at most $k-2$ such
pairs connecting $\mathcal P'$, with the bound for $\ell'$.  Put
$E=E'\cup\{uv\}$.  A union of blocks of $\mathcal P$ that does not
separate $u$ from $v$ is a union of blocks of $\mathcal P'$ and is
crossed by $E'$; otherwise it is crossed by $uv$.  Moreover
\[
 \sum_{e\in E}\ell_e\leq\ell_0+\sum_{e\in E'}(\ell'_e+\ell_0)
 \leq\sum_{e\in E'}\ell'_e+(k-1)\ell_0.
\]
The total weight $W$ of the pairs crossing $\mathcal P$ is half the sum
of the $k$ block cuts, so $W\geq k\lambda/2$.  Pairs that cross
$\mathcal P$ but not $\mathcal P'$ have $w_H\ell'\geq0$.  Hence
\[
 \frac2\lambda\sum_{\text{crossing }\mathcal P}w_H\ell
 =\frac2\lambda\sum_{\text{crossing }\mathcal P}w_H\ell'+\frac{2W}\lambda\,\ell_0
 \geq\frac2\lambda\sum_{\text{crossing }\mathcal P'}w_H\ell'+k\ell_0
 \geq\sum_{e\in E}\ell_e.\qedhere
\]
\end{proof}

\begin{lemma}[A deterministic tree cover]\label{lem:tree-cover}
From an explicit nonnegative weighted graph $H$ with $\lambda(H)>0$,
one can deterministically compute $O(\log n)$ spanning trees on $V$
such that every cut of weight at most $(5/4)\lambda(H)$ crosses at
most two edges of at least one tree.  No queries to $G$ are required.
\end{lemma}
\begin{proof}
Put $\lambda=\lambda(H)$ and let $\mathcal F$ be the set of nontrivial
cuts of weight at most $(5/4)\lambda$, each counted once up to
complementation.  We first show that for every nonempty
$\mathcal F'\subseteq\mathcal F$ some spanning tree $T$ crosses at most
two edges of at least $|\mathcal F'|/4$ of its cuts.

Let $\ell_{ab}$ be the number of cuts in $\mathcal F'$ that separate
$a$ from $b$.  Then $\sum_{\{a,b\}}w_H(a,b)\ell_{ab}=\sum_{S\in\mathcal F'}c_H(S)$,
and \cref{lem:light} gives a spanning set $E$ with
\[
 \sum_{S\in\mathcal F'}|\delta_E(S)|=\sum_{e\in E}\ell_e
 \leq\frac2\lambda\sum_{S\in\mathcal F'}c_H(S)\leq\frac52\,|\mathcal F'|.
\]
Each term on the left is at least one because $E$ is spanning.  If $b$
cuts have $|\delta_E(S)|\geq3$, the left side is at least
$|\mathcal F'|+2b$, so $b\leq\frac34|\mathcal F'|$.  The graph $(V,E)$
is connected; let $T$ be any spanning tree of it, for instance a
breadth-first tree.  Since $T\subseteq E$, we have
$|\delta_T(S)|\leq|\delta_E(S)|$, and $T$ crosses at most two edges of
at least $|\mathcal F'|/4$ cuts of $\mathcal F'$.

A fixed spanning tree crosses at most two edges of at most
$\binom n2$ nontrivial cuts, up to complementation: choose one or two
of its edges.  Applying the first step to $\mathcal F'=\mathcal F$
therefore gives $|\mathcal F|\leq4\binom n2$.  Now choose trees
greedily, each time applying the first step to the cuts not yet
covered.  After $t$ choices at most $(3/4)^t\cdot4\binom n2$ cuts
remain, so $t=O(\log n)$ trees cover $\mathcal F$.

Every step concerns only the explicit graph $H$, so this is a
deterministic computation in the query model.  The proof of
\cref{lem:light} is itself a greedy procedure: repeatedly add a
shortest positive-weight pair joining two different components.
\end{proof}

\begin{remark}\label{rem:packing-dual}
The inequality \eqref{eq:light} is the minimum-spanning-tree bound
that is dual to the fractional spanning-tree packing theorem.  Using
it directly with lengths that count uncovered cuts avoids computing
an optimal packing.  All choices in this cover use only the explicit
skeleton.
The construction has no random choices and does not use the original
graph's unqueried weights.
\end{remark}

\begin{theorem}[Two-respecting cuts]\label{thm:2respecting}
Given an explicit tree $T$ on $V(G)$, a minimum $G$-cut crossing
at most two edges of $T$ can be found deterministically with
$O(n\log n)$ cut queries.  The edges of $T$ need not be edges of $G$.
\end{theorem}

\Cref{app:2respecting} proves this theorem by implementing the
interested-path reduction of Gawrychowski, Mozes, and
Weimann~\cite{GMW}.  Boundary-measure searches locate all interest
paths with $O(n\log n)$ queries, and SMAWK~\cite{SMAWK} searches
the resulting full Monge rectangles within the same budget.

\begin{corollary}\label{cor:skeleton-search}
An algorithm constructing a weak skeleton in $q(n)$ queries yields
a deterministic minimum-cut algorithm using $q(n)+O(n\log^2 n)$
queries.  The same holds for an explicit graph satisfying
\eqref{eq:target} for a single minimum cut.
\end{corollary}
\begin{proof}
Compute the tree cover of \cref{lem:tree-cover} on the explicit
skeleton and apply \cref{thm:2respecting} to each tree.  Every
minimum cut of $G$ satisfying \eqref{eq:target} is among the covered
cuts, and every returned candidate is an actual cut of $G$.
\end{proof}

\section{A weak skeleton from a supplied hierarchy}\label{sec:hierarchy}

The certificate-to-skeleton connection used full input knowledge to
select measurements.  From this section onward, every query is
selected from supplied data and previous answers.  We first isolate
what a known hierarchy must guarantee about the unknown minimum cuts;
\cref{sec:discovery} will construct such a hierarchy.

A hierarchy $\cT$ is a rooted tree whose leaves are precisely $V$
and whose internal nodes have at least two children.  A node represents
its descendant leaf set.  We also admit, as the formalization does,
unary nodes and leaves that are not in $V$, provided that the nodes
of $V$ have no children.  A nonroot node $u$ then represents the
cluster $A_u=\operatorname{sub}(u)\cap V$ of terminals below it;
distinct nodes may represent the same cluster, and a cluster may be
$\emptyset$ or $V$.  Every hierarchy constructed below has at most
$2n+1$ nodes, and the hierarchy supplied at the start has $O(n)$
nodes, which affects only the constant in the first round.  Write
$\cA$ for the nonroot clusters, indexed by nodes, so $|\cA|=O(n)$.
The \emph{canonical} capacities for $G$ are $d_A=c_G(A)$ on the edge
from a cluster $A$ to its parent, so $d_A=0$ when $A\in\{\emptyset,V\}$
and $d_A\geq\lambda$ otherwise; they are learned in $O(n)$ queries.
Define the representation cost
\begin{equation}\label{eq:R}
 R(S)=\min_{\substack{\text{binary labels on tree nodes}\\
                       \text{labels on }V\text{ equal }\ind_S}}
       \sum_{A\in\cA}d_A
           |\ell(A)-\ell(\operatorname{parent}(A))|.
\end{equation}
It is the minimum cut in the tree extending the prescribed labels on
$V$; leaves outside $V$ are unconstrained.

\begin{theorem}[Supplied-hierarchy bound]\label{thm:depthfree}
Suppose $\lambda(G)>0$, $\cT$ and $B\geq1$ are supplied, and every
minimum cut $S^*$ of $G$ satisfies $R(S^*)\leq B\lambda(G)$.  A weak
skeleton can be constructed deterministically in $O(nB^2\log n)$ cut
queries.  The bound is independent of hierarchy depth, and neither
the minimum cuts nor $\lambda(G)$ are needed as input.
\end{theorem}

\subsection{Quantiles of boundary measures}

\begin{lemma}[Exact checkpoints for a nonnegative measure]\label{lem:quantile}
Let $\mu$ be a nonnegative measure on an ordered set $[n]$, with known
total mass $M>0$.  Suppose prefix masses can be queried.  For an integer
$q\geq1$, $O(q\log n)$ queries give $O(q)$ exact prefix constraints
such that every nonnegative measure $\nu$ satisfying those constraints
obeys $|\mu([x])-\nu([x])|\leq M/q$ for every prefix $[x]$, $0\leq x\leq n$.
Consequently
\begin{equation}\label{eq:interval-error}
 |\mu(I)-\nu(I)|\leq\frac{2M}{q}
 \qquad\text{for every interval }I\subseteq[n].
\end{equation}
More precisely, at most $(q-1)(\lfloor\log_2n\rfloor+3)$ prefix queries and at most
$2q$ recorded prefixes suffice.
\end{lemma}
\begin{proof}
For $j=1,\ldots,q-1$, find by binary search the first index $k_j$ with
\[
 \mu([k_j])\geq jM/q.
\]
Record the exact masses of the prefixes ending at $k_j-1$ and $k_j$,
as well as the masses at indices $0$ and $n$.  Put $k_0=0$ and
$k_q=n+1$; recorded indices need not be distinct.

Fix $0\leq x\leq n$ and let $j$ be the largest index with $k_j\leq x$.
Then $k_j\leq x\leq k_{j+1}-1$, and both endpoints are recorded
(for $j=q-1$ the right endpoint is $n$).  Their $\mu$-masses lie in
$[jM/q,(j+1)M/q]$: the left one by the choice of $k_j$, and the right
one because $k_{j+1}$ is the first index reaching $(j+1)M/q$, or
because the total is $M$.  Since $\nu$ agrees with $\mu$ at both
endpoints and both measures are nonnegative, $\mu([x])$ and
$\nu([x])$ lie in this interval of length $M/q$.  Recording both sides
of each crossing is essential when there is a large atom.  An interval
is the difference of two prefixes, which proves
\eqref{eq:interval-error}.
\end{proof}

Choose a depth-first order of the leaves in $V$ of the supplied
hierarchy.  Every cluster is an interval in this order.  For each $A\in\cA$, use
the two boundary measures
\begin{align*}
 \mu_A^{\rm out}(X)&=w_G(A,X\setminus A),\\
 \mu_A^{\rm in}(X)&=w_G(A\cap X,V\setminus A).
\end{align*}
Both have total mass $d_A$; when $d_A=0$ both vanish identically and
no queries are needed.  Each prefix mass is obtained in at most
two additional cut queries using
\begin{align*}
 2\mu_A^{\rm out}(X)
   &=d_A+c_G(X\setminus A)-c_G(A\cup X),\\
 2\mu_A^{\rm in}(X)
   &=d_A+c_G(A\cap X)-c_G(A\setminus X).
\end{align*}

\begin{lemma}[Completion without a depth factor]\label{lem:depthfree-completion}
For $0<\eta<1$, $O(n\eta^{-1}\log n)$ deterministic cut queries
construct a nonnegative weighted graph $F$ with $c_F(A)=d_A$ for
every cluster and
\begin{equation}\label{eq:M-error}
 |M_F(A,B)-M_G(A,B)|\leq\eta\min(d_A,d_B),
 \qquad M_G(A,B)=\ind_A^{\mathsf T}L_G\ind_B.
\end{equation}
The graph $F$ can be chosen with $O(n\eta^{-1})$ positive-weight edges.
\end{lemma}
\begin{proof}
Apply Lemma~\ref{lem:quantile} to both measures of every cluster, with
$q=\lceil2/\eta\rceil$.  Find $F$ by nonnegative linear feasibility,
requiring all cluster cuts to equal $d_A$ and all the recorded boundary
prefix masses to have their recorded values.  These are linear equality
constraints in the weights of $F$, and $G$ is a feasible solution.

For fixed $A$, every interaction $M_G(A,B)$ is one of
\[
 -\mu_A^{\rm out}(B),\qquad
 \mu_A^{\rm in}(B),\qquad
 d_A-\mu_A^{\rm out}(B),
\]
according as $B$ is disjoint from, contained in, or contains $A$;
the diagonal is already exact.  The same identities hold for $F$.
Since $B$ is an interval, Lemma~\ref{lem:quantile} bounds the error by
$\eta d_A$.  Swapping $A,B$ bounds it by $\eta d_B$, giving
\eqref{eq:M-error}.

There are $O(nq)$ equality constraints.  The feasible polytope is bounded
because it fixes every singleton weighted degree.  An extreme point has at most
the rank of the constraint matrix many positive coordinates: otherwise
there is a nonzero dependence supported on its positive coordinates,
and sufficiently small perturbations in both directions preserve
feasibility.  Choose an extreme point deterministically to obtain
$O(nq)$ positive edge weights.
\end{proof}

\subsection{Representation cost controls completion error}

\begin{lemma}[Error and domination]\label{lem:R-error}
For every nontrivial cut $S$,
\begin{equation}\label{eq:R-error}
 |c_F(S)-c_G(S)|\leq\eta\frac{R(S)^2}{\lambda}.
\end{equation}
Moreover $R(S)\geq c_G(S)$.
\end{lemma}
\begin{proof}
Take a minimizing binary labeling in \eqref{eq:R}.  Let
$A_1,\ldots,A_k$ be the clusters whose parent edge has different
endpoint labels, and let $\sigma_i\in\{-1,1\}$ be the change of label
across that edge, from parent to child.  Telescoping along each root
path gives
\begin{equation}\label{eq:indicator}
 \ind_S=\ell(\text{root})\ind_V+
                  \sum_{i=1}^k\sigma_i\ind_{A_i},
 \qquad R(S)=\sum_{i=1}^k d_{A_i}.
\end{equation}
A cut edge whose cluster is $\emptyset$ or $V$ has $d_{A_i}=0$ and
$\ind_{A_i}\in\{0,\ind_V\}$; absorb its term into the constant
multiple of $\ind_V$ and drop it from the list, which changes neither
$R(S)$ nor the sums below.  Every remaining $A_i$ is a nontrivial
shore of $G$, so $d_{A_i}\geq\lambda$ and $k\leq R(S)/\lambda$.
The Laplacians annihilate $\ind_V$, so \eqref{eq:M-error} gives
\begin{align*}
 |c_F(S)-c_G(S)|
 &\leq\eta\sum_{i,j=1}^k\min(d_{A_i},d_{A_j})\\
 &\leq\eta k\sum_{i=1}^k d_{A_i}
 \leq\eta R(S)^2/\lambda.
\end{align*}
Finally, every edge of $G$ crossing $S$ must cross at least one of the
sets $A_i$, by \eqref{eq:indicator}.  Summing nonnegative edge weights
gives $c_G(S)\leq\sum_i c_G(A_i)=R(S)$.
\end{proof}

\subsection{An explicit graph preventing artificial small cuts}

\begin{lemma}[Terminal retraction of a tree]\label{lem:tree-retraction}
Let $T$ be a finite tree with a nonempty terminal set, nonnegative
edge capacities $c_e$, and nonnegative edge lengths $\ell_e$.
There is a map $f$ from its vertices to its terminals, fixing every
terminal, such that
\begin{equation}\label{eq:aggregate-retraction}
 \sum_{xy\in E(T)}c_{xy}d_\ell(f(x),f(y))
 \leq42\sum_{e\in E(T)}c_e\ell_e.
\end{equation}
\end{lemma}
\begin{proof}
First assume all lengths are positive.  Root the tree, let $h(x)$
be the length of the root path, and let $r(x)$ be the distance from
$x$ to its nearest terminal.  Fix an order for breaking nearest
terminal ties.  Choose independent uniform shifts $s,t\in[0,1)$.
For a nonterminal $x$ set
\[
 j(x)=\lfloor\log_2r(x)+s\rfloor,\qquad
 \Delta(x)=2^{j(x)-s},\qquad
 b_\Delta(x)=\lfloor2h(x)/\Delta+t\rfloor.
\]
Thus $\Delta(x)\leq r(x)<2\Delta(x)$.  At this fixed scale,
climb from $x$ to the highest ancestor in its same height band,
and map $x$ to a nearest terminal of that ancestor.  Map every
terminal to itself.  The climb has length less than $\Delta(x)/2$;
the triangle inequality and the fact that $r$ is 1-Lipschitz give
\[
 d_\ell(x,f(x))\leq2r(x).
\]

Consider a tree edge $xy$ of length $\delta$, and put
$r_0=\min\{r(x),r(y)\}$.  Every choice of shifts satisfies
\[
 d_\ell(f(x),f(y))\leq2r(x)+\delta+2r(y)
                         \leq4r_0+3\delta.
\]
If $\delta\geq r_0$ this is at most $7\delta$.  Otherwise both
endpoints are nonterminals.  Their scale indices differ with
probability at most
\[
 |\log_2r(x)-\log_2r(y)|\leq2\delta/r_0.
\]
Conditional on equal scale indices, their common scale satisfies
$\Delta>r_0/2$.  The independent band shift separates their bands
with probability at most $2\delta/\Delta\leq4\delta/r_0$.
Here we used the elementary fact that a uniformly shifted unit grid
separates two real numbers with probability at most their distance.
If neither event occurs, the endpoints have the same highest
ancestor in their band and hence the same image under $f$.
The union bound and $4r_0+3\delta<7r_0$ therefore give
\[
 \mathbb E\,d_\ell(f(x),f(y))\leq42\delta.
\]
Multiplying by capacities and summing proves that some map satisfies
\eqref{eq:aggregate-retraction}.  Only finitely many maps are possible.

For nonnegative lengths replace $\ell$ by $\ell+\varepsilon$.
The minimum, over the same finite set of terminal-fixing maps, of
the left side measured in the original metric is at most
$42\sum_e c_e\ell_e+42\varepsilon\sum_e c_e$ for every
$\varepsilon>0$.  Letting $\varepsilon$ tend to zero proves
\eqref{eq:aggregate-retraction}.  Zero capacities require no division
and cause no exception.
\end{proof}

\begin{lemma}[A graph approximating the terminal cuts of a tree]\label{lem:guard}
For the known capacitated hierarchy tree there is an explicit
nonnegative weighted graph $\Jprot$ on the terminal set $V$ such that
\begin{equation}\label{eq:terminal-guard}
 R(S)\leq c_{\Jprot}(S)\leq42R(S)\qquad(S\subseteq V).
\end{equation}
The graph can be found deterministically without queries to $G$.
More generally we write $\rho\geq1$ for a valid guard quality;
we may always take the absolute constant $\rho_0=42$.
\end{lemma}
\begin{proof}
For each terminal-fixing map $f$, replace a tree edge $xy$ by the
terminal edge $f(x)f(y)$ with the same capacity, dropping loops.
Call this terminal graph $J_f$, and route its edges on their unique
tree paths.  Let $\operatorname{load}_f(e)$ be the load on tree
edge $e$.  Exchanging two finite sums gives
\[
 \sum_e\ell_e\operatorname{load}_f(e)
   =\sum_{xy\in E(T)}c_{xy}d_\ell(f(x),f(y)).
\]
Apply finite minimax to the matrix with rows $f$, columns $e$,
and entries $\operatorname{load}_f(e)-42c_e$.
For every probability distribution on columns,
\cref{lem:tree-retraction}, with those probabilities as lengths,
supplies a row of nonpositive average.  Consequently there is a
distribution $p$ on the finite set of maps with
$\sum_f p_f\operatorname{load}_f(e)\leq42c_e$ for every edge.
This statement includes edges of capacity zero.

Put $\Jprot=\sum_f p_fJ_f$.  Pulling a terminal cut back through $f$
gives a feasible tree labeling of cost $c_{J_f}(S)$, so averaging
gives $R(S)\leq c_{\Jprot}(S)$.  Conversely fix a minimum tree extension
of $S$.  Every routed terminal edge crossing $S$ uses at least one
tree edge cut by this extension.  Its expected load bounds give
$c_{\Jprot}(S)\leq42R(S)$.  A tree with no edges is immediate.

For deterministic computation, enumerate the terminal-fixing maps
and solve the finite feasibility linear program for $p$ just
described, or enumerate terminal cuts and solve directly for edge
weights satisfying \eqref{eq:terminal-guard}.  All its data are
known.  The random shifts prove existence only; the query algorithm
uses deterministic finite optimization.  The construction and its
analysis follow the framework of Englert et
al.~\cite[Sections~3--4]{EGKRTT}, who combine terminal-fixing maps
with the same transfer from distance bounds to congestion bounds; their
Theorem~3.7 gives constant-quality flow sparsifiers of weighted trees,
and hence \eqref{eq:terminal-guard} with some absolute constant.  We
include the argument
to obtain an explicit constant and a cut-only statement.
\end{proof}

\begin{proof}[Proof of \cref{thm:depthfree}]
Query the cluster capacities and compute $\Jprot,\rho$ from
\cref{lem:guard}.  Put
\[
 \eta=\frac{1}{100\rho^2B^2},\qquad H_0=F+\sqrt\eta\,\Jprot,
\]
where $F$ is the completion from \cref{lem:depthfree-completion}.
For a nontrivial cut put $x=\sqrt\eta\,R(S)/\lambda(G)$.
The completion error, nonnegativity of $F$, and the lower bound
in \eqref{eq:terminal-guard} imply
\[
 \frac{c_{H_0}(S)}{\lambda(G)}
 \geq\max\{0,1-x^2\}+x\geq1.
\]
For every minimum cut $S^*$, which satisfies the promise, the upper
bounds give
\[
 \frac{c_{H_0}(S^*)}{\lambda(G)}
 \leq1+\eta B^2+\sqrt\eta\,\rho B
 =1+\frac1{100\rho^2}+\frac1{10}\leq\frac{111}{100}.
\]
Thus $H_0$ satisfies \eqref{eq:target} and is a weak skeleton.  Since $\rho=O(1)$, the number of
queries is $O(n\eta^{-1}\log n)=O(nB^2\log n)$.
\end{proof}

\begin{remark}[An elementary alternative]\label{rem:schur}
For a hierarchy of height $h$, one can replace \cref{lem:guard}
by eliminating the internal vertices of its Laplacian.  The resulting
Schur-complement graph $J_0$ on the leaves satisfies
\[
 R(S)/(2h)\leq c_{J_0}(S)\leq R(S).
\]
Indeed, its cut value is the minimum energy of real internal potentials
extending the leaf labels.  A binary minimizing labeling gives the
upper bound.  For the lower bound, decompose a maximum flow of value
$R(S)$ between the terminal shores into paths.  Each path has length
at most $2h$ and endpoint potential difference one, so its sum of
squared potential differences is at least $1/(2h)$ by
Cauchy--Schwarz.  Summing over the capacity-respecting flow paths
bounds the total energy below by $R(S)/(2h)$.
Thus $\Jprot=2hJ_0$ satisfies
$R(S)\leq c_{\Jprot}(S)\leq2hR(S)$, giving guard quality $\rho=2h$.
The same proof gives $O(nh^2B^2\log n)$ queries without a terminal
sparsification theorem.
\end{remark}

\section{Coarse completion and hierarchy refinement}\label{sec:refinement}

We improve the quadratic dependence on the supplied hierarchy bound
in \cref{thm:depthfree}.  The refinement uses a tree approximation of
an explicit graph.  We state that offline ingredient first, use it to
prove the refinement bound, and discuss it in \cref{sec:offline-hierarchy}.

\subsection{Coarse completion of the current hierarchy}

The one-step construction in Theorem~\ref{thm:depthfree} uses
quadratically many queries in $B$.  A coarser completion
can instead be used to improve the hierarchy.  The important choice
is to give the guard more weight than the minimal value $\sqrt\eta$
used for the final skeleton.

The offline ingredient is the following lemma, due to R\"acke and
Shah~\cite{RS14}; see \cref{sec:offline-hierarchy}.  Put
\[
 L(r)=\lceil\log_2(2r)\rceil\quad(r\geq1),\qquad L(0)=1,
 \qquad K(r)=L(L(r)).
\]
Thus $K(r)=\lceil\log_2(2L(r))\rceil$, and $L(r)K(r)\geq1$ always.
In natural logarithms, $L(r)\leq3\log(2r)$ and
$K(r)\leq7\log\log(2r)$ for $r\geq2$, so
$K(r)=O(\log\log(2r))$.
\begin{lemma}[An offline hierarchy]\label{lem:offline-hierarchy}
There is an absolute constant $a_0$ such that for every explicit
nonnegative weighted graph $J$ on $r\geq1$
vertices there is a canonical hierarchy $\cT_J$ with at most
$2r+1$ nodes, height at most $2L(r)$, and
\begin{equation}\label{eq:offline-tree}
 c_J(S)\leq R_{\cT_J}(S)\leq a(r)c_J(S)
 \quad(S\subseteq V(J)),\qquad a(r)=a_0L(r)K(r).
\end{equation}
It can be chosen deterministically from explicit $J$ without queries
to any other graph.  No connectivity or positive-weight assumption
is required.
\end{lemma}

Fix a known uniform bound $a=a(n)\geq1$ in
Lemma~\ref{lem:offline-hierarchy}, with
$a(n)=a_0L(n)K(n)$.  Fix also the absolute constant $\rho_0=42$
such that the guard of \cref{lem:guard} can always be
chosen to satisfy
\begin{equation}\label{eq:refinement-guard}
 R(S)\leq c_{\Jprot}(S)\leq\rho_0 R(S)
 \qquad(S\subseteq V).
\end{equation}
These bounds concern explicit graphs and trees.  They are independent
of the weights and of the input minimum-cut value.

\subsection{Transferring a new hierarchy back to the input}

Suppose the current canonical hierarchy has terminal function $R$
and satisfies $R(S^*)\leq B\lambda$ for every minimum cut $S^*$; fix
one such $S^*$ for the analysis.
Obtain a nonnegative completion $F$ by
Lemma~\ref{lem:depthfree-completion} with parameter $\eta$ and put
\[
 H=F+\gamma\Jprot,\qquad \gamma>0.
\]
The graph $H$ is explicit and has connected positive support, since
$\Jprot$ dominates $R$, and $R(S)\geq c_G(S)>0$ for every nontrivial cut.
An offline hierarchy of $H$ has canonical capacities $c_H(A)$ and
terminal function $R_H$ satisfying $R_H(S)\leq a c_H(S)$.
Keep its topology, query $c_G(A)$ for each of its nonroot clusters,
and denote the resulting canonical terminal function by $R_{\rm new}$.

\begin{lemma}[Hierarchy transfer]\label{lem:hierarchy-transfer}
For every nontrivial cut $S$,
\begin{align}
 c_G(S)&\leq c_H(S)+\frac{\eta}{\gamma^2\lambda}c_H(S)^2,
       \label{eq:refinement-cut-transfer}\\
 R_{\rm new}(S)&\leq R_H(S)
                  +\frac{\eta}{\gamma^2\lambda}R_H(S)^2.
       \label{eq:refinement-tree-transfer}
\end{align}
In particular, if
\begin{equation}\label{eq:refinement-x}
 x=a\bigl(1+\eta B^2+\gamma\rho_0B\bigr),
\end{equation}
then
\begin{equation}\label{eq:refinement-promise}
 R_{\rm new}(S^*)\leq
       \left(x+\frac{\eta}{\gamma^2}x^2\right)\lambda.
\end{equation}
The new hierarchy requires at most $2n$ additional cut queries.
\end{lemma}
\begin{proof}
Nonnegativity of $F$ and \eqref{eq:refinement-guard} give
$c_H(S)\geq\gamma R(S)$.
Lemma~\ref{lem:R-error} therefore implies
\[
 c_G(S)\leq c_F(S)+\eta R(S)^2/\lambda
          \leq c_H(S)+\eta c_H(S)^2/(\gamma^2\lambda),
\]
proving \eqref{eq:refinement-cut-transfer}.

Take an optimal labeling of the $H$-hierarchy for $S$.  Let
$A_1,\ldots,A_m$ be the clusters whose parent edges it cuts and write
$h_i=c_H(A_i)\geq0$.  Use this same labeling after querying the
$G$-capacities.  Applying the preceding inequality to each $A_i$ gives
\[
 \begin{aligned}
 R_{\rm new}(S)
 &\leq\sum_{i=1}^m c_G(A_i)\\
 &\leq\sum_{i=1}^m h_i+
              \frac{\eta}{\gamma^2\lambda}\sum_{i=1}^m h_i^2\\
 &\leq R_H(S)+\frac{\eta}{\gamma^2\lambda}R_H(S)^2.
 \end{aligned}
\]
This is a bound on the sum of the queried capacities; no term is
multiplied by the number of clusters or the hierarchy depth.

For the fixed minimum cut $S^*$, the old completion bound gives
\[
 c_H(S^*)\leq
       \bigl(1+\eta B^2+\gamma\rho_0 B\bigr)\lambda,
 \qquad R_H(S^*)\leq x\lambda.
\]
Substitute this in \eqref{eq:refinement-tree-transfer}.  There are at
most $2n$ nonroot clusters, proving the query count.
\end{proof}

\subsection{Halving distortion with linearly many queries in its bound}

\begin{lemma}[One refinement round]\label{lem:refinement-round}
Set
\begin{equation}\label{eq:refinement-parameters}
 \gamma=\frac{1}{100a^2\rho_0^2}.
\end{equation}
If $B\geq1/\gamma$, a canonical hierarchy with promise $B\lambda$
for every minimum cut can be replaced by one with promise
$(B/2)\lambda$ for every minimum cut, using $O(na^2B\log(2n))$ cut
queries.
\end{lemma}
\begin{proof}
Use the preceding construction with $\eta=\gamma/B$.
In particular $0<\eta<1$.  Since $\gamma B\geq1$ and $\rho_0\geq1$,
the value in \eqref{eq:refinement-x} satisfies
\[
 x=a\bigl(1+\gamma B+\gamma\rho_0B\bigr)
      \leq 3a\rho_0\gamma B.
\]
Consequently
\[
 \begin{aligned}
 x+\frac{\eta}{\gamma^2}x^2
 &=x+\frac{x^2}{\gamma B}\\
 &\leq (3a\rho_0+9a^2\rho_0^2)\gamma B\\
 &\leq12a^2\rho_0^2\gamma B
   =\frac{12}{100}B<\frac{B}{2}.
 \end{aligned}
\]
Lemma~\ref{lem:hierarchy-transfer} proves the new promise.
The completion costs
$O(n\eta^{-1}\log(2n))=O(na^2B\log(2n))$ queries, since
$\rho_0$ is an absolute constant.  The guard and the $H$-hierarchy
use explicit data only, and querying the new cluster capacities
adds $O(n)$ queries.  Neither $S^*$ nor $\lambda$ is queried or
used to choose the parameters.
\end{proof}

\begin{theorem}[Linear dependence on the hierarchy bound]\label{thm:linear-hierarchy}
Suppose $\lambda>0$ and a canonical hierarchy on $n\geq2$ vertices
is supplied, together with $B\geq1$ such that
$R(S^*)\leq B\lambda$ for every minimum cut $S^*$.
A weak skeleton can be constructed deterministically
using
\[
 O\!\left(nB\log^3(2n)(\log\log(2n))^2\right)
\]
cut queries, independently of hierarchy depth and weight spread.
Offline computation may be exponential.
\end{theorem}
\begin{proof}
Query the initial cluster capacities if they are not already known.
Keep $a,\rho_0,\gamma$ fixed.  While the current promised bound is at
least $1/\gamma$, apply Lemma~\ref{lem:refinement-round}, and replace
$B$ by $B/2$.  Every original minimum cut satisfies every new
promise, so the process is valid without ever identifying these cuts.
The sum of the successive promised bounds is at most twice the
initial one.  Refinement therefore costs
$O(na^2B\log(2n))$ queries in total.

At termination let $b<1/\gamma$ be the remaining bound.  Apply
Theorem~\ref{thm:depthfree} to this last hierarchy.  Its cost is
\[
 O(nb^2\log(2n))
   =O(nb\gamma^{-1}\log(2n))
   =O(na^2B\log(2n)).
\]
This also covers the case where the initial $B$ is below the stopping
threshold and no refinement is needed.  Finally
$a^2=a_0^2L(n)^2K(n)^2=O(\log^2(2n)(\log\log(2n))^2)$, giving the stated bound.  The final application
of Theorem~\ref{thm:depthfree} supplies all the weak-skeleton
guarantees; intermediate completions need not themselves be weak
skeletons.

The query parameters can be integers throughout: use
$q=\lceil2B/\gamma\rceil$ in a refinement round and
$q=\lceil200\rho_0^2b^2\rceil$ in the final completion.
Then the completion error parameter $2/q$ is no larger than the
one used in the corresponding proof.  The ceilings change only
absolute constants.  Because the promised bounds sum to at most
$2B$, no factor $\log B$ is charged for the number of rounds.
With the fixed value $\gamma^{-1}=100a_0^2\rho_0^2L(n)^2K(n)^2$,
this is a public $O(nBL(n)^3K(n)^2)$ budget determined by $n,B$ and the
supplied hierarchy, before the hidden graph is known.
\end{proof}

\begin{remark}[All minimum cuts at once]\label{rem:all-promise}
The universal promise costs nothing.  The construction never reads a
minimum cut: every parameter depends on $B$ and explicit data, and
each refinement round and the final completion are fixed before the
analysis picks an $S^*$.  The proofs of \cref{thm:depthfree},
\cref{lem:refinement-round} and \cref{thm:linear-hierarchy} are
therefore pointwise in $S^*$, and a single run yields
\eqref{eq:target} for every minimum cut simultaneously.  The tree cut
sparsifier of \cref{thm:tree-sparsifier-discovery} satisfies
$R(S)\leq\alpha\,c(S)$ for every shore, so the promise holds for every
minimum cut of the quotient.
\end{remark}

\Cref{thm:conversion-overview} follows.  Bring the given $B$-tree cut
sparsifier into normal form as in \cref{sec:conversion-overview};
this is a canonical hierarchy with $O(n)$ nodes and
$c_G(S)\leq R(S)\leq Bc_G(S)$ for every shore, so the promise
$R(S^*)\leq B\lambda$ holds for every minimum cut, and
\cref{thm:linear-hierarchy} gives the weak skeleton.

\begin{remark}[Calibrating the constants from explicit data]\label{rem:calibrate}
The constants $a_0$ and $\rho_0=42$ above suffice.  Alternatively,
an implementation can calibrate smaller constants from explicit data.  Use the guard's computed quality
$\rho$, maintain a trial hierarchy distortion $\widehat a\geq1$,
initially $1$, and set
$\gamma=(100\widehat a^2\rho^2)^{-1}$.
If $B<1/\gamma$, finish with Theorem~\ref{thm:depthfree}.
Otherwise construct $F,H$ as above and measure the chosen
$H$-hierarchy's exact distortion offline.  If it exceeds
$\widehat a$, double $\widehat a$ and retry; if it does not, query
the new cluster capacities and halve $B$.
Choose the $H$-hierarchy by the minimization in
Lemma~\ref{lem:offline-hierarchy}, so the trial bound never exceeds
$2\max\{1,a(n)\}$.  Also $\rho\leq\rho_0$.
The retry costs are geometric in $\widehat a^2$, while accepted
rounds are geometric in $B$, giving the same total query bound.
All tests in this implementation use known explicit graphs.
\end{remark}

\subsection{Offline hierarchies of explicit graphs}\label{sec:offline-hierarchy}

R\"acke and Shah~\cite{RS14} show that every graph has a tree cut
sparsifier of quality $O(\beta\log n\log\log n)$, where $\beta$ is
the approximation factor of the sparsest-cut routine used in their
construction.  With exact sparsest cuts, $\beta=1$, and this is the
existential bound of \cref{lem:offline-hierarchy}; their
polynomial-time construction uses $\beta=O(\sqrt{\log n})$.  Their
tree is a hierarchy with singleton leaves in which the children of
each cluster partition it into parts of at most half its size, so it
has at most $2r-1$ nodes and height less than $L(r)$, and the
normalization of \cref{sec:conversion-overview} gives it canonical
capacities without increasing its distortion.  A formalized, cut-only
version of their argument, with exact finite minimization in place
of sparsest-cut approximation, gives $a_0=268$; it is stated with the
slack of \cref{lem:offline-hierarchy}, at most $2r+1$ nodes and
height at most $2L(r)$, and its nodes of $V(J)$ are childless but
need not be the only leaves.  Since $J$ is
explicit, the lemma can be applied deterministically: enumerate all
hierarchies with these labeled leaves and no unary nodes, and choose
one minimizing its worst cut distortion, with fixed tie-breaking.  A
positive tree cost on a zero $J$-cut has infinite distortion; a zero
cost on both sides imposes no constraint.  The lemma guarantees a
finite answer obeying \eqref{eq:offline-tree}, even for disconnected
positive support or the all-zero graph.

\section{Discovering a hierarchy for arbitrary weights}\label{sec:discovery}

We now construct an initial hierarchy to which
\cref{thm:linear-hierarchy} applies.  Weighted degree normalization gives a
polynomial ratio between the largest weighted degree and a witnessed cut bound.
Residual rounding then either verifies the required hierarchy bound
or supplies a smaller witnessed cut bound that forces the next
normalization to halve the graph.  This test is computed from explicit
data and queried cut values; it is not the minimum-cut certificate
of \cref{sec:certificates}.

\subsection{Terminal compression of explicit trees}

\begin{lemma}[Terminal compression of a tree]\label{lem:tree-compression}
Let $T$ be a finite tree with nonnegative edge capacities and let
$C$ be a nonempty set of its vertices, with $q=|C|$.  There is a
fixed contraction map $\pi:V(T)\to V(T')$, injective on $C$, with
\[
 |V(T')|\leq\max\{1,2q-2\},
\]
such that, for every assignment of binary labels to $C$, some
minimum-cost extension to $T$ is constant on each fiber of $\pi$.
The contraction is computable from the explicit tree and is independent
of the terminal labels.
\end{lemma}
\begin{proof}
Repeatedly contract leaves outside $C$ into their neighbors.  This
prunes the tree down to the minimal subtree joining $C$; every removed
branch can be labeled uniformly with its attachment vertex at zero
cost.  If $q=1$, this leaves one vertex and proves the assertion.

For $q\geq2$, every leaf of the pruned tree lies in $C$.  On each
maximal path whose internal vertices lie outside $C$ and have degree
two, choose a minimum-capacity edge, breaking ties deterministically,
and contract all other edges of the path.  For fixed endpoint labels,
the optimum path cost is zero if the labels agree and the minimum edge
capacity if they differ.  Both optima are attained with all contracted
edges uncut.  The choices on different paths are compatible, so the
fixed contraction preserves the minimum extension cost for every
labeling of $C$.  In the resulting tree $T'$ with $n'$ vertices, every
vertex outside $C$ has degree at least three and every vertex of $C$
has degree at least one.  Summing degrees gives
$2(n'-1)\geq q+3(n'-q)$, so $n'\leq2q-2$.  The argument also applies
when some capacities are zero, and the terminals need not be leaves.
\end{proof}

\subsection{Safe contractions and weighted degree normalization}

We work on a current quotient of the input.  Write $N$ for its
number of vertices, $G$ for this quotient, and $\delta=\min_v d_v$.
Every query to $G$ is one original query on a union of contracted
classes.  We maintain all weighted degrees and a known nontrivial cut of value
$\Lambda$, with
\begin{equation}\label{eq:general-upper}
 0<\lambda\leq \Lambda\leq\delta.
\end{equation}
Initially query the weighted degrees and use a singleton of minimum weighted degree.
An individually learned edge with weight strictly greater than $\Lambda$
can be contracted: no cut of value at most $\Lambda$, including every
minimum cut and the witness for $\Lambda$, can separate its endpoints.

\begin{lemma}[Weighted degree normalization]\label{lem:degree-normalization}
Using safe contractions, one can obtain a quotient with $N'$ vertices
and an updated witnessed bound $\Lambda'\leq \Lambda$ such that
\begin{equation}\label{eq:normalized-degree}
 0<\lambda\leq \Lambda'\leq\min_v d_v,
 \qquad \max_v d_v\leq(N'-1)\Lambda'.
\end{equation}
Every original minimum cut survives.  With the weighted degrees already known,
the query cost is $O((N-N')\log(2N))$.  Across all invocations in
one algorithm, the total normalization cost is $O(n\log(2n))$.
\end{lemma}
\begin{proof}
While some vertex $v$ has $d_v>(N-1)\Lambda$, find an edge $vu$ of weight
greater than $\Lambda$ as follows.  Start with candidates
$W=V\setminus\{v\}$ and known total mass $w_G(v,W)=d_v$.
Split $W$ into two nearly equal nonempty parts.  Evaluate one part's
mass using
\[
 2w_G(v,X)=d_v+c_G(X)-c_G(X\cup\{v\}),
 \qquad v\notin X,
\]
and obtain the other's by subtraction.  Retain the part with larger
average mass per candidate, with fixed tie-breaking.  Its average
is at least the old average, which is strictly greater than $\Lambda$.
After $O(\log(2N))$ queries, the remaining singleton gives
$w_G(v,u)>\Lambda$.

Contract $u,v$.  No cut of value at most $\Lambda$ separates them, so all
minimum cuts and the current witness survive.  The new vertex has
weighted degree
\begin{equation}\label{eq:free-degree-update}
 d_{uv}=d_u+d_v-2w_G(u,v).
\end{equation}
Every other vertex keeps exactly the same weighted degree.  Thus no additional
queries are needed to update the entire weighted degree list.  Replace $\Lambda$
by $\min\{\Lambda,d_{uv}\}$, using the new singleton shore as witness if
its weighted degree is smaller, and decrement $N$.  The new witnessed bound
still satisfies $\lambda\leq \Lambda\leq\min_v d_v$.

Repeat with the new values of $N$ and $\Lambda$.  The witness remains
nontrivial, so at least two vertices remain.  The stopping condition
is exactly \eqref{eq:normalized-degree}.  There is one search per
contraction, proving the query bound.  Over the entire algorithm
there are at most $n-2$ contractions, even when the witnessed bound
changes many times.
\end{proof}

The lemma does not assert that every remaining edge has weight at
most $\Lambda'$.  For example, on vertices $0,1,2,3$ give edges $01,02,12,23$
weights $2,1,1,1$, respectively.  The minimum weighted degree and the witnessed
bound can both be $1$, and every weighted degree is at most $3$, so normalization
does nothing although edge $01$ has weight $2$.  Such edges may be
retained in the explicit graph $K$ during the next phase.

\begin{lemma}[A large bound decrease forces halving]\label{lem:normalization-halves}
Suppose the current graph has $N\geq2$ vertices and all weighted degrees are
at least a witnessed bound $\Lambda$.  If a new nontrivial cut gives a bound
$u<\Lambda/(4N)$, as in \cref{lem:general-certificate}, then
normalization started with $u$ leaves at most
$\lfloor N/2\rfloor$ vertices.
\end{lemma}
\begin{proof}
Throughout this normalization, the current bound $\Lambda'$ is at most $u$
and the number of vertices is at most $N$.  Every witnessed bound is
the value of a nontrivial cut, so $\Lambda'\geq\lambda>0$.  Consider an old
vertex that never participates in a contraction.  Its weighted degree is
unchanged and at least $\Lambda$.  At termination the weighted degree condition would
require
\[
 \Lambda\leq d_v\leq(N'-1)\Lambda'\leq(N-1)u<\Lambda,
\]
a contradiction; the third inequality uses $\Lambda'\geq0$.  Every final vertex therefore contains at least
two vertices from the graph at the start of this invocation.  These
classes partition the old vertices, proving the claim.  Further
decreases of the bound during normalization only strengthen it.
\end{proof}

\subsection{Learning a residual graph down to a weighted degree floor}

Start each phase after weighted degree normalization, so
\begin{equation}\label{eq:normalized-phase}
 \Lambda\leq d_v\leq(N-1)\Lambda\qquad(v\in V(G)).
\end{equation}
Fix an integer $2\leq k\leq N-1$ and set
\begin{equation}\label{eq:normalized-floor}
 t=\frac{\Lambda}{16N^2},\qquad
 L=16\lceil N/k\rceil\lceil\log_2(2N)\rceil.
\end{equation}
The parameter $t$ is a residual weighted degree floor, and $L$ bounds the
number of centers.  Both remain fixed during the phase.
Contractions occur only during normalization between phases.

\begin{lemma}[Residual preparation]\label{lem:residual-preparation}
Using $O(Nk\log^2(2N))$ cut queries, one can learn an exact
subgraph $K\subseteq G$ such that, for $R=G-K$ and its known
weighted degrees $r_v$,
\begin{equation}\label{eq:residual-light}
 r_v\leq t\quad\text{or}\quad
 w_R(v,u)\leq r_v/k\quad\text{for every }u.
\end{equation}
The graph $K$ may contain edges heavier than $\Lambda$.
\end{lemma}
\begin{proof}
Start with $K$ empty.  Cut queries to $R$ are simulated by subtracting
the known contribution of $K$ from a query to $G$.  Its weighted degrees are
updated without further queries whenever an edge is added to $K$.

A scan of a vertex $v$ with $r_v>t$ sets $q_v=r_v$ and, keeping
$K$ fixed during the search, finds all residual incident edges heavier
than $q_v/(2k)$.  By \cref{lem:row-search}, there are fewer than
$2k$ such edges and the scan costs $O(k\log(2N))$ queries.
Add their full weights to $K$ and update the residual weighted degrees.
Thus each edge of $R$ is either an entire original edge or zero.

After a scan every remaining incident edge has weight at most
$q_v/(2k)$, and subsequent removals preserve this upper bound.
The relative lightness condition holds while $r_v\geq q_v/2$.
Repeatedly scan any vertex with $r_v>t$ that has not been scanned
or whose weighted degree has fallen below half its last reference value.
At termination this proves \eqref{eq:residual-light}.

Every first reference is at most $(N-1)\Lambda$.  Successive references
for a vertex decrease by a factor greater than two and stay above
$t$.  Since $(N-1)\Lambda/t=16N^2(N-1)$, there are at most
$1+\log_2(16N^2(N-1))=O(\log(2N))$ scans per vertex.  This proves
both finite termination and the query bound, independently of the
weight spread.  The argument uses only $\Lambda>0$ and the upper bound
$d_v\leq(N-1)\Lambda$ in \eqref{eq:normalized-phase}, not the lower bound
$\Lambda\leq d_v$; the count of edges found by one scan is likewise not
needed for the query bound.
\end{proof}

\subsection{Selecting centers with weighted degree buckets}

Freeze $K,R$, and the residual weighted degrees after preparation.
Learning center rows does not remove further edges from $R$.
Let $H_C^R$ be the subgraph of $R$ containing exactly the edges
incident to the selected center set $C$.

For possibly overlapping vertex sets $A,W$, write
\[
 m_R(A,W)=\sum_{v\in A}\sum_{u\in W}w_R(v,u).
\]
This counts an edge inside $A\cap W$ twice.  With
$I=A\cap W$, $X=A\setminus W$, $Y=W\setminus A$, and
$Z=V\setminus(A\cup W)$,
\begin{equation}\label{eq:overlapping-mass}
 m_R(A,W)=\sum_{v\in I}r_v-w_R(I,Z)+w_R(X,Y).
\end{equation}
The weighted degrees are known and the two disjoint-set masses cost at most
six cut queries in total by \eqref{eq:rectangle}.

\begin{lemma}[Bucketed center selection]\label{lem:bucketed-centers}
Let $2\leq k\leq N-1$ be as in \eqref{eq:normalized-floor}, so in
particular $k\leq8N$.
Using $O((N^2/k)\log(2N))$ further queries, one can learn all
edges incident to a set $C$ of at most $L$ centers such that
\begin{equation}\label{eq:center-coverage}
 v\notin C\quad\Longrightarrow\quad
 r_v\leq t\ \text{ or }\ w_R(v,C)\geq r_v/k.
\end{equation}
\end{lemma}
\begin{proof}
Call a vertex active when
\[
 v\notin C,\qquad r_v>t,\qquad b_v:=w_R(v,C)<r_v/k.
\]
For an active vertex put $a_{vu}=kw_R(v,u)/(2r_v)$ and
$p_v=\prod_{c\in C}(1-a_{vc})$.  Relative lightness gives
$0\leq a_{vu}\leq1/2$ and
$p_v\geq1-\sum_{c\in C}a_{vc}>1/2$.
With $\Phi=\sum_{v\text{ active}}p_v$ and
$\sigma(u)=\sum_{v\text{ active}}p_va_{vu}$, we have
\begin{equation}\label{eq:center-total-score}
 \sum_{u\notin C}\sigma(u)
 =\frac{k}{2}\sum_{v\text{ active}}p_v(1-b_v/r_v)
 \geq\frac{k}{4}\Phi.
\end{equation}

To evaluate group scores cheaply, use
$t<r_v\leq(N-1)\Lambda$ and $1/2<p_v\leq1$ to bound
$\theta_v=p_v/r_v$ by
\[
 \frac{1}{2(N-1)\Lambda}<\theta_v<\frac1t,
 \qquad \frac{2(N-1)\Lambda}{t}<32N^3.
\]
Round each coefficient down to a power of two times
$[2(N-1)\Lambda]^{-1}$, so
$\widetilde\theta_v\leq\theta_v<2\widetilde\theta_v$.
There are $O(\log(2N))$ buckets $A_j$ with common coefficient
$\theta_j$.  The surrogate scores
\[
 \widetilde\sigma(u)
   =\frac{k}{2}\sum_{v\text{ active}}
                         \widetilde\theta_v w_R(v,u)
\]
satisfy $\sigma(u)/2\leq\widetilde\sigma(u)\leq\sigma(u)$.
For a candidate set $W$, their total is
\begin{equation}\label{eq:bucketed-score}
 \sum_{u\in W}\widetilde\sigma(u)
    =\frac{k}{2}\sum_j\theta_j m_R(A_j,W),
\end{equation}
which costs $O(\log(2N))$ queries by \eqref{eq:overlapping-mass}.
The total over $V\setminus C$ is also known directly from the
$r_v$ and $b_v$.

Keep the active set and the buckets fixed during a balanced search.
Split the candidate set into two nearly equal parts, evaluate one
part's total score, and obtain the other's by subtraction.  Retain
the part with larger average surrogate score, with fixed tie-breaking.
The average never decreases.  Since the surrogate scores are
nonnegative, the initial average over the at most $N$ candidates is at
least $\frac1N\sum_{x\notin C}\widetilde\sigma(x)$, so the selected
vertex $u$ satisfies
\[
 \sigma(u)\geq\widetilde\sigma(u)
 \geq\frac1N\sum_{x\notin C}\widetilde\sigma(x)
 \geq\frac{k}{8N}\Phi.
\]
This search costs $O(\log^2(2N))$ queries.  Add $u$ to $C$ and
learn its full original row with $O(N)$ queries, subtracting $K$
to obtain its residual row.

A previously active vertex that stays active has its potential
multiplied by $1-a_{vu}$, so it decreases by $p_va_{vu}$.  Other
previously active vertices leave the sum; since $0\leq a_{vu}\leq1/2$,
their potentials $p_v$ are nonnegative and at least $p_va_{vu}$.  No
new vertex becomes active because $C$ and each $b_v$ only increase.
Thus $\Phi$ decreases by at least $\sigma(u)\geq k\Phi/(8N)$.  Initially $\Phi\leq N$, whereas
any nonempty active set has $\Phi>1/2$.  At most
$8(N/k)\ln(2N)+1\leq L$ selections therefore suffice.
The total query cost is
\[
 O\!\left(\frac{N^2}{k}\log(2N)
           +\frac{N}{k}\log^3(2N)\right)
 =O\!\left(\frac{N^2}{k}\log(2N)\right).
\]
When the active set is empty, \eqref{eq:center-coverage} holds.
\end{proof}

\subsection{Keeping known edges and rounding only the residual}

Suppose preparation and center selection both finish.  The explicit
graph
\[
 H=K+H_C^R\subseteq G
\]
has disjoint contributions from $K$ and $H_C^R$.
Use Lemma~\ref{lem:offline-hierarchy} to find a canonical hierarchy
$T_H$ approximating all its cuts with distortion
$a_1=O(\log(2N)\log\log(2N))$, $a_1\geq1$.
Choose an arbitrary vertex $v_0$ and apply
Lemma~\ref{lem:tree-compression} with terminal set
\[
 T_0=C\cup\{v_0\}.
\]
The extra vertex ensures that this set is nonempty; no additional
row is learned for it.  The nonempty fibers on original vertices give
a fixed partition $P_1,\ldots,P_p$, where
\begin{equation}\label{eq:general-partition-size}
 p\leq N,\qquad
 p\leq\max\{1,2|T_0|-2\}
   =O\!\left(1+\frac{N}{k}\log(2N)\right).
\end{equation}

For every cut $S$, minimum-cost extension of its labels on $T_0$
gives a union $\widehat S$ of these classes with
\begin{equation}\label{eq:general-H-rounding}
 \widehat S\cap T_0=S\cap T_0,\qquad
 c_H(\widehat S)\leq a_1 c_H(S).
\end{equation}
To verify this, fix only the labels on $T_0$.
The original labeling of all leaves costs at most $a_1 c_H(S)$,
terminal compression permits a minimizing extension constant on
fibers, and the canonical tree dominates $H$.

Put $X=S\mathbin\triangle\widehat S$.
No vertex in $X$ belongs to $C$.  Charge the vertices with
$r_v\leq t$ at most $Nt$ in total, and use
\eqref{eq:center-coverage} on the others.  Since $V\setminus C$
is independent in $H_C^R$,
\begin{align}
 \vol_R(X)
 &\leq k\vol_{H_C^R}(X)+Nt
   =k c_{H_C^R}(X)+Nt\notag\\
 &\leq k\bigl(c_{H_C^R}(S)+c_{H_C^R}(\widehat S)\bigr)+Nt\notag\\
 &\leq k(1+a_1)c_G(S)+Nt.
 \label{eq:general-residual-rounding}
\end{align}
There is no corresponding bound claimed for $\vol_G(X)$ or
$\vol_K(X)$.  The known edges will be retained explicitly instead.

Learn the quotient $Q=G/\{P_1,\ldots,P_p\}$ with $O(p^2)$ queries.
This partition is used for rounding and defining the auxiliary graph.
Safe contractions in the algorithm are performed only by
\cref{lem:degree-normalization} on the current graph $G$.
Construct a graph $\Jaux$ with terminal set $V(G)$ and one additional
vertex $z_i$ for each part $P_i$, as follows:
\begin{enumerate}
 \item Keep the edges of $K$ between the original vertices.
 \item Put the quotient graph $Q$ on $z_1,\ldots,z_p$.
 \item For $v\in P_i$, add an edge $vz_i$ of weight $r_v$.
\end{enumerate}
All its weights are known.  For $S\subseteq V(G)$, let $\Psi(S)$
be its minimum terminal-extension cut value in $\Jaux$.  Equivalently,
\begin{equation}\label{eq:general-J-formula}
 \Psi(S)=c_K(S)+
 \min_{\widehat S\text{ a union of }P_i}
       \bigl(c_G(\widehat S)
                 +\vol_R(S\mathbin\triangle\widehat S)\bigr).
\end{equation}

\begin{lemma}[Auxiliary graph with an additive error]\label{lem:general-auxiliary}
Set $A=(1+a_1)(2k+1)$, so that $A=O(k\log(2N)\log\log(2N))$.  For
every cut $S$,
\begin{equation}\label{eq:general-J-bounds}
 c_G(S)\leq\Psi(S)\leq A c_G(S)+2Nt.
\end{equation}
The explicit graph $\Jaux$ has at most $2N$ vertices and requires only
$O((N^2/k^2)\log^2(2N))$ queries after residual preparation and
center selection.
\end{lemma}
\begin{proof}
For any union $\widehat S$ in \eqref{eq:general-J-formula}, changing
the labels in $X=S\mathbin\triangle\widehat S$ changes an $R$-cut
by at most $\vol_R(X)$.  Thus
\[
 c_G(\widehat S)+\vol_R(X)
 \geq c_R(\widehat S)+\vol_R(X)\geq c_R(S).
\]
Adding $c_K(S)$ and minimizing proves the lower bound.

For the upper bound, use the rounded cut from
\eqref{eq:general-H-rounding}.  We have
\[
 c_K(\widehat S)\leq a_1 c_G(S),\qquad
 c_R(\widehat S)\leq c_R(S)+\vol_R(X).
\]
Substituting $c_G(\widehat S)=c_K(\widehat S)+c_R(\widehat S)$ in
\eqref{eq:general-J-formula} and using $c_K(S)+c_R(S)=c_G(S)$, then
applying \eqref{eq:general-residual-rounding}, gives
\[
 \Psi(S)\leq(1+a_1)c_G(S)+2\vol_R(X)
             \leq A c_G(S)+2Nt.
\]
The size and query claims follow from
\eqref{eq:general-partition-size}.  All hierarchy choices and tree
compression steps operate on already explicit graphs.
\end{proof}

\subsection{A test that does not require knowing the optimum}

Compute offline the minimum terminal-separating cut of $\Jaux$:
\begin{equation}\label{eq:general-mu}
 \mu=\min_{\varnothing\ne S\subsetneq V(G)}\Psi(S),
 \qquad \Psi(S_{\mathrm{aux}})=\mu.
\end{equation}
The terminal side $S_{\mathrm{aux}}$ is nontrivial.  The unrestricted global
minimum cut of $\Jaux$ would not suffice here, since it might isolate
only added vertices.  Enumeration of terminal labelings and their
extensions is one valid deterministic implementation.
Lemma~\ref{lem:general-auxiliary} implies
\begin{equation}\label{eq:general-mu-bounds}
 \lambda\leq\mu\leq A\lambda+2Nt.
\end{equation}

\begin{lemma}[Certify a hierarchy or halve the graph]\label{lem:general-certificate}
If $\mu\geq4Nt$, then $O(N)$ additional queries construct a canonical
hierarchy of $G$ whose representation cost $R_G$ satisfies
\[
 c_G(S)\leq R_G(S)\leq Bc_G(S)\quad\text{for every }S\subseteq V(G),
 \qquad B=O(k\log^2(2N)(\log\log(2N))^2),
\]
with an explicit valid value of $B$ supplied to the algorithm.
If $\mu<4Nt$, one additional query gives a new witnessed bound
\begin{equation}\label{eq:halve-bound}
 \Lambda'=c_G(S_{\mathrm{aux}})\leq\mu<\frac{\Lambda}{4N}.
\end{equation}
Weighted degree normalization with this new bound leaves at most
$\lfloor N/2\rfloor$ vertices before the next phase.
\end{lemma}
\begin{proof}
Suppose first that $\mu\geq4Nt$.  Combining this with
\eqref{eq:general-mu-bounds} gives $2Nt\leq A\lambda$.
Since $c_G(S)\geq\lambda$ for every nontrivial cut, the
additive error is absorbed uniformly:
\begin{equation}\label{eq:general-good-optimum}
 \Psi(S)\leq A c_G(S)+2Nt\leq2A c_G(S)
 \qquad(\varnothing\ne S\subsetneq V(G)).
\end{equation}
For trivial cuts, $\Psi(S)=c_G(S)=0$.

Apply Lemma~\ref{lem:offline-hierarchy} to the explicit graph $\Jaux$
on all its vertices, obtaining a tree of distortion
$a_2=O(\log(2N)\log\log(2N))$, $a_2\geq1$.  Every vertex of $\Jaux$
is a leaf of this tree, since in a hierarchy the nodes of the vertex
set have no children; in particular every terminal in $V(G)$ already
has degree one.  In this tree leave the leaves
corresponding to the added vertices of $\Jaux$ free, and apply
Lemma~\ref{lem:tree-compression} with terminal set $V(G)$.
The resulting tree $T'$ has $O(N)$ vertices and preserves these
terminal-extension costs.  Minimizing the original tree inequalities
over the labels of the added leaves proves
\begin{equation}\label{eq:general-terminal-tree}
 c_G(S)\leq\Psi(S)\leq R_{T'}(S)
                         \leq a_2\Psi(S).
\end{equation}
Compression contracts only leaves outside $V(G)$ and shortens
terminal-free paths, so each terminal keeps degree one and every
remaining leaf is a terminal: the leaves of $T'$ are exactly $V(G)$.
Root it at an internal vertex; for
two terminals subdivide its single edge into two edges of the same
capacity and use the new vertex as root.

For an edge $e$ of this rooted tree, let $A_e$ be its descendant
terminal set.  Query $c_G(A_e)$ and replace the edge capacity by this
number.  This never increases a capacity: cutting $e$ alone is a
feasible extension of the terminal cut $A_e$, so
\[
 c_G(A_e)\leq R_{T'}(A_e)\leq\operatorname{cap}_{T'}(e).
\]
The resulting canonical hierarchy dominates every $G$-cut by the
usual cut triangle inequality, and its extension costs are at most
those of $T'$.  By \eqref{eq:general-good-optimum} and
\eqref{eq:general-terminal-tree}, the explicit choice $B=2a_2 A$
works.  The values $a_1,a_2$ can be computed by exhaustive
evaluation of the distortions of the chosen explicit hierarchies.
There are only $O(N)$ newly queried cluster capacities.

Now suppose $\mu<4Nt$.  The lower inequality in
\eqref{eq:general-J-bounds} gives $c_G(S_{\mathrm{aux}})\leq\mu$.
Since $4Nt=\Lambda/(4N)$, querying this nontrivial cut proves
\eqref{eq:halve-bound}.  Before the decrease every weighted degree is at
least $\Lambda$, so \cref{lem:normalization-halves} applies.  The next
phase therefore starts on at most half as many vertices, independently
of the weight spread.
\end{proof}

\subsection{The tree cut sparsifier tradeoff}\label{sec:tree-sparsifier-proof}

\begin{proof}[Proof of \cref{thm:tree-sparsifier-discovery}]
Fix an absolute constant $D\geq1$ large enough that a successful
phase on $N$ vertices with parameter $k$ produces a canonical
hierarchy of distortion at most $DkL(N)^2K(N)^2$, by
\cref{lem:general-certificate}, and that the offline hierarchy of
\cref{lem:offline-hierarchy} has distortion at most $DL(N)K(N)$.
Choose the constant $C$ in the theorem large enough that its lower
bound on $\alpha$ implies $\alpha\geq2DL(n)^2K(n)^2$.

Query the weighted degrees, initialize $\Lambda$ with a singleton of minimum weighted degree,
and apply \cref{lem:degree-normalization}.  On the current quotient
with $N$ vertices, set
\[
 k=\left\lfloor\frac{\alpha}{DL(N)^2K(N)^2}\right\rfloor\geq2.
\]
If $k\geq N$, reconstruct this quotient using $O(N^2)$ queries
and compute its offline hierarchy.  Its distortion is at most
$DL(N)K(N)\leq\alpha$, and the query cost is $O(N\alpha)$.
Otherwise run a discovery phase with this $k$: residual
preparation (\cref{lem:residual-preparation}), center selection
(\cref{lem:bucketed-centers}), reconstruction of the rounding
quotient and assembly of the auxiliary graph
(\cref{lem:general-auxiliary}), and the test of
\cref{lem:general-certificate}.  On success, its
hierarchy has distortion at most $DkL(N)^2K(N)^2\leq\alpha$.
On failure, \cref{lem:general-certificate} supplies a smaller
witnessed bound; weighted degree normalization then at least halves $N$.

For the query bound, write $L=L(N)$ and $K=K(N)$.  Adding the costs stated in
the four lemmas, one phase through its test and canonicalization uses
\begin{equation}\label{eq:phase-query-cost}
 O\!\left(NkL^2+\frac{N^2}{k}L
                  +\frac{N^2}{k^2}L^2+N\right)
\end{equation}
queries, the third term being the $O(p^2)$ reconstruction of the
rounding quotient on $p$ parts.  By
\eqref{eq:general-partition-size}, the rounding partition has
$p\leq N$ parts and $p=O((N/k)L)$.  Thus
$p^2\leq Np=O(N^2L/k)$.
Using this bound for quotient reconstruction in
\eqref{eq:phase-query-cost}, one phase costs
\[
 O\!\left(NkL^2+\frac{N^2L}{k}+N\right)
 =O\!\left(N\alpha+\frac{N^2L^3K^2}{\alpha}\right),
\]
since $\alpha/(2DL^2K^2)\leq k\leq\alpha/(DL^2K^2)$.
The successive phase sizes decrease geometrically, so their sum
is at most $2n$ and the sum of their squares is $O(n^2)$.
Since $L(N)\leq L(n)$ and $K(N)\leq K(n)$, including a possible final reconstruction
and the $O(n\log(2n))$ queries for all normalization calls gives
\[
 O\!\left(n\alpha+\frac{n^2L(n)^3K(n)^2}{\alpha}+nL(n)\right)
 =O\!\left(\frac{n^2}{\alpha}\log^3(2n)(\log\log(2n))^2\right)
\]
queries in total, where the last step uses $\alpha^2\leq n$.

Let $\mathcal P$ be the final partition into contracted classes.
Every contraction preserves every minimum cut, so $\mathcal P$
is safe.
The final canonical hierarchy has at most $2N-1\leq8n$ vertices
and is the required $\alpha$-tree cut sparsifier of $G/\mathcal P$.
The formalization omits the compression of \cref{lem:tree-compression}
in \cref{lem:general-certificate} and keeps the hierarchy of
$\Jaux$ restricted to the terminals $V(G)$, which has at most
$2(N+p)+1\leq8N$ nodes; this is the source of the bound $8n$.
\end{proof}

\subsection{The complete algorithm and its query bound}\label{sec:complete-algorithm}

\begin{proof}[Proof of \cref{thm:all-mincuts}]
For $n=2$, a singleton query returns a minimizing shore and its
value, and the two singletons are the only shores.  If $n<2$, there is no proper shore; one may return no shore
and value zero by convention, without queries.  The theorem concerns
$n\geq2$.

First find a maximal forest of the
positive support with $O(n\log(2n))$ queries by \cref{lem:forest}.
If the support is disconnected, return value zero, a proper component
as a minimizing shore, and the partition of $V$ into components.  A cut
has value zero exactly when it is a union of components, so the proper
unions are exactly the minimum-cut shores, including in the all-zero
graph.  Otherwise $\lambda=\lambda(G)>0$.

Let $C$ be the constant of \cref{thm:tree-sparsifier-discovery}, and
fix an absolute integer $N_0\geq2$ such that
\[
 C\log^2(2n)(\log\log(2n))^2\leq\lfloor\sqrt n\rfloor
 \qquad(n\geq N_0).
\]
Such an $N_0$ exists because every fixed power of a logarithm is
$o(\sqrt n)$, and it is fixed before any oracle answer is read.  For
$n<N_0$, reconstruct $G$ from its singleton and pair cuts with
$O(N_0^2)=O(1)$ queries and compute a minimum cut offline.

For $n\geq N_0$, apply \cref{thm:tree-sparsifier-discovery} with
$\alpha=\lfloor\sqrt n\rfloor$.  Using
$O((n^2/\alpha)\log^3(2n)(\log\log(2n))^2)=O(n^{3/2}\log^3(2n)(\log\log(2n))^2)$ queries, it returns
a safe partition $\mathcal P$ of $V(G)$ and a canonical hierarchy of
the quotient $G_q=G/\mathcal P$, on $N\leq n$ vertices and with at
most $8n$ nodes, with
$c_{G_q}(S)\leq R(S)\leq\alpha c_{G_q}(S)$ for every quotient shore
$S$ (\cref{sec:tree-sparsifier-proof}).  Since $\mathcal P$ is safe,
$N\geq2$, $\lambda(G_q)=\lambda>0$, and every minimum cut $S_q^*$ of
$G_q$ satisfies $R(S_q^*)\leq\alpha\lambda$.  Now apply
\cref{thm:linear-hierarchy} to $G_q$ with $B=\alpha$.  Using
$O(N\alpha\log^3(2N)(\log\log(2N))^2)=O(n^{3/2}\log^3(2n)(\log\log(2n))^2)$ further queries, each
of them one query to $G$ on a union of classes, it constructs a weak
skeleton $H_q$ of $G_q$: $\lambda(H_q)\geq\lambda$ and
$c_{H_q}(S_q^*)\leq(5/4)\lambda$ for every minimum cut $S_q^*$ of
$G_q$.  The parameter balance happens here: learning costs
$O((n^2/\alpha)\log^3(2n)(\log\log(2n))^2)$ and conversion
$O(n\alpha\log^3(2n)(\log\log(2n))^2)$, and $\alpha=\Theta(\sqrt n)$
equalizes them.

To produce a skeleton on the original $n$ vertices, choose a fixed
representative of each contracted class, realize the edges of $H_q$
between these representatives, and connect the vertices inside each
class by any explicit tree.  Give every such internal tree edge
capacity $2\lambda(H_q)$, a known positive number.  Every cut that
splits a class has value at least $2\lambda(H_q)$; cuts respecting
all classes have their exact $H_q$-values.  Therefore the lifted
graph $H$ satisfies
\[
 \lambda(H)=\lambda(H_q)\geq\lambda.
\]
Since $\mathcal P$ is safe, every original minimum cut is a union of
classes and has its exact $H_q$-value, so every original minimum cut
has $H$-value at most $(5/4)\lambda$.  Thus $H$ is a weak skeleton
of $G$, obtained without further oracle queries.

Finally, Lemma~\ref{lem:tree-cover} supplies $O(\log n)$ explicit
trees covering every $(5/4)$-minimum cut of $H$, hence every minimum
cut of $G$.  The algorithm of \cref{thm:2respecting} finds the minimum
$G$-cut over these trees with $O(n\log^2 n)$ more queries.  Query the
returned shore once more, if its value has not already been recorded,
to learn $\lambda$ exactly; this adds at most one query.  Then apply
\cref{thm:all-2respecting} with $\ell=\lambda$ to each tree and return
$\lambda$ together with the reported shores and their complements.
Every minimum cut of $G$, or its complement, crosses at most two edges
of some cover tree, so every minimum cut is reported, and every
reported shore is a cut of $G$ of value $\lambda$.  A cut may be
reported by several of the $O(\log n)$ trees, or by one tree as a
shore and by another as its complement; the final list is
deduplicated offline, at no query cost.  This uses
$O(n\log^2n)$ further queries, independent of the number of minimum
cuts.

All choices depend only on $n$, explicit graphs, and previous answers.
The estimates give a fixed cap $C'n(\lfloor\sqrt n\rfloor+1)L(n)^3K(n)^2$
for a universal constant $C'$.  If desired, stop any inconsistent
answer path at this cap and return failure.  Correctness on every
nonnegative weighted graph shows that an honest path never fails.
Thus one deterministic decision tree is selected for each $n$ before
the input graph, and its cap holds for every answer function.
\end{proof}

\begin{remark}[Where the logarithmic factors are saved]\label{rem:log-saving}
Suppose the offline distortion is $a=a(N)$ and the guard quality is
constant.  The two offline hierarchies in discovery give
$B=O(ka^2)$.  Refinement costs $O(NBa^2L)=O(Nka^4L)$, while
learning center rows costs $O(N^2L/k)$.  Balancing these terms gives
\[
 k\asymp\sqrt N/a^2,\qquad
 Q(N)=O(N^{3/2}a^2L).
\]
A cubic-distortion bound $a=O(L^3)$ gives $L^7$, a quadratic
bound $a=O(L^2)$ gives $L^5$, and the R\"acke--Shah bound $a=O(LK)$
gives $L^3K^2$.  The constant guard
and geometric phase sum are essential to this accounting: neither
contributes a hidden factor of $L$.
\end{remark}

\begin{remark}[The exponent is independent of the terminal sparsifier]\label{rem:guard-independent}
\Cref{lem:offline-hierarchy} also supplies a canonical tree of height
$O(\log N)$.  Choose the initial hierarchy of the explicit auxiliary
graph and each subsequent refinement hierarchy among such trees.
Terminal compression cannot increase diameter, so the initial
canonical hierarchy also has height $O(\log N)$.

The elementary guard of \cref{rem:schur} then has a uniform
quality bound $\rho_0=O(\log(2N))$.  The same refinement proof,
with $\gamma=(100a^2\rho_0^2)^{-1}$, costs
$O(Na^2\rho_0^2B\log(2N))$ until the remaining bound satisfies
$b<1/\gamma$.  The final elementary completion costs
\[
 O(N\rho_0^2b^2\log(2N))
   =O(Na^2\rho_0^4B\log(2N))
   =O(NB\log^7(2N)(\log\log(2N))^2).
\]
With $B=O(k\log^2(2N)(\log\log(2N))^2)$ and $k=\lceil\sqrt N\rceil$, this
gives $O(N^{3/2}\log^9(2N)(\log\log(2N))^4)$ queries, so the exponent
$3/2$ is unchanged.  The constant-quality terminal sparsifier improves
the logarithmic factor.
\end{remark}

\section{Deterministic two-respecting cuts}\label{app:2respecting}

We prove \cref{thm:2respecting} using the interested-path and
Monge-matrix structure of Gawrychowski, Mozes, and Weimann~\cite{GMW}.
All input-dependent measurements will be ordinary cut queries.
The given tree is any tree on $V$, including one with edges absent
from $G$.  A cut two-respects the tree when it crosses at most two
tree edges; one-edge cuts are included.

Root $T$ at any vertex $r$.  For a vertex $v$, let $T_v=\sub(v)$ be its
set of descendants, including $v$.  For an edge $e=(\operatorname{parent}(v),v)$,
write $A_e=T_v$ and $a_e=\cut(A_e)$.  Two distinct tree edges are
\emph{independent} if their subtrees are disjoint; otherwise one is a
proper ancestor of the other.

\begin{lemma}[Cut-query identities]\label{lem:identities}
For disjoint $A,B\subseteq V$,
\begin{equation}\label{eq:bipartite}
  2w(A,B)=\cut(A)+\cut(B)-\cut(A\cup B).
\end{equation}
For distinct tree edges $e,f$, the unique cut, up to complementation,
crossing precisely $e$ and $f$ has the side $A_e\mathbin{\triangle}A_f$.
Its weight
\[
  K(e,f)=\cut(A_e\mathbin{\triangle}A_f)
\]
can therefore be evaluated with one cut query.  Moreover,
\begin{align}
 K(e,f)&=a_e+a_f-2w(A_e,A_f)
   &&\text{if $e,f$ are independent},\label{eq:independent}\\
 K(e,f)&=a_e+a_f-2w(A_f,V\setminus A_e)
   &&\text{if $f$ is below $e$}.\label{eq:nested}
\end{align}
\end{lemma}

\begin{proof}
In $\cut(A)+\cut(B)$, edges between $A$ and $B$ are counted twice,
and all other contributions are exactly $\cut(A\cup B)$, proving
\eqref{eq:bipartite}.  The indicator of $A_e$ changes along exactly the
tree edge $e$.  Taking the exclusive-or of the indicators of $A_e$ and
$A_f$ gives the assertion about their symmetric difference.  Conversely,
after choosing the side containing the root, membership in a cut is
determined by the tree edges it crosses, so the cut is unique up to
complementation.  The two displayed formulas follow by cancelling the
graph edges counted twice in $a_e+a_f$.
\end{proof}

Query all $a_e$, using $n-1$ queries, and store
\begin{equation}\label{eq:baseline}
  \lambda_1=\min_{e\in E(T)}a_e
\end{equation}
together with a minimizing edge.  If $\lambda_1=0$, return that cut:
nonnegative weights make it optimal.  Henceforth assume $\lambda_1>0$,
so every $a_e$ is positive.  We only need to ensure that every two-edge
cut of weight \emph{strictly less} than $\lambda_1$ is considered.
If there is no such cut, the stored one-edge cut is already an answer.
This observation handles all equality cases in the strict interest tests.

\subsection{Interest paths and their deterministic discovery}

\subsubsection{Two boundary measures for each tree edge}

For each tree edge $e$, put $A=A_e$ and define two nonnegative additive
measures on vertex subsets:
\begin{align}
 \mu_e^{\out}(X)&=w(A,X\setminus A),\label{eq:outmeasure}\\
 \mu_e^{\inn}(X)&=w(A\cap X,V\setminus A).\label{eq:inmeasure}
\end{align}
They distribute the weight of the boundary of $A$ over its outside
endpoints and its inside endpoints, respectively.  Both measures have
total mass $a_e$.  The following identities show that each evaluation
costs at most two additional cut queries, since $a_e$ is already stored:
\begin{align}
 2\mu_e^{\out}(X)
   &=a_e+\cut(X\setminus A)-\cut(A\cup X),\label{eq:outoracle}\\
 2\mu_e^{\inn}(X)
   &=a_e+\cut(A\cap X)-\cut(A\setminus X).\label{eq:inoracle}
\end{align}
The first follows directly from \eqref{eq:bipartite}.  For the second,
apply \eqref{eq:bipartite} to $A\cap X$ and $V\setminus A$, and use
the symmetry of the cut function.

For any nonnegative measure $\mu$ of total mass $D>0$, call a vertex $v$
\emph{heavy for $\mu$} if
\[
  \mu(T_v)>D/2.
\]
The root is heavy.  Two incomparable vertices cannot both be heavy,
because their subtrees are disjoint.  Also, every ancestor of a heavy
vertex is heavy.  Consequently the heavy vertices are precisely the
vertices on a root-to-vertex path.

Let $C_e$ be this path for $\mu_e^{\out}$, and let $D_e$ be this path
for $\mu_e^{\inn}$; write $c_e,d_e$ for their last vertices.  We use
the same symbols for the edge sets of these paths.  Thus an edge $f$
with lower endpoint $v$ belongs to $C_e$ or $D_e$ exactly when the
corresponding measure of $T_v$ is greater than $a_e/2$.  A path ending
at the root has no edges.

Say that $e$ is \emph{interested in} $f$ if $f\in C_e\cup D_e$.
For independent edges this is exactly the cross-interest inequality
from~\cite{GMW}, because
\[
 \mu_e^{\out}(A_f)=w(A_e,A_f),\qquad
 \mu_e^{\inn}(A_f)=0.
\]
If $f$ is below $e$, membership in $D_e$ is exactly the down-interest
inequality from that paper.  If $f$ is an ancestor of $e$, then
$\mu_e^{\inn}(A_f)=a_e$, so $f\in D_e$ automatically.
The measure definition also specifies harmless extra cross-interest
tests on ancestors.  Its advantage is that the two sets really are
entire root paths, without any ambiguity at ancestors or at the root.

\begin{lemma}[Interest captures every improvement]\label{lem:interest}
If $K(e,f)<\lambda_1$, then $e$ is interested in $f$ and $f$ is
interested in $e$.
\end{lemma}

\begin{proof}
If $e,f$ are independent, then $K(e,f)<\lambda_1\leq a_f$ and
\eqref{eq:independent} imply
$2w(A_e,A_f)>a_e$.  Hence $f\in C_e$.  Interchanging $e$ and $f$
gives $e\in C_f$.

If $f$ is below $e$, then $K(e,f)<\lambda_1\leq a_f$ and
\eqref{eq:nested} imply
$2w(A_f,V\setminus A_e)>a_e$.  Hence $f\in D_e$.
In the other direction, $A_f\subset A_e$ gives
$\mu_f^{\inn}(A_e)=a_f>a_f/2$, so $e\in D_f$.
The other nesting order is symmetric.
\end{proof}

\subsubsection{Finding the last heavy vertex without seeing the graph}

The centroid search in~\cite[Section~3.1]{GMW} is one way of locating
interest-path endpoints.  In the cut-query model, the following search
is convenient: it works directly on a tree of arbitrary degree and
requires no auxiliary vertices or infinite-weight edges.

\begin{lemma}[Majority-subtree search]\label{lem:majority}
Let $\mu$ be an additive nonnegative measure on $V$, with known total
mass $D>0$, and suppose $\mu(X)$ can be evaluated for any specified
subset $X$.  The deepest vertex $c$ with $\mu(T_c)>D/2$ can be found
deterministically with $O(\log n)$ measure evaluations.
\end{lemma}

\begin{proof}
Fix a depth-first preorder $v_1,\ldots,v_n$ of $T$; every rooted
subtree is an interval in this order.  Let $F(k)=\mu(\{v_1,\ldots,v_k\})$,
with $F(0)=0$.  Binary search finds the smallest $k$ with $F(k)>D/2$
using $O(\log n)$ evaluations, since $F$ is nondecreasing and $F(n)=D$.

Every subtree of mass greater than $D/2$ contains $v_k$.  Indeed, if
the subtree occupies the interval $[\ell,u]$, then
\[
 F(\ell-1)\leq D-\mu(T_v)<D/2,
 \qquad F(u)\geq\mu(T_v)>D/2,
\]
so $\ell\leq k\leq u$.  Its root must therefore be an ancestor of
$v_k$.  Along the known path from $r$ to $v_k$, subtree masses are
nonincreasing.  A second binary search finds the last vertex whose
subtree mass is greater than $D/2$, again in $O(\log n)$ evaluations.
The root satisfies the test, so this last vertex always exists.
\end{proof}

Apply this lemma to each of $\mu_e^{\out}$ and $\mu_e^{\inn}$,
using \eqref{eq:outoracle} and \eqref{eq:inoracle} for the evaluations.
Thus all endpoints $c_e,d_e$ are found with $O(n\log n)$ cut queries.
Only these endpoints are stored; enumerating every edge of every
interest path is unnecessary.

\subsection{A small collection of path-pair problems}

Construct a heavy-path decomposition $\paths$ of the \emph{edges} of
$T$ using only its topology.  One precise convention is as follows.
At each nonleaf vertex choose a child of maximum subtree size, marking
the edge to that child heavy and the others light.  Continue each
incoming edge through the heavy outgoing edge, if one exists.
The resulting maximal downward paths partition all tree edges;
a nonroot path may begin with a light edge.  Omit empty paths.
Every root-to-vertex path intersects $O(\log n)$ paths of $\paths$:
each light edge reduces the remaining subtree size by at least a
factor of two, and all consecutive heavy edges stay on one path.

For $e\in P\in\paths$, let $\mathcal I(e)$ consist of the paths
$Q\in\paths\setminus\{P\}$ that contain an edge of $C_e\cup D_e$.
Because $C_e,D_e$ are root paths,
\begin{equation}\label{eq:interest-count}
 |\mathcal I(e)|=O(\log n),\qquad
 \sum_{e\in E(T)}|\mathcal I(e)|=O(n\log n).
\end{equation}
The endpoints $c_e,d_e$ and the decomposition determine these sets
without further cut queries.  They can be enumerated by jumping upward
between heavy paths and deduplicating the two resulting lists.

For distinct $P,Q\in\paths$, define the ordered lists
\[
 L_{P,Q}=\{e\in P:Q\in\mathcal I(e)\},
 \qquad
 L_{Q,P}=\{f\in Q:P\in\mathcal I(f)\},
\]
where each list is initially ordered from the root toward the leaves.
Retain the unordered pair $\{P,Q\}$ only when both lists are nonempty.
These are the \emph{interesting path pairs}.
They are constructed by producing a record $(P,Q,e)$ for each
$Q\in\mathcal I(e)$ and sorting/grouping the records by their path
identifiers.  This is computation on known data, with no oracle calls.
Equation~\eqref{eq:interest-count} gives
\begin{equation}\label{eq:list-budget}
 \sum_{\{P,Q\}\text{ retained}}
       \bigl(|L_{P,Q}|+|L_{Q,P}|\bigr)=O(n\log n).
\end{equation}
In particular there are only $O(n\log n)$ retained pairs.

If $K(e,f)<\lambda_1$ and $e,f$ lie on different paths $P,Q$,
Lemma~\ref{lem:interest} ensures both that $\{P,Q\}$ is retained and
that $e\in L_{P,Q}$ and $f\in L_{Q,P}$.
We will search the \emph{entire} Cartesian product of these lists.
Pairs in this product need not themselves be mutually interested;
including such extra pairs is harmless and allows a full Monge matrix
to be searched.

\subsection{Searching the pair-cut matrices}

We use the convention that a matrix $M$ is \emph{Monge} when
\begin{equation}\label{eq:monge}
 M[i,j]+M[i',j']\leq M[i,j']+M[i',j]
 \qquad (i<i',\ j<j').
\end{equation}
The orientations below give this sign, appropriate for finding minima.

\begin{lemma}[Full Monge rectangles]\label{lem:monge}
Let $e_1,\ldots,e_a$ and $f_1,\ldots,f_b$ be two nonempty lists of
tree edges, each taken from a downward path, and put
$M[i,j]=K(e_i,f_j)$.  The matrix is Monge in either of these cases:
\begin{enumerate}
 \item All edges in one list are independent of all edges in the other,
 and both lists are ordered from the root toward the leaves.
 \item Every $e_i$ is an ancestor of every $f_j$, the $e_i$ are ordered
 from the root toward the leaves, and the $f_j$ are ordered toward
 the root.
\end{enumerate}
The lists may have gaps along their paths.
\end{lemma}

\begin{proof}
Write $A_i=A_{e_i}$ and $B_j=A_{f_j}$.
In the first case, for $i<i'$ and $j<j'$ we have
$A_{i'}\subset A_i$ and $B_{j'}\subset B_j$, and every $A$-set is
disjoint from every $B$-set.  The row and column terms in
\eqref{eq:independent} cancel, leaving
\begin{align*}
 &M[i,j]+M[i',j']-M[i,j']-M[i',j]\\
 &\hspace{2em}=-2w(A_i\setminus A_{i'},B_j\setminus B_{j'})\leq 0.
\end{align*}
In the second case $A_{i'}\subset A_i$ and $B_j\subset B_{j'}$,
with $B_{j'}\subset A_{i'}$.
Put $R_i=V\setminus A_i$, so $R_i\subset R_{i'}$.
Equation~\eqref{eq:nested} reads
$M[i,j]=a_{e_i}+a_{f_j}-2w(B_j,R_i)$, and therefore
\begin{align*}
 &M[i,j]+M[i',j']-M[i,j']-M[i',j]\\
 &\hspace{2em}=-2w(B_{j'}\setminus B_j,R_{i'}\setminus R_i)\leq 0.
\end{align*}
Both inequalities use nonnegative graph weights.  The differences
are disjoint sets, and nothing requires consecutive edges in a list.
\end{proof}

\begin{lemma}[Matrix searching: SMAWK~\cite{SMAWK}]\label{lem:smawk}
The minimum entry of a full $a\times b$ Monge matrix, together with
its row and column indices, can be found deterministically using
$O(a+b)$ entry evaluations.
\end{lemma}

To apply the standard theorem, note that leftmost row-minimum indices
are nondecreasing in every submatrix of a Monge matrix.  Such a matrix
is totally monotone for row minima.  SMAWK finds those minima using
$O(a+b)$ entry evaluations; comparing their values finds the minimum
of the matrix.  Each entry evaluation in our application is one cut
query by Lemma~\ref{lem:identities}.

\subsubsection{Two different heavy paths}

Consider a retained pair $P,Q$ and its two lists.  Split their Cartesian
product into at most two full rectangles so that each rectangle is
either wholly independent or wholly nested, as in
Lemma~\ref{lem:monge}.  Here is why this is possible.

For two edge-disjoint downward paths, either all their edges are
independent, or, after possibly swapping their names, an edge of $P$
is above an edge of $Q$.  Such an edge must be above the first edge
of $Q$, since otherwise it would itself lie on $Q$.  Thus a prefix
of $P$ is above every edge of $Q$.  The remainder of $P$, if nonempty,
is independent of every edge of $Q$: it has branched away from $Q$,
since following the branch into or beyond $Q$ would make the paths
share an edge.
Thus split the list from $P$ into the ancestor part and the independent
part, and pair each nonempty part with the entire list from $Q$.
This argument also covers paths meeting at an endpoint or lying
consecutively on the same branch.  The split is determined by tree
ancestry alone.

Orient each rectangle as in Lemma~\ref{lem:monge} and apply
Lemma~\ref{lem:smawk}.  The sum of its row and column counts, over
the at most two rectangles, is at most
$2(|L_{P,Q}|+|L_{Q,P}|)$.  Consequently the total number of cut
queries over all retained path pairs is, by \eqref{eq:list-budget},
\begin{equation}\label{eq:different-cost}
 O\!\left(\sum_{\{P,Q\}\text{ retained}}
     (|L_{P,Q}|+|L_{Q,P}|)\right)=O(n\log n).
\end{equation}

\subsubsection{Two edges on the same heavy path}

Let $P$ have $k$ edges in root-to-leaf order.  If $k\leq 1$, there is
nothing to search.  Otherwise split $P$ into an upper half $U$ and a
lower half $W$, of sizes differing by at most one.  Every edge of
$U$ is an ancestor of every edge of $W$.  Search all pairs in
$U\times W$ using the nested orientation in
Lemma~\ref{lem:monge} and SMAWK, at a cost of $O(k)$ cut queries.
Then recurse on each half.

Every pair of distinct edges is separated at exactly one split, so it
occurs in one of the searched rectangles.  At each recursion depth
the subpaths have total length at most $k$, and there are
$O(\log(k+1))$ depths.  This costs $O(k\log(k+1))$ queries for $P$,
and hence
\begin{equation}\label{eq:same-cost}
 \sum_{P\in\paths}O(|P|\log(|P|+1))=O(n\log n)
\end{equation}
over all heavy paths.  This avoids searching a partial matrix or
assigning artificial entries to its diagonal.

\subsection{The algorithm, correctness, and query count}

The complete algorithm is the following.
\begin{enumerate}
 \item Root the given tree and compute its preorder, ancestry
 information, and heavy-path decomposition.  Query all $a_e$, store
 a minimum one-edge cut, and return it immediately if its weight is zero.
 \item Compute $c_e,d_e$ for every edge using the two measure oracles
 and Lemma~\ref{lem:majority}.
 \item Construct the interested-edge lists and retained path pairs.
 Search their full Monge rectangles, retaining a minimum entry and its
 two defining tree edges.
 \item Search pairs on each individual heavy path by the balanced
 recursion above, again retaining the best pair.
 \item Return the lightest candidate among the stored one-edge cut
 and all pairs found.  For a pair $e,f$, output
 $A_e\mathbin{\triangle}A_f$; for a single edge $e$, output $A_e$.
\end{enumerate}

\begin{proof}[Proof of Theorem~\ref{thm:2respecting}]
Every candidate returned by a search is a valid nontrivial cut crossing
one or two tree edges, so no candidate is lighter than the optimum
$\OPT$ among those cuts.  If $\OPT=\lambda_1$, the stored one-edge
cut is optimal.  Otherwise an optimal cut is determined by distinct
edges $e,f$ and has $K(e,f)=\OPT<\lambda_1$.
If $e,f$ lie on the same heavy path, their pair occurs in a rectangle
of the same-path recursion.  If they lie on different paths, then
Lemma~\ref{lem:interest} puts them in the two lists of a retained
pair, and the splitting step puts them together in one of its
rectangles.  That rectangle's minimum is at most $K(e,f)$ and is
itself a feasible cut.  In either case the algorithm finds a cut of
weight $\OPT$.

The query bounds are
\begin{center}
\begin{tabular}{@{}ll@{}}
\toprule
Stage & Cut queries\\
\midrule
One-edge cut values & $n-1$\\
All interest-path endpoints & $O(n\log n)$\\
Heavy paths, lists, and rectangle splits & $0$\\
Pairs on different heavy paths & $O(n\log n)$\\
Pairs on the same heavy path & $O(n\log n)$\\
Recovering the chosen cut side & $0$\\
\bottomrule
\end{tabular}
\end{center}
Their sum is $O(n\log n)$.
\end{proof}

\subsection{All minimum cuts respecting a tree}\label{sec:all-2respecting}

The algorithm above discards ties on purpose: it only has to consider
two-edge cuts strictly lighter than $\lambda_1$.  To report every cut
at a given level, three changes are needed: a search for the weakly
heavy vertices of a measure, a weak form of interest that captures
ties, and a listing of all minima of a Monge matrix.

\subsubsection{Weak heaviness and weak interest}

For a nonnegative measure $\mu$ of total mass $D>0$, call $v$
\emph{weakly heavy} if $\mu(T_v)\geq D/2$.  Unlike heavy vertices, two
incomparable vertices can both be weakly heavy: a root with two
children each carrying mass $D/2$.  Every ancestor of a weakly heavy
vertex is weakly heavy.

\begin{lemma}[Weak majority search]\label{lem:weak-majority}
Let $\mu$ be an additive nonnegative measure with known total mass
$D>0$ that can be evaluated on any specified set.  The weakly heavy
vertices are the vertices of two root-to-vertex paths, and their
endpoints can be found deterministically with $O(\log n)$
evaluations.
\end{lemma}

\begin{proof}
Use the preorder and the prefix masses $F(k)$ of \cref{lem:majority}.
Binary search finds the least $k_1$ with $F(k_1)\geq D/3$ and the
least $k_2$ with $F(k_2)\geq2D/3$.  A subtree occupying the interval
$[\ell,u]$ that contains neither $k_1$ nor $k_2$ has mass less than
$D/3$ if $u<k_1$, mass $F(u)-F(\ell-1)<2D/3-D/3$ if $k_1<\ell$ and
$u<k_2$, and mass at most $D-F(\ell-1)\leq D/3$ if $\ell>k_2$.  Since
$D>0$, these are all below $D/2$.  So the root of every weakly heavy
subtree is an ancestor of $v_{k_1}$ or of $v_{k_2}$.  Subtree masses
are nonincreasing along each root path and the root is weakly heavy,
so a binary search along each of the two root paths finds the deepest
weakly heavy vertex on it.
\end{proof}

Let $C'_e$ and $D'_e$ be the sets of weakly heavy vertices for
$\mu_e^{\out}$ and $\mu_e^{\inn}$, each a union of at most two root
paths, and say that $e$ is \emph{weakly interested in} $f$ if the
lower endpoint of $f$ lies in $C'_e\cup D'_e$.  Since this set is a
union of at most four root paths, \eqref{eq:interest-count} holds for
weak interest with a larger constant, and \cref{lem:weak-majority}
finds the four endpoints with $O(\log n)$ cut queries per edge.

\begin{lemma}[Weak interest captures every tie]\label{lem:weak-interest}
If $a_e,a_f>0$ and $K(e,f)\leq\min\{a_e,a_f\}$, then $e$ is weakly
interested in $f$ and $f$ is weakly interested in $e$.
\end{lemma}

\begin{proof}
This is the proof of \cref{lem:interest} with strict inequalities
relaxed.  If $e,f$ are independent, $K(e,f)\leq a_f$ and
\eqref{eq:independent} give $2w(A_e,A_f)\geq a_e$, that is,
$\mu_e^{\out}(A_f)\geq a_e/2$, so the lower endpoint of $f$ is weakly
heavy for $\mu_e^{\out}$; interchanging $e$ and $f$ gives the other
direction.  If $f$ is below $e$, $K(e,f)\leq a_f$ and \eqref{eq:nested}
give $2w(A_f,V\setminus A_e)\geq a_e$, that is,
$\mu_e^{\inn}(A_f)\geq a_e/2$, while $A_f\subset A_e$ gives
$\mu_f^{\inn}(A_e)=a_f\geq a_f/2$.
\end{proof}

\subsubsection{All minima of a Monge matrix}

\begin{lemma}[Opposite minimum corners]\label{lem:opposite-corners}
Let $M$ be Monge, $\ell\leq M[i,j]$ for all entries, $i<k$,
$l\leq j\leq r$, and $M[i,r]=M[k,l]=\ell$.  Then $M[i,j]=M[k,j]$.
\end{lemma}

\begin{proof}
If $l=r$ then $j=l=r$ and both entries equal $\ell$.  Otherwise the
Monge inequality for rows $i<k$ and columns $l<r$ gives
$M[i,l]+M[k,r]\leq M[i,r]+M[k,l]=2\ell$, so $M[i,l]=M[k,r]=\ell$.  If
$j<r$, the inequality for columns $j<r$ gives $M[i,j]\leq M[k,j]$; if
$j=r$ both entries equal $\ell$.  Symmetrically, columns $l<j$ give
$M[k,j]\leq M[i,j]$, and $j=l$ gives equality directly.
\end{proof}

\begin{lemma}[All minima of a Monge matrix]\label{lem:monge-all}
Let $M$ be a full $a\times b$ Monge matrix and $\ell$ a known number
with $\ell\leq M[i,j]$ for all entries.  The set of entries equal to
$\ell$ can be listed deterministically with $O(a+b)$ entry
evaluations, independently of its size.
\end{lemma}

\begin{proof}
Find the leftmost minimum $L_i$ of every row by SMAWK, and the
rightmost minimum $R_i$ by SMAWK applied with both the row and the
column orders reversed, which keeps the Monge property (reversing only
the columns would not).  Evaluate $M[i,L_i]$ for every row and call the
row \emph{active} if the value is $\ell$.  For each column $j$, let
$A_j$ be the active rows with $L_i\leq j\leq R_i$.  If
$A_j\neq\varnothing$, evaluate one entry $M[i_0,j]$ with $i_0\in A_j$
and report all of $A_j\times\{j\}$ if it equals $\ell$.  By
\cref{lem:opposite-corners}, all entries of $A_j\times\{j\}$ are
equal, so every reported entry is $\ell$.  Conversely, if
$M[i,j]=\ell$, then row $i$ is active and $L_i\leq j\leq R_i$, so
$(i,j)$ is reported.  The evaluations are two SMAWK runs, $a$ row
values and at most $b$ column representatives.
\end{proof}

The minima of one row need not be contiguous: two equal rows
$(0,1,0)$ form a Monge matrix whose row minima are separated by a
gap.  The representative evaluation in each column is what handles
this.

\subsubsection{The reporting theorem}

\begin{theorem}[All two-respecting cuts at a lower level]\label{thm:all-2respecting}
Given an explicit tree $T$ on $V(G)$ and a number $\ell>0$ such that
every cut crossing at most two edges of $T$ has weight at least
$\ell$, all such cuts of weight exactly $\ell$ can be listed
deterministically with $O(n\log n)$ cut queries, independently of
their number.
\end{theorem}

\begin{proof}
Run the algorithm of this section with three changes.  First, in
step~2 compute the weakly heavy paths of \cref{lem:weak-majority} for
both measures of every edge, and build the interested-edge lists and
retained path pairs from weak interest.  Second, in steps~3 and~4
replace SMAWK by \cref{lem:monge-all} at level $\ell$ in every full
Monge rectangle, including those of the same-path recursion; every
entry of such a rectangle is a cut crossing two tree edges, so $\ell$
is a lower bound on all entries, as that lemma requires.  Third,
report $A_e$ for every one-edge cut whose value $a_e$, queried in
step~1, equals $\ell$, and report $A_e\mathbin\triangle A_f$ for every
pair $(e,f)$ listed by \cref{lem:monge-all}; no candidate is queried
again.  Distinct listed pairs give distinct shores, since a shore
crossing exactly the tree edges $e$ and $f$ determines the pair, so
no shore is listed twice.

Every reported shore is a cut crossing at most two tree edges whose
weight is exactly $\ell$.  Conversely, let $e\neq f$ with
$K(e,f)=\ell$.  Then $a_e,a_f\geq\ell=K(e,f)$ and $a_e,a_f>0$, so
\cref{lem:weak-interest} makes $e$ and $f$ weakly interested in each
other.  If they lie on different heavy paths, they belong to the two
lists of a retained pair and the splitting step puts them in one
rectangle; if they lie on the same path, the recursion separates them
in one rectangle.  In both cases \cref{lem:monge-all} lists the pair.
One-edge cuts of weight $\ell$ are reported directly.  The query
counts are those of the table above with larger constants:
$O(\log n)$ evaluations per edge for the four path endpoints, and
$O(a+b)$ evaluations per $a\times b$ rectangle.
\end{proof}

\begin{corollary}\label{cor:all-2respecting}
\begin{enumerate}
\item With $\ell=\lambda(G)>0$, the algorithm lists every minimum cut
  of $G$ that crosses at most two edges of $T$.
\item If every cut crossing at most two edges of $T$ has positive
  weight, for instance if $\lambda(G)>0$, all cuts of least weight
  among them can be listed with $O(n\log n)$ queries: take $\ell$ to
  be the value found by \cref{thm:2respecting}.
\end{enumerate}
\end{corollary}

\begin{proof}
For the first part, every cut of $G$ has weight at least $\lambda(G)$,
so the hypothesis of \cref{thm:all-2respecting} holds with
$\ell=\lambda(G)$.  For the second, the least weight $\ell$ of a cut
crossing at most two edges is found by \cref{thm:2respecting} with
$O(n\log n)$ queries and is positive by assumption, so
\cref{thm:all-2respecting} applies.
\end{proof}

\section*{Statement on AI use}

I started with the problem of finding an upper bound on the certificate
complexity of global minimum cut.  Before using AI, I had a proof of an
$\widetilde O(n^{3/2})$ upper bound on certificate complexity for the complete
graph, based on a two-case argument handling small and large shores
separately.  I gave this proof to GPT-6 Astra, which produced an
$\widetilde O(n^{3/2})$ upper bound for a general graph, again using a
two-case argument.  As part of this argument, it developed an $O(n)$
certificate for the maximum $s$--$t$ flow value.  When I saw this, I suggested
applying it within the isolating-cuts framework of Li and
Panigrahi~\cite{LP25}. Astra then came up with the terminal-halving argument 
to obtain the $O(n\log^2 n)$ certificate.

My work on the deterministic algorithm for global minimum cut began
before I used AI, in a conversation with Pawe{\l} Gawrychowski.  He told
me that the minimum two-respecting-cut algorithm in his joint work with
Shay Mozes and Oren Weimann~\cite{GMW} also worked deterministically in the cut query model, 
using $\widetilde O(n)$ queries.  I wondered whether this could form the basis of a
deterministic cut-query algorithm, rather than the flow-based approach
of Anand, Saranurak, and Wang~\cite{ASW}.  I asked GPT-6 Astra to try this
approach.  It quickly wrote up the minimum 2-respecting cut proof in the cut query model and understood the obstacle
of coming up with a small set of trees such that a minimum cut of the input graph 2-respected at
least one of them.  With further guided prompting, it produced the argument to construct a weak skeleton.

I used Claude Opus 5.5 and GPT-6 Astra to assist with drafting and reviewing the manuscript.
The Lean formalization was done with Claude Opus 5.5 as the main driver and using GPT-6 Astra
to review the code and documentation after each session.

\newcommand{\etalchar}[1]{$^{#1}$}

\clearpage
\appendix
\crefalias{section}{appendix}\crefalias{subsection}{subappendix}
\section{Correspondence with the Lean formalization}\label{app:lean}

The Lean library is at
\url{https://github.com/troyjlee/cut-query-complexity}, and the
correspondence below refers to its version~0.1.0~\cite{Lee26}.  Declaration
names below are relative to the namespace \lean{CutQueryComplexity}.
Every algorithmic statement is a decision tree (\lean{DTree}) or a
query program (\lean{Search.Comp}) selected from $n$ and the public
parameters before the hidden graph; its query cap holds for every
answer function, and its correctness for every nonnegative weighted
graph on the stated vertex set.  Certificate statements quantify over
every nonnegative weighted competitor on the same vertex set.  The
library's integer logarithms are
$L(n)=\max\{1,\lceil\log_2(2n)\rceil\}$ (\lean{DetMinCut.logL}) and
$K(n)=L(L(n))$; the statements in natural logarithms follow from
$L(n)\leq3\log(2n)$ and $K(n)\leq7\log\log(2n)$ for $n\geq2$.
Each endpoint is restated literally and audited in
\lean{CutQueryComplexityChecks/Endpoints.lean}; the only axioms in
their transitive closure are \lean{propext}, \lean{Classical.choice}
and \lean{Quot.sound}.

The quantity $\cert(G)$ of \eqref{eq:certificate-definition} is
\lean{Certificate.certificateComplexity}, the least cardinality of a
family of shores satisfying \lean{Certificate.IsMinCutCertificate};
the minimum is attained
(\lean{Certificate.exists_certificate_card_eq}), and for $n\geq2$ the
empty family certifies nothing
(\lean{Certificate.one_le_certificateComplexity}).  The lower bound
$\cert(G)\leq D(\lambda)$ on the deterministic cut-query complexity of
the minimum-cut value is
\lean{Certificate.certificateComplexity_le_minCutValueComplexity},
with the instance-wise form
\lean{Certificate.certificateComplexity_le_cost}: a correct decision
tree asks at least $\cert(G)$ shores on input $G$.
\Cref{rem:sparse,rem:calibrate} are not formalized: no Lean statement
bounds the number of edges of a skeleton, and all constants are
explicit, so no calibration is needed.

{\small
\begin{longtable}{@{}p{2.3cm}p{7.2cm}p{6.1cm}@{}}
\toprule
Paper & Lean & Notes\\
\midrule
\endfirsthead
\toprule
Paper & Lean & Notes\\
\midrule
\endhead
\bottomrule
\endlastfoot
\eqref{eq:certificate-definition}, $\cert(G)$ &
\lean{Certificate.certificateComplexity}\newline
\lean{Certificate.one_le_certificateComplexity}\newline
\lean{Certificate.certificateComplexity_le_minCutValueComplexity} &
Least size of a value certificate; attained, $\cert(G)\geq1$ for
$n\geq2$, and $\cert(G)\leq D(\lambda)$.\\
\cref{thm:certificates} &
\lean{Certificate.exists_minCut_witnessCertificate}\newline
\lean{Certificate.exists_minCut_value_certificate}\newline
\lean{Certificate.certificateComplexity_le_bound}\newline
\lean{Certificate.certificateComplexity_eq_one} &
$18n\lceil\log_2(2n)\rceil^2$ shores; the witness shore is in the
family.  $\cert(G)\leq18n\lceil\log_2(2n)\rceil^2$, and
$\cert(G)=1$ when $\lambda(G)=0$.\\
\cref{lem:flow} &
\lean{Certificate.TreePattern.public_flow_patterns}\newline
\lean{Certificate.TreePattern.exists_terminal_tree_certificate}\newline
\lean{Certificate.exists_cutCert} &
$4n-1$ shores fixed by a labelled tree; $4n$ for an $s$--$t$ shore and
value; $8n+1$ for terminal sets $A,B$.\\
\cref{def:weak-skeleton} &
\lean{AllMinCuts.AllMinCutSkeleton}\newline
\lean{TwoRespecting.IsWeakSkeleton} &
The first is the definition with $111/100$ in place of $5/4$:
$\lambda(G)\leq\lambda(H)$ and $c_H(S)\leq\frac{111}{100}\lambda(G)$
for every minimum cut $S$ of $G$.  The second is the weaker form used
by the single-shore theorems, $\lambda(H)>0$ and
$4c_H(S^*)\leq5\lambda(H)$ for some minimum cut $S^*$, which suffices
for the final search.\\
\cref{thm:all-mincuts}, one shore &
\lean{UndirectedClosure.deterministic_mincut_shore_value_log3loglog2}\newline
\lean{DetMinCut.deterministic_mincut_log3loglog2_real_log}\newline
\lean{TreeCutSparsifier.deterministic_mincut_two_stage} &
A single minimum-cut shore and the exact value; cap
$Cn(\lfloor\sqrt n\rfloor+1)L(n)^3K(n)^2$.
The last is a second proof through
\cref{thm:tree-sparsifier-discovery,thm:conversion-overview}.\\
\cref{thm:all-mincuts} &
\lean{AllMinCuts.deterministic_all_mincuts_log3loglog2}\newline
\lean{AllMinCuts.deterministic_all_mincuts_log3loglog2_real_log} &
Output \lean{out} with \lean{out.value} $=\lambda(G)$ and
$S\in$ \lean{out.cuts} $\iff S$ is a minimum-cut shore
(\lean{AllMinCuts.Correct}); one decision tree per $n$ with the same
cap.  When $\lambda(G)=0$ the Lean output
lists all $2^c-2$ shores explicitly rather than the $c$ components;
the two are equivalent by \lean{AllMinCuts.cut_eq_zero_iff_sat}.  The
quotient is handled by \lean{AllMinCuts.isMinCut_iff_exists_quot} and
\lean{AllMinCuts.Correct.pullOutput}, and the final search by
\lean{AllMinCuts.coverReport_spec}; the composed program is
\lean{AllMinCuts.allTwoStageAuto_spec}.\\
\cref{def:tree-cut-sparsifier} &
\lean{TreeCutSparsifier.TreeData.Approximates}\newline
\lean{TreeCutSparsifier.TreeData.Good} &
The second adds canonical capacities, childless terminals, and at most
$8n$ nodes.\\
\cref{thm:conversion-overview} &
\lean{TreeCutSparsifier.exists_conversion_log3loglog2_real_log}\newline
\lean{AllMinCuts.exists_conversion_allMinCuts} &
The input is an explicit tree with arbitrary nonnegative capacities
whose representation costs approximate every cut of $G$ within $B$;
no bound on its size is needed, and the program computes the normal
form and queries the canonical capacities itself.  The first returns
the weaker \lean{IsWeakSkeleton} form; the second, in integer form,
returns a weak skeleton (\lean{AllMinCuts.AllMinCutSkeleton}), with
$111/100$ in place of $5/4$.\\
\cref{thm:tree-sparsifier-discovery} &
\lean{TreeCutSparsifier.theorem5LL}\newline
\lean{TreeCutSparsifier.learn_tree_cut_sparsifier_LL}\newline
\lean{TreeCutSparsifier.learnLL_spec} &
Integer form: $2\cdot3892968\,L(n)^2K(n)^2\leq\alpha$ and
$\alpha^2\leq n$.  The output is a safe partition and a \lean{Good}
tree of the quotient.\\
\cref{lem:row-search} &
\lean{Discovery.residualRow_spec} &
$2b\lceil\log_2(N-1)\rceil$ queries, for residual rows of $G-K$.\\
\cref{lem:forest} &
\lean{SupportForest.exists_support_forest}\newline
\lean{SupportForest.exists_connectivity_test} &
$3n+4n\lceil\log_2 n\rceil\leq7nL(n)$ queries.\\
\cref{rem:local-flow} &
\lean{Certificate.exists_flowCert} &
$8|W|$ shores.\\
\cref{lem:isolating} &
\lean{Certificate.exists_isoCert} &
$9n(\lceil\log_2|R|\rceil+1)$ shores.\\
\cref{lem:splitting} &
\lean{Certificate.exists_splitOff} &
Literal.\\
\cref{lem:halving} &
\lean{Certificate.exists_halving}\newline
\lean{Certificate.exists_halvingCert} &
$8n$ shores.\\
\cref{rem:sparse} & --- & Not formalized.\\
\cref{prop:bridge} &
\lean{UndirectedClosure.certificate_to_skeleton}\newline
\lean{UndirectedClosure.witnessCertificate_to_skeleton} &
Zero further queries
(\lean{UndirectedClosure.sparseCompletionProg_cost}).\\
\cref{lem:light} &
\lean{TwoRespecting.exists_light_spanning} &
Written over ordered pairs:
$(\sum_{e\in E}\ell_e)\lambda\leq\sum_{a,b}w_H(a,b)\ell_{ab}$.\\
\cref{lem:tree-cover} &
\lean{TwoRespecting.exists_tree_cover} &
Exactly $5(\lceil\log_2 n\rceil+3)$ trees.\\
\cref{rem:packing-dual} & --- & Commentary on \cref{lem:light}.\\
\cref{thm:2respecting} &
\lean{TwoRespecting.twoRespecting_queryBound}\newline
\lean{TwoRespecting.twoRespectingProg_cost} &
$72n(\lceil\log_2 n\rceil+1)$ queries.\\
\cref{cor:skeleton-search} &
\lean{TwoRespecting.minCut_of_weakSkeleton}\newline
\lean{TwoRespecting.weakSkeleton_minCut} &
$q+5(\lceil\log_2 n\rceil+3)\bigl(72n(\lceil\log_2 n\rceil+1)+1\bigr)$
queries.\\
\cref{thm:depthfree} &
\lean{Hierarchy.Hier.supplied_weakSkeleton} with the guard
\lean{Hierarchy.Hier.ConstantCutGuard.guardFor} of
\lean{Hierarchy.constantCutGuard_42} &
$|\cA|\bigl(1+4(q-1)(\lfloor\log_2 n\rfloor+3)\bigr)$ queries with
$q=\lceil200\rho^2B^2\rceil$ and $\rho=42$.  Promise for some minimum
cut and the weaker \lean{IsWeakSkeleton} conclusion; the pointwise
form is \lean{AllMinCuts.refine_preserves_cut}.\\
\cref{lem:quantile} &
\lean{Discovery.quantileCheckpoints_spec} &
Interval form $2M/q$ only.\\
\cref{lem:depthfree-completion} &
\lean{Hierarchy.Hier.completion_spec} &
$\eta=2/q$; the edge count of $F$ is not stated.\\
\cref{lem:R-error} &
\lean{Hierarchy.Hier.cut_error_le}\newline
\lean{Hierarchy.Hier.cut_le_rep} &
Literal.\\
\cref{lem:tree-retraction} &
\lean{Hierarchy.treeZeroExt_42} &
For hierarchies whose terminals are childless nodes.\\
\cref{lem:guard} &
\lean{Hierarchy.constantCutGuard_42} &
$\rho_0=42$, literal.\\
\cref{rem:schur} &
\lean{Hierarchy.Hier.guardFor_of_leafChildless} &
Quality twice the height.\\
\cref{lem:offline-hierarchy} &
\lean{Hierarchy.RackeShah.rackeShah_hierarchyBound} &
$a_0=268$, at most $2r+1$ nodes, height at most $2L(r)$.\\
\cref{lem:hierarchy-transfer} &
\lean{Hierarchy.Hier.transfer_cut}\newline
\lean{Hierarchy.Hier.transfer_rep} &
Literal.\\
\cref{lem:refinement-round} &
\lean{Hierarchy.Hier.round_promise}\newline
\lean{Hierarchy.refine_spec} &
Literal.\\
\cref{thm:linear-hierarchy} &
\lean{UndirectedClosure.supplied_hierarchy_log3loglog2}\newline
\lean{UndirectedClosure.suppliedHierarchyProgLL_spec} &
Hierarchies with at most $2n+1$ nodes and childless terminals; the
program queries the canonical capacities itself.  These take the
promise for some minimum cut and return the weaker
\lean{IsWeakSkeleton} form; the statement of the paper, with the
promise for every minimum cut, is
\lean{AllMinCuts.suppliedHierarchyProgLL_allMinCuts} in the next row.
Larger supplied hierarchies are covered by
\lean{TreeCutSparsifier.exists_conversion_log3loglog2_real_log},
which has no size bound.\\
\cref{rem:all-promise} &
\lean{AllMinCuts.refine_preserves_cut}\newline
\lean{AllMinCuts.suppliedHierarchyProgLL_allMinCuts}\newline
\lean{AllMinCuts.conversion_allMinCuts} &
Same program and budget as \cref{thm:linear-hierarchy}, with the
pointwise argument of the remark; the constant is $111/100$ in place
of $5/4$.  \lean{AllMinCuts.AllMinCutSkeleton.covers} is
\cref{lem:tree-cover} applied to a weak skeleton.\\
\cref{rem:calibrate} & --- & Not formalized.\\
\cref{lem:tree-compression} &
\lean{Discovery.tree_compression} &
Literal.\\
\cref{lem:degree-normalization} &
\lean{Discovery.heavyNeighbor_spec}\newline
\lean{DetMinCut.normalize_spec} &
$2\lceil\log_2(N-1)\rceil$ queries per contraction; amortized total
$(2\lceil\log_2 n\rceil+1)n$.\\
\cref{lem:normalization-halves} &
\lean{Discovery.normalization_halves} &
Literal.\\
\cref{lem:residual-preparation} &
\lean{Discovery.residualPrep_normalized} &
$N(4+3\lceil\log_2N\rceil)\cdot4k\lceil\log_2(N-1)\rceil$ queries.\\
\cref{lem:bucketed-centers} &
\lean{Discovery.centerSelection_normalized} &
At most $\lceil(8N/k)\ln(2N)\rceil$ centers.\\
\cref{lem:general-auxiliary} &
\lean{Discovery.psi_lower}\newline
\lean{Discovery.psi_upper}\newline
\lean{Discovery.auxGraph} &
Literal.\\
\cref{lem:general-certificate} &
\lean{TreeCutSparsifier.certifyLL}\newline
\lean{TreeCutSparsifier.phaseLL_spec} &
The certified hierarchy is not compressed and has at most $8N$
nodes.\\
\cref{rem:log-saving}, \cref{rem:guard-independent} & --- &
Commentary.  The height-guard route is
\lean{DetMinCut.deterministic_mincut_log5}.\\
\cref{lem:identities} &
\lean{WeightedGraph.edge_eq_three_cuts}\newline
\lean{TwoRespecting.cut_union_of_disjoint}\newline
\lean{TwoRespecting.cut_sdiff_of_subset} &
Literal.\\
\cref{lem:interest} &
\lean{TwoRespecting.interest_indep}\newline
\lean{TwoRespecting.interest_nested_below}\newline
\lean{TwoRespecting.interest_nested_above} &
Literal.\\
\cref{lem:majority} &
\lean{TwoRespecting.majoritySearch_spec} &
$4(\lfloor\log_2 n\rfloor+1)$ queries.\\
\cref{lem:monge} &
\lean{TwoRespecting.pairCut_monge_indep}\newline
\lean{TwoRespecting.pairCut_monge_nested} &
Literal.\\
\cref{lem:smawk} &
\lean{Search.smawk_spec}\newline
\lean{Search.matrixMin_spec} &
Proved rather than cited: $4|C|+12|R|$ entry evaluations for all row
minima; $4|C|+13|R|$ for a minimum entry of the matrix.\\
\cref{lem:weak-majority} &
\lean{TwoRespecting.weakMajoritySearch_spec}\newline
\lean{TwoRespecting.weak_heavy_cover} &
Stated for sums of nonnegative vertex weights; at most
$8(\lfloor\log_2 n\rfloor+1)$ evaluations.\\
\cref{lem:weak-interest} &
\lean{TwoRespecting.mutual_weak_interest} &
Literal.\\
\cref{lem:opposite-corners}, \cref{lem:monge-all} &
\lean{Search.monge_equal_between_opposite_minima}\newline
\lean{Search.allMinima_spec}\newline
\lean{Search.monge_allMinima_queryBound} &
At most $9b+25a$ entry evaluations.\\
\cref{thm:all-2respecting} &
\lean{TwoRespecting.allTwoRespecting_correct_of_le}\newline
\lean{TwoRespecting.allTwoRespecting_spec} &
At most $250n(\lceil\log_2 n\rceil+1)$ queries.\\
\cref{cor:all-2respecting} &
\lean{TwoRespecting.allMinTwoRespecting_spec}\newline
\lean{TwoRespecting.allMinTwoRespecting_queryBound} &
At most $322n(\lceil\log_2 n\rceil+1)$ queries; the level comes from
the existing search with no extra query.\\
\end{longtable}
}


\begin{thebibliography}{GPRW20}

\bibitem[ADK26]{ADK26}
Daniel Agassy, Dani Dorfman, and Haim Kaplan.
\newblock Improved tree sparsifiers in near-linear time.
\newblock In \emph{53rd International Colloquium on Automata, Languages, and
  Programming (ICALP 2026)}, volume 374 of \emph{LIPIcs}, pages 7:1--7:23,
  2026.
\newblock \href{https://doi.org/10.4230/LIPIcs.ICALP.2026.7}{doi:10.4230/LIPIcs.ICALP.2026.7}.

\bibitem[AEG{\etalchar{+}}22]{AEG+22}
Simon Apers, Yuval Efron, Pawe{\l} Gawrychowski, Troy Lee, Sagnik Mukhopadhyay,
  and Danupon Nanongkai.
\newblock Cut query algorithms with star contraction.
\newblock In {\em 2022 IEEE 63rd Annual Symposium on Foundations of Computer
  Science (FOCS)}, pages 507--518. IEEE, 2022.

\bibitem[AKM{\etalchar{+}}87]{SMAWK}
Alok Aggarwal, Maria~M. Klawe, Shlomo Moran, Peter~W. Shor, and Robert Wilber.
\newblock Geometric applications of a matrix-searching algorithm.
\newblock {\em Algorithmica}, 2:195--208, 1987.

\bibitem[AL21]{AL21}
Arinta Auza and Troy Lee.
\newblock On the query complexity of connectivity with global queries.
\newblock \href{https://arxiv.org/abs/2109.02115}{arXiv:2109.02115}, 2021.

\bibitem[ASW25]{ASW}
Aditya Anand, Thatchaphol Saranurak, and Yunfan Wang.
\newblock Deterministic edge connectivity and max flow using subquadratic cut
  queries.
\newblock In {\em Proceedings of the 2025 Annual ACM-SIAM Symposium on Discrete
  Algorithms (SODA)}, pages 124--142. SIAM, 2025.

\bibitem[BSS12]{BSS}
Joshua Batson, Daniel~A. Spielman, and Nikhil Srivastava.
\newblock Twice-{Ramanujan} sparsifiers.
\newblock {\em SIAM Journal on Computing}, 41(6):1704--1721, 2012.

\bibitem[Ber91]{Bertsekas91}
Dimitri~P. Bertsekas.
\newblock \href{https://www.mit.edu/~dimitrib/LNets_Full_Book.pdf}{\emph{Linear
  Network Optimization: Algorithms and Codes}}.
\newblock MIT Press, Cambridge, MA, 1991.

\bibitem[EGK{\etalchar{+}}14]{EGKRTT}
Matthias Englert, Anupam Gupta, Robert Krauthgamer, Harald R{\"a}cke, Inbal
  Talgam-Cohen, and Kunal Talwar.
\newblock Vertex sparsifiers: New results from old techniques.
\newblock {\em SIAM Journal on Computing}, 43(4):1239--1262, 2014.

\bibitem[GK00]{GK00}
Vladimir Grebinski and Gregory Kucherov.
\newblock Optimal reconstruction of graphs under the additive model.
\newblock {\em Algorithmica}, 28(1):104--124, 2000.

\bibitem[GMW21]{GMW}
Pawe{\l} Gawrychowski, Shay Mozes, and Oren Weimann.
\newblock A note on a recent algorithm for minimum cut.
\newblock In {\em Symposium on Simplicity in Algorithms (SOSA 2021)}, pages
  74--79. SIAM, 2021.

\bibitem[GPRW20]{GPRW20}
Andrei Graur, Tristan Pollner, Vidhya Ramaswamy, and S.~Matthew Weinberg.
\newblock New query lower bounds for submodular function minimization.
\newblock In {\em 11th Innovations in Theoretical Computer Science Conference
  (ITCS 2020)}, volume 151 of {\em Leibniz International Proceedings in
  Informatics (LIPIcs)}, pages 64:1--64:16. Schloss Dagstuhl--Leibniz-Zentrum
  f{\"u}r Informatik, 2020.

\bibitem[Kar00]{Karger00}
David~R. Karger.
\newblock Minimum cuts in near-linear time.
\newblock {\em Journal of the ACM}, 47(1):46--76, 2000.

\bibitem[KM26]{KM26}
Yotam Kenneth-Mordoch.
\newblock Deterministic global minimum cut using {$\widetilde{O}(n^{3/2})$} cut
  queries.
\newblock \href{https://arxiv.org/abs/2610.23461}{arXiv:2610.23461}, 2026.

\bibitem[Lee26]{Lee26}
Troy Lee.
\newblock Cut query complexity in {L}ean, version 0.1.0.
\newblock Lean~4 library, 2026.
\newblock \href{https://doi.org/10.5281/zenodo.23885654}{doi:10.5281/zenodo.23885654}.

\bibitem[LLSZ21]{LLSZ21}
Troy Lee, Tongyang Li, Mikl{\'o}s Santha, and Shengyu Zhang.
\newblock On the cut dimension of a graph.
\newblock In {\em 36th Computational Complexity Conference (CCC 2021)}, volume
  200 of {\em Leibniz International Proceedings in Informatics (LIPIcs)}, pages
  15:1--15:35. Schloss Dagstuhl--Leibniz-Zentrum f{\"u}r Informatik, 2021.

\bibitem[Lov76]{Lov76}
L{\'a}szl{\'o} Lov{\'a}sz.
\newblock On some connectivity properties of {E}ulerian graphs.
\newblock {\em Acta Mathematica Academiae Scientiarum Hungaricae}, 28:129--138,
  1976.

\bibitem[LP25]{LP25}
Jason Li and Debmalya Panigrahi.
\newblock Deterministic minimum cut in poly-logarithmic maximum flows.
\newblock {\em Journal of the ACM}, 72(4):29:1--29:18, 2025.

\bibitem[Mad78]{Mad78}
Wolfgang Mader.
\newblock A reduction method for edge-connectivity in graphs.
\newblock In {\em Advances in Graph Theory}, volume~3 of {\em Annals of
  Discrete Mathematics}, pages 145--164. North-Holland, 1978.

\bibitem[MN20]{MN20}
Sagnik Mukhopadhyay and Danupon Nanongkai.
\newblock Weighted min-cut: Sequential, cut-query and streaming algorithms.
\newblock In {\em Proceedings of the 52nd Annual ACM SIGACT Symposium on Theory
  of Computing (STOC 2020)}, pages 496--509. ACM, 2020.

\bibitem[R{\"a}c02]{Racke02}
Harald R{\"a}cke.
\newblock Minimizing congestion in general networks.
\newblock In {\em Proceedings of the 43rd Annual IEEE Symposium on Foundations
  of Computer Science (FOCS 2002)}, pages 43--52. IEEE, 2002.

\bibitem[RS14]{RS14}
Harald R{\"a}cke and Chintan Shah.
\newblock Improved guarantees for tree cut sparsifiers.
\newblock In {\em Algorithms -- ESA 2014}, volume 8737 of {\em Lecture Notes
  in Computer Science}, pages 774--785. Springer, 2014.

\bibitem[RSW18]{RSW18}
Aviad Rubinstein, Tselil Schramm, and S.~Matthew Weinberg.
\newblock Computing exact minimum cuts without knowing the graph.
\newblock In {\em 9th Innovations in Theoretical Computer Science Conference
  (ITCS 2018)}, volume~94 of {\em Leibniz International Proceedings in
  Informatics (LIPIcs)}, pages 39:1--39:16. Schloss Dagstuhl--Leibniz-Zentrum
  f{\"u}r Informatik, 2018.

\end{thebibliography}
\end{document}